\documentclass[twocolumn,twocolappendix]{aastex701}
\usepackage{xspace}
\usepackage{amsmath}
\usepackage{bm}
\usepackage{soul}
\usepackage{booktabs}
\usepackage{longtable}

\newcommand{\chandra}{{Chandra}\xspace}
\newcommand{\xmmnewton}{{\em{XMM-Newton}}\xspace}
\newcommand{\xmm}{{XMM}\xspace}

\newcommand{\hst}{{HST}\xspace}
\newcommand{\ciao}{\textit{CIAO}\xspace}
\newcommand{\marx}{\textit{MARX}\xspace}
\newcommand{\saotrace}{\textit{SAOTrace}\xspace}
\newcommand{\sherpa}{\textit{Sherpa}\xspace}
\newcommand{\ObsID}{\texttt{ObsID}\xspace}

\newcommand{\kev}{\ensuremath{\;\mathrm{keV}}\xspace}
\newcommand{\kms}{\ensuremath{\mathrm{km\,s^{-1}}\xspace}
\newcommand{\km}{\ensuremath{\;\mathrm{km}}\xspace}
\newcommand{\pc}{\ensuremath{\;\mathrm{pc}}\xspace}
\newcommand{\kpc}{\ensuremath{\;\mathrm{kpc}}\xspace}
\newcommand{\cmthrees}{\ensuremath{\mathrm{cm}^3\,\mathrm{s}^{-1}}\xspace}
\newcommand{\AU}{\mbox{AU}\xspace}
\newcommand{\yr}{\ensuremath{\mathrm{yr}}\xspace}
\newcommand{\Myr}{\ensuremath{\mathrm{Myr}}}\xspace}
\newcommand{\degree}{$^{\circ}$\xspace}

\newcommand{\msol}{\ensuremath{\mbox{$\mathrm{M}_{\sun}$}}\xspace}

\newcommand{\xspec}{\texttt{Xspec}\xspace}

\newcommand{\epocha}{Epoch\,\textit{A}\xspace}
\newcommand{\epochb}{Epoch\,\textit{B}\xspace}

\newcommand{\qeratio}[2]{\ensuremath{\mathcal{Q}(#1;#2)}\xspace}

\shorttitle{N132D Forward shock expansion}
\shortauthors{Xi et al.}

\begin{document}

\title{Chandra Large Project Observations of the Supernova Remnant N132D: Measuring the Expansion of the Forward Shock}

\author[0000-0003-3350-1832]{Xi Long}
\affiliation{Department of Physics, The University of Hong Kong, Pokfulam Road, Hong Kong}
\affiliation{Center for Astrophysics $|$ Harvard \& Smithsonian, 60 Garden St., MS-3, Cambridge, MA 02138, USA}
\email{xilong@hku.hk}

\author[0000-0003-1415-5823]{Paul P. Plucinsky}
\affiliation{Center for Astrophysics $|$ Harvard \& Smithsonian, 60 Garden St., MS-3, Cambridge, MA 02138, USA}
\email{pplucinsky@cfa.harvard.edu}

\author[0000-0002-5115-1533]{Terrance J. Gaetz}
\affiliation{Center for Astrophysics $|$ Harvard \& Smithsonian, 60 Garden St., MS-3, Cambridge, MA 02138, USA}
\email{tgaetz@cfa.harvard.edu}

\author[0000-0002-3869-7996]{Vinay L.\ Kashyap}
\affiliation{Center for Astrophysics $|$ Harvard \& Smithsonian, 60 Garden St., MS-3, Cambridge, MA 02138, USA}
\email{vkashyap@cfa.harvard.edu}

\author[0000-0003-0890-4920]{Aya Bamba}
\affiliation{Department of Physics, Graduate School of Science,
The University of Tokyo, 7-3-1 Hongo, Bunkyo-ku, Tokyo 113-0033, Japan}
\affiliation{Research Center for the Early Universe, School of Science, The University of Tokyo, 7-3-1
Hongo, Bunkyo-ku, Tokyo 113-0033, Japan}
\affiliation{Trans-Scale Quantum Science Institute, The University of Tokyo, Tokyo  113-0033, Japan}
\email{bamba@phys.s.u-tokyo.ac.jp}

\author[0000-0003-2379-6518]{William P.\ Blair}
\affiliation{The William H. Miller III Department of Physics and Astronomy, Johns Hopkins University, 3400 N. Charles Street, Baltimore, MD 21218, USA}
\email{wpb@pha.jhu.edu}

\author[0000-0002-0394-3173]{Daniel Castro}
\affiliation{Center for Astrophysics $|$ Harvard \& Smithsonian, 60 Garden St., MS-3, Cambridge, MA 02138, USA}
\email{daniel.castro@cfa.harvard.edu}

\author[0000-0003-3462-8886]{Adam R. Foster}
\affiliation{Center for Astrophysics $|$ Harvard \& Smithsonian, 60 Garden St., MS-3, Cambridge, MA 02138, USA}
\email{afoster@cfa.harvard.edu}


\author[0000-0003-1413-1776]{Charles J. Law}
\altaffiliation{NASA Hubble Fellowship Program Sagan Fellow}
\affiliation{Department of Astronomy, University of Virginia, Charlottesville, VA 22904, USA}
\email{cjl8rd@virginia.edu}

\author[0000-0002-0763-3885]{Dan Milisavljevic}
\affiliation{Purdue University, Department of Physics and Astronomy, 525 Northwestern Ave, West Lafayette, IN 47907, USA}
\affiliation{Integrative Data Science Initiative, Purdue University, West Lafayette, IN 47907, USA}
\email{dmilisav@purdue.edu}

\author[0000-0002-3031-2326]{Eric Miller}
\affiliation{Kavli Institute for Astrophysics and Space Research, Massachusetts Institute of Technology, Cambridge, MA 02139, USA}
\email{milleric@mit.edu}

\author[0000-0002-7507-8115]{Daniel J.Patnaude}
\affiliation{Center for Astrophysics $\vert$ Harvard \& Smithsonian, 60 Garden Street, Cambridge, MA 02138, USA}
\email{dpatnaude@cfa.harvard.edu}

\author[0000-0001-5302-1866]{Manami Sasaki}
\affiliation{Dr. Karl Remeis Observatory, Erlangen Centre for Astroparticle Physics, Friedrich-Alexander-Universität Erlangen-Nürnberg, Sternwartstraße 7, 96049 Bamberg, Germany}
\email{manami.sasaki@fau.de}

\author[0000-0003-2062-5692]{Hidetoshi Sano}
\affiliation{Faculty of Engineering, Gifu University, 1-1 Yanagido, Gifu 501-1193, Japan}
\affiliation{Center for Space Research and Utilization Promotion (c-SRUP), Gifu University, 1-1 Yanagido, Gifu 501-1193, Japan}
\email{sano.hidetoshi.w4@f.gifu-u.ac.jp}

\author[0000-0003-3347-7094]{Piyush Sharda}
\affiliation{Leiden Observatory, Leiden University, PO Box 9513, NL-2300 RA Leiden, the Netherlands}
\email{sharda@strw.leidenuniv.nl}

\author[0000-0002-7502-0597]{Benjamin F. Williams}
\affiliation{Department of Astronomy, University of Washington, Box 351580, Seattle, WA 98195, USA}
\email{ben@astro.washington.edu}

\author[0000-0003-2063-381X]{Brian J. Williams}
\affiliation{NASA Goddard Space Flight Center, Code 662, Greenbelt, MD 20771, USA}
\email{brian.j.williams@nasa.gov}

\author[0000-0002-5092-6085]{Hiroya Yamaguchi}
\affiliation{ISAS/JAXA, 3-1-1 Yoshinodai, Chuo-ku, Sagamihara, Kanagawa 252-5210, Japan}
\email{yamaguchi@astro.isas.jaxa.jp}



\begin{abstract}
We present results from the Chandra X-ray Observatory Large Project (878~ks in 28 observations) of the Large Magellanic Cloud supernova remnant N132D. We measure the expansion of the forward shock in the bright southern rim to be $0\farcs10\pm0\farcs02$ over the $\sim14.5$~yr baseline, which corresponds to a velocity of $1620\pm400~\kms$ after accounting for several instrumental effects.  We measure an expansion of $0\farcs23\pm0\farcs02$ and a  shock velocity of $3840\pm260~\kms$ for two features in an apparent blowout region in the northeast. The emission-measure-weighted average temperature inferred from X-ray spectral fits to regions in the southern rim is $0.95\pm0.17$~keV, consistent with the electron temperature implied by the shock velocity after accounting for Coulomb equilibration and adiabatic expansion. In contrast, the emission-measure-weighted average temperature for the northeast region is $0.77\pm0.04$~keV, which is significantly lower than the value inferred from the shock velocity. We fit 1-D evolutionary models for the shock in the southern rim and northeast region, using the measured radius and propagation velocity  into a constant density and power-law profile circumstellar medium. We find good agreement with the age of $\sim2500$ years derived from optical expansion measurements for explosion energies of $1.5-3.0 \times 10^{51}\,\mathrm{erg}$, ejecta masses of $2-6 \,\msol$ and ambient medium densities of $\sim0.33-0.66$ $\mathrm{amu~cm}^{-3}$  in the south and $\sim0.01-0.02$ $\mathrm{amu~cm}^{-3}$ in the northeast assuming a constant density medium. These results are consistent with previous studies that suggested the progenitor of N132D was an energetic supernova that exploded into an pre-existing cavity.

\end{abstract}

\vfill


\section{Introduction} \label{sec:intro}

 Massive stars and their supernovae (SNe) are important contributors to the structure and evolution of the interstellar medium (ISM) \citep{cox1974,mckee1977,oey1996,silich2005,Gent2013,kim2015,kim2017,brady2019} and determine the chemical evolution of galaxies \citep{kobayashi2006,nomoto2013}. 
They are the dominant sources of the warm and hot ionized gas in the ISM and a significant contributor to the turbulence in the ISM~\citep{avillez2004,joung2006}.
 In addition, most of the metals in the Universe are produced by nucleosynthesis in massive stars \citep{nomoto2006,sukhbold2016} and these products are distributed throughout the ISM by SNe and supernova remnants (SNRs)  \citep{hughes2000,bhalerao2019}. 

Massive stars form in dense clumps in molecular cloud (MC) complexes or OB associations \citep{zinnecker2007,2012A&A...545A.122P,2018MNRAS.475.5659W} as shown in  surveys of star-forming regions in the Milky Way \citep{motte2018,2022Univ....8..111C}. The winds of massive stars may be powerful enough to  excavate a cavity in the molecular cloud complex over the lifetime of the star  \citep{mckee1984,dwarkadas2023}, producing a region of low density, ionized or partially-ionized gas surrounded by a denser, neutral shell of material.
Stars more massive than 8\,\msol will end their lives as core-collapse SNe (CCSNe) or directly collapse to a black hole  \citep{heger2003}.  Given the relatively short lifetimes of these massive stars, it is likely that the star will be close to the molecular cloud in which it formed at the time of explosion \citep{chevalier1999,slane2016,2020NewAR..9001549W}.  
  
After the star explodes, the expanding shock from the supernova will propagate through the relatively low density in the cavity before encountering denser material at the edge of the cavity \citep{chevalier1989,tenorio-tagle1990,tenorio-tagle1991,dwarkadas2005}.  The structure of the surrounding medium may be further complicated by multiple episodes of eruptive mass loss \citep{dwarkadas2007,patnaude2015} leading up to the explosion.
The SNRs of such events will evolve differently than explosions into a homogeneous medium with relatively fainter emission while the shock is propagating in the low density medium followed by relatively brighter emission after the shock interacts with the denser  material at the edge of the cavity. 
The evolution of the SNR depends on the details of the interior of the cavity and the shell that defines it \citep{tenorio-tagle1990,dwarkadas2005}.  A strong stellar wind may create a low density cavity with a radius of $\sim15$~pc \citep{dwarkadas2007,chen2013} devoid of dense cores of molecular  material  while a weaker stellar wind might produce a smaller cavity or no cavity at all in which the dense cores from the MC complex survive passage of the forward shock of the SNR \citep{chevalier1999}. The observed characteristics of a SNR will differ dramatically depending on the structure of the medium that surrounded the star at the time of explosion which in turn depends on the stellar wind and mass-loss history of the progenitor.  Therefore, observations of SNRs in their current state provide important constraints on the type of star that exploded and how that star shaped its environment over its lifetime.

The Large Magellanic Cloud (LMC) supernova remnant N132D is the most luminous SNR in X-rays in the Local Group with an  X-ray luminosity of $L_\mathrm{X}(0.3--10.0~\mathrm{keV}) \sim 1\times10^{38}\,\mathrm{erg\,s^{-1}}$; only the SNR in NGC~4449 is more luminous in X-rays \citep{patnaude2003}. 
It was first classified as a core-collapse supernova (CCSN) by \cite{1966MNRAS.131..371W}, and has been subsequently studied in detail over the last few decades \citep{1997A&A...324L..45F,2008AdSpR..41..416X,2018ApJ...854...71B}. Based on optical observations, it has been classified as an Oxygen-rich remnant \citep{1976ApJ...207..394D,1978ApJ...223..109L,1980ApJ...237..765L}, thought to have exploded inside a low density cavity in the interstellar medium. \cite{1995ApJ...439..365S} discuss the origin of this cavity, which might have formed due to a wind bubble mechanism common to Wolf-Rayet stars \citep{dwarkadas2007}.  
\cite{1987ApJ...314..103H} analyzed the X-ray data from the {\em Einstein Observatory} to conclude that a cavity explosion model could provide reasonable values of the explosion energy and  age.
It has been proposed by \cite{2000ApJ...537..667B} that this remnant might be the outcome of a Type Ib supernova (core collapse) and is believed to be roughly $2500\,\mathrm{yr}$ old \citep{morse1995,1998ApJ...505..732H,2003ApJ...595..227C,2011Ap&SS.331..521V}.  \citet{law2020} performed a 3D reconstruction of the optically-emitting O ejecta to refine the age estimate to $2450\pm195\,\mathrm{yr}$ and \cite{banovetz2023}  derived a consistent age of $2770\pm500\,\mathrm{yr}$ from proper motion measurements of the O-rich ejecta based on multi-epoch observations with the {\em Hubble Space Telescope} (HST).

\textit{Chandra X-ray Observatory} (hereafter \chandra) observations \citep{2007ApJ...671L..45B} reveal a well-structured rim running along the southern part of the remnant.
This well-defined rim is associated with dense molecular clouds in this direction \citep{1997ApJ...480..607B,2015ASPC..499..257S,sano2020} and is also present in the infrared (IR) observations of dust continuum emission in N132D taken by \textit{Spitzer} \citep{2006ApJ...652L..33W}. 
N132D is a luminous GeV and TeV gamma-ray source \citep{ackermann2016,hess2021,vink2022n132d} with a location and spectrum that would be consistent with a hadronic origin due to an interaction with the molecular cloud complex.
The X-ray emission also shows a bright arc-shaped structure close to the outermost shell in the south and south-east that may be attributed to the reverse shock encountering the ejecta or face-on filaments produced by the forward shock interacting with density enhancements in the surrounding medium. \cite{sharda2020} presented maps based on the \chandra\, data in narrow energy bands centered on the prominent K shell emission lines of  O, Ne, Mg, Si, S, and Fe that showed the Fe-K emission is distributed throughout the southern half of the remnant with a particularly bright region close to, but interior to, the southeastern shell.  High resolution spectra from the X-ray calorimeter on {\textit{XRISM}} \citep{xrism2024} show that the Fe~He$\alpha$ emission is broader than the Si \& S emission and the Fe~Ly$\alpha$ emission has a significantly higher redshift than the local ISM in the LMC, both consistent with an ejecta origin for the Fe~K emission.  \xmmnewton~(\xmm) observations \citep{foster2025} demonstrate that a plasma with a temperature of $\sim4.5$~keV is required to explain the Fe~K emission and confirm the distribution of the Fe~K emission in the southern half of the remnant.

\cite{sano2020} detect evidence of shocked and clumpy CO material in the southern region and also towards the center of the remnant with the Atacama Large Millimeter/submillimeter Array (ALMA).
Towards the north, there are filament-like structures protruding outwards that are relatively faint in X-rays compared to the rest of the remnant. 
The overall shape of the remnant resembles an ellipse, but the northern regions deviate from an elliptical shape with a fragmented and even box-like morphology in one region.  The northern regions are significantly farther from the center of expansion (COE) as estimated by \citet{law2020} and \cite{banovetz2023}, consistent with a higher average expansion velocity for these regions over the lifetime of the remnant.  This morphology is consistent with a SNR shock encountering a roughly constant but higher density medium in the south and a lower density medium in the north.

Motivated by this morphology and the suggestion that N132D was produced by a SNe in a pre-existing cavity, \cite{2003ApJ...595..227C} modeled the evolution of a SNR shock across a density jump, in which the medium before the jump has a lower density to represent the cavity and the medium after the jump has a higher density to represent the material swept up by the stellar wind. They applied their semi-analytic model to N132D and found solutions that are consistent with an explosion energy of $\epsilon_\mathit{o}\sim3.0\times10^{51}~\mathrm{ergs}$, a pre-shock density of $n_\mathit{o}\sim3.0~\mathrm{cm^{-3}}$, an age of $\sim3,000$~yr, and a shock velocity of $v_s\sim1.9\times10^{3}~\mathrm{km~s^{-1}}$ before the shock encounters the cavity wall.  They suggested that the forward shock may have decelerated to a velocity of $\sim8.0\times10^{2}~\mathrm{km~s^{-1}}$ after encountering the cavity wall in order to match the analysis of the X-ray data discussed in \cite{morse1996}.  \cite{sharda2020} 
used the high angular resolution \chandra\/ data to extract spectra from narrow regions near the shock front to determine an average temperature of $kT = 0.85 \pm0.20\,\mathrm{keV}$ which corresponds to a shock velocity of $v_\mathit{s}\sim8.6\times10^{2}~\mathrm{km~s^{-1}}$ assuming full equilibration between electrons and ions.  They also estimated a progenitor mass of $15\pm5\,M_{\odot}$ based on the size of the cavity \citep{chen2013} and the abundance ratios of ejecta-rich regions. \cite{foster2025} estimate a progenitor mass of $13-15\, M_{\odot}$ based on the Ca/Fe and Ni/Fe ratios for the ejecta component in the \xmm data. 

The determination of the forward shock velocity, $v_s$, from X-ray spectral fits alone is subject to various difficulties that results in discrepancies with other methods, as described in \cite{raymond2023}.  Among these difficulties are the fact that the extraction regions typically used for the X-ray spectral analysis are larger than the forward shock region thereby including plasmas with different conditions, the unknown level of electron-ion equilibration, the moderate or low resolution spectra, and the inadequacy of the spectral models.  \cite{Shimoda2015} point out that shocks propagating into an inhomogeneous medium will be rippled or oblique leading to an underestimate of the $v_s$ based on the X-ray temperature alone. Given these challenges, any estimate of the forward shock velocity from X-ray spectral fits alone must be interpreted with these caveats in mind.  Although the size of the cavity in N132D is well-constrained by the existing \chandra\/ data, the $v_s$ is still uncertain.

A more direct method of determining the shock velocity is to measure the proper motion of a feature or features associated with the shock front, if the distance to the object is known or assumed. The high angular resolution of \chandra\, has been exploited to measure the proper motion of features in SNR shocks and to measure the global expansion of SNRs by comparing observations of the same remnant at different epochs.  \chandra\, has measured the expansion rate at various locations in the youngest Galactic SNR G1.9+0.3 \citep{borkowski2017} to vary from $0\farcs09~\mathrm{yr^{-1}}$ to $0\farcs44~\mathrm{yr^{-1}}$  which corresponds to velocities of $3,600~\kms$  to $17,000~\kms$  assuming a distance of 8.5~kpc. \cite{vink2022casA} have used \chandra\, to measure the proper motion of the forward shock ($0.15$ to $0.28\%~\mathrm{yr^{-1}}$, $v_s=4,000$ to $7,000~\kms$) and the reverse shock 
($-0.10$ to $0.23\%~\mathrm{yr^{-1}}$, $v_s=-1,900$ to $4,200~\kms$) at different locations in Cassipoeia~A assuming a distance of 3.4~kpc.  \chandra\, data have even been used to measure the expansion of SNRs in the Magellanic Clouds which is considerably more difficult given the larger distance.  \cite{xi2019} measured an expansion of $0.025\%~\mathrm{yr^{-1}}$ or a $v_s$ of $1,600~\kms$ for 1E~0102.2-7219 in the Small Magellanic Cloud assuming a distance of 60.6~kpc and \cite{williams2018} measure velocities ranging from $2,860$ to $5,450~\kms$ in the LMC remnant SNR~0509-68.7 (N103B) assuming a distance of 50~kpc. These measurements are not subject to the systematic uncertainties of estimating the shock velocity from the fitted X-ray temperature, but have their own systematic uncertainties.

In this paper, we use new X-ray observations of N132D obtained as part of a Chandra Large Project (LP) to measure the proper motion of the forward shock and hence the $v_s$ in the south and northeast, 
thereby circumventing the difficulties in determining the shock velocity based on the spectral fits.  These values are then used in 1D SNR evolution models to refine the estimates of the explosion energy, 
ejecta mass, and ambient medium density and structure.
The paper is organized as follows:  Section~\ref{sec:data} discusses the data and the initial reduction and our registration method, Section~\ref{sec:exp_analysis} describes our expansion analysis and the estimate of the shock velocity, $v_s$,
Section~\ref{sec:spec_analysis}
presents the spectral analysis of narrow regions near the shock front,
Section~\ref{sec:discussion} presents the results of the 1D SNR models, the constraints on the explosion energy, ejecta mass, and ambient medium properties, and a comparison of the plasma temperature inferred from the X-ray spectral fits to the temperature inferred from the measured shock velocities,
and finally in Section~\ref{sec:conclusions} we present our conclusions.  The three appendices (\ref{sec:app_src_selection},\ref{sec:app_qe_difference}, \& \ref{sec:app_psf_diff}) include more information on the details of the analysis.
We assume the distance to N132D to be $50\,\mathrm{kpc}$ in all calculations hereafter \citep{2003AJ....125.1309C,2013Natur.495...76P,2019Natur.567..200P}. At this distance, $1\arcsec = 0.24\,\mathrm{pc}$. All uncertainties reported are the $1.0\sigma$ uncertainties unless noted otherwise.

\section{Data and Reduction} \label{sec:data}

\subsection{Chandra Large Project} \label{sec:lp}

N132D was observed by \chandra for a total of 878~ks from 27 March 2019 to 16 July 2020 in 28 separate observations as part of a \chandra\, LP (Proposal Number:20500554, PI:Plucinsky); see Table~\ref{tab:obslist} for details.  The goals of the Large Project include studying the late stages of massive star evolution, the SNe explosion mechanism, SNe elemental abundances, ejecta properties and distribution, and the physical mechanisms for the interaction of shocks with molecular clouds and cavities.  This is the first paper in a series of papers from this \chandra legacy data set and focuses on the expansion of the forward shock and the evolution of the SNR. The data acquired in 2019/2020 (labeled as \epochb in Table~\ref{tab:obslist}) are compared to the previous \chandra\ observations executed in 2006 (labeled as \epocha), to measure the proper motion of different features in the remnant over a $\sim14.5$~yr interval.  A significant challenge for this analysis is the reduction in the low-energy sensitivity of the {\em {Advanced CCD Imaging Spectrometer}} (ACIS) instrument from 2006 to 2020 due to the accumulation of a contamination layer on the optical-blocking filter \citep{paul2016,plucinsky2022}. The analysis presented in this paper is restricted to the energy range of 1.2-7.0~keV unless otherwise stated to minimize the effect of the contamination correction on the results.  Another challenge to this analysis is the fact that the 2019/2020 observations were executed in 28 distinct observations.  Not all of these observations are suitable for the expansion analysis described in this paper as explained later but may be used for future analyses.  We describe in detail in the following sections our efforts to register these observations to take full advantage of \chandra's exquisite angular resolution.

\begin{table}[htb]
\caption{Observations List}
\begin{center}
\label{tab:obslist}
\begin{tabular}{ l l c c c}
\hline\hline
ObsID & Start Date & Exposure & Roll Angle & Used to measure \\ & & [ks] & [deg] & Expansion?\\
\hline 
\multicolumn{5}{c}{\epocha: January 2006} \\
\hline
\phantom{0}5532 & 2006-01-09 & 44.6 & 330.16 & Y\\
\phantom{0}7259 & 2006-01-10 & 24.8 & 330.16 & Y\\
\phantom{0}7266 & 2006-01-15 & 19.9 &330.16 & Y\\
\hline
\multicolumn{5}{c}{\epochb: March~2019$-$July~2020} \\
\hline
21362 & 2019-03-27 & 34.4 & 255.62 & Y$^a$\\
21363 & 2019-08-29 & 46.0 & 105.15 & Y$^a$\\
21364 & 2019-09-01 & 20.8 & 105.15 & \\
22687 & 2019-09-02 & 34.4 & 102.66 & Y$^a$\\
22094 & 2019-09-10 & 36.2 & \phantom{0}93.15 & Y$^a$\\
21687 & 2019-09-11 & 24.7 & \phantom{0}93.15 & \\
22841 & 2019-09-12 & 36.5 & \phantom{0}93.15 & Y$^a$\\
22853 & 2019-09-22 & 19.8 & \phantom{0}83.30 & \\
22740 & 2019-09-26 & 19.8 & \phantom{0}78.15 & \\
22858 & 2019-09-27 & 19.8 & \phantom{0}78.15 & \\
22859 & 2019-09-28 & 18.8 & \phantom{0}78.15 & \\
21881 & 2019-10-04 & 23.3 & \phantom{0}60.15 & \\
22860 & 2019-10-06 & 17.8 & \phantom{0}60.15 & \\
23270 & 2020-05-29 & 27.7 & 178.14 & \\
21882 & 2020-05-30 & 34.6 & 178.14 & Y$^a$\\
21883 & 2020-05-31 & 32.6 & 178.14 & Y$^a$\\
23044 & 2020-06-02 & 52.9 & 175.15 & Y$^a$\\
21886 & 2020-06-05 & 43.0 & 175.14 & Y$^a$\\
21365 & 2020-06-07 & 56.3 & 175.14 & Y$^a$\\
23277 & 2020-06-08 & 14.9 & 183.13 & \\
21884 & 2020-06-09 & 42.5 & 183.14 & Y$^a$\\
21887 & 2020-06-10 & 51.4 & 183.14 & Y$^a$\\
21885 & 2020-06-25 & 21.3 & 167.15 & \\
23286 & 2020-06-27 & 14.9 & 167.14 & \\
21888 & 2020-07-11 & 24.7 & 152.15 & \\
23303 & 2020-07-12 & 24.7 & 152.14 & \\
21361 & 2020-07-13 & 31.1 & 160.14 & Y$^a$\\
23317 & 2020-07-16 & 43.1 & 160.14 & Y$^a$\\
\hline
\multicolumn{5}{l}{$a:$ The observations used in the spectral analysis.} \\
\end{tabular}
\end{center}
\end{table}

\subsection{Data Processing} \label{sec:data_processing}
\begin{figure*}
    \centering
    \includegraphics[width=1.0\textwidth]{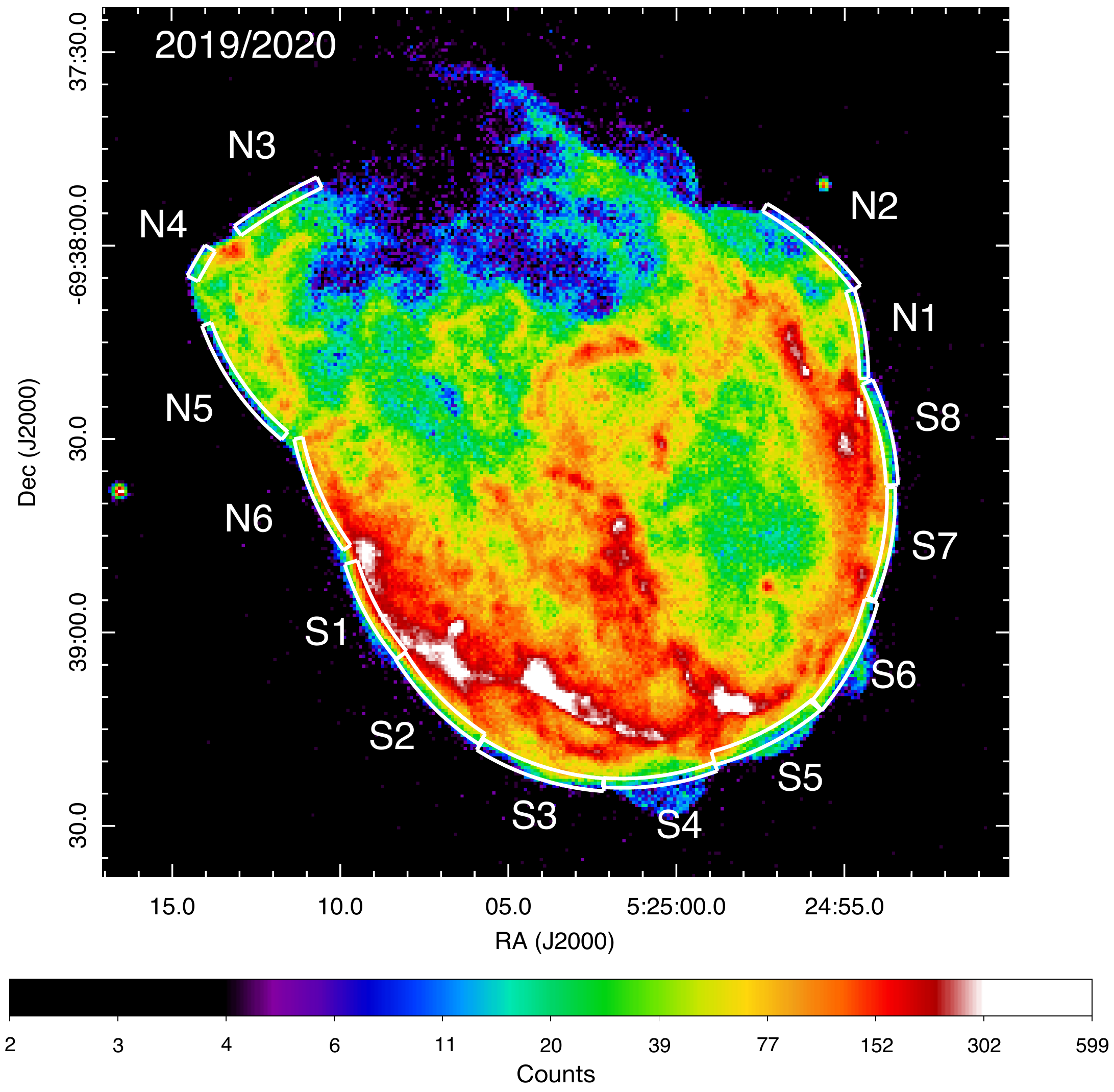}
    \caption{\chandra ACIS-S image of N132D in the energy band 1.2--7.0 keV produced from the LP data (\epochb). The white arc regions (labeled S1$\ldots$S8 along the southern rim and N1$\ldots$N6 along the northern boundary) indicate the regions used for the expansion measurement and the spectral analysis. The binszie of the image is 0\farcs492. The units of the color bar is counts.\label{fig:n132d}}
\end{figure*}

We reprocessed each observation (see Table~\ref{tab:obslist}) generating new level~2 event lists using \ciao 4.13 \citep{fruscione2006} and CALDB 4.9.5, and the \ciao tool \texttt{chandra\_repro}\footnote{\url{https://cxc.harvard.edu/ciao/ahelp/chandra\_repro.html}}. The default \texttt{pix\_adj=EDSER} was applied. Spectra were extracted for the point sources, forward shock and background regions, and the corresponding response files were created using \texttt{specextract} and analyzed in \xspec version 12.11.0k. The \texttt{FTOOLS} command 
\texttt{fkeyprint} was used to set the \texttt{AREASCAL} of the background spectra to the ratio of the source and background \texttt{BACKSCAL} to scale the background area to the source area.
\saotrace~2.0.5 and \marx~5.5.1 were used for simulating point-spread functions (PSFs) of the point sources (see Section \ref{sec:pointsource} below for details.)
To correct the Quantum Efficiency (QE) changing from the first 2006 observation to later observations, we use the \ciao tool \texttt{eff2evt} to get the QE for every event in each observation, by setting the option \texttt{detsubsysmod} to the start time of the corresponding observation. The added columns were renamed, and \texttt{eff2evt} was rerun with \texttt{detsubsysmod} set to the start time of the reference observation \ObsID 23317. The \ciao tool \texttt{dmtcalc} was used to evaluate the event QE correction corresponding to the time of the reference observation. 
To merge the events lists of the 2006 observations and the 2019/2020 observations, we used the \ciao tool \texttt{reproject\_events} to project each event
list to the tangent point of the observation \ObsID 5532, then we used \texttt{dmmerge} to merge the event lists. Figure~\ref{fig:n132d} displays the merged image from the 2019/2020 data in the 1.2--7.0~keV band with the regions used for the expansion analysis indicated.
\subsection{Registration of the Observations} \label{sec:registration}
We first obtain a list of common sources in the vicinity of the remnant (see Section~\ref{sec:pointsource} below) in each observation, and reposition the pointing to minimize the scatter in these point source positions (see Section~\ref{sec:regis_method} below).  \epocha observations are each first individually registered to \ObsID 5532, and \epochb observations to \ObsID 23317.  The merged \epocha dataset is then matched to \epochb by re-registering them to \ObsID 23317.

\subsubsection{Point source positions} \label{sec:pointsource}
To produce a list of point sources around the remnant for registration purposes, we run \texttt{wavdetect} \citep{2002ApJS..138..185F} on each observation. We select point-like sources
that are detected in at least two observations, since sources detected in only 
one observation cannot be used in the registration of multiple observations. 
This results in a total of 50 point sources in total. 
For each pair of observations, we select point sources which are present in both observations.  Of the 50 total sources detected, the number of common point sources for pairs of observations ranged from 10 to 22.

The off-axis angles of the detected sources range from $0.9\arcmin$--$8\arcmin$. The position reported by \texttt{wavdetect} is not accurate enough for our purposes because the tool assumes a symmetric PSF.  The PSF varies with off-axis angle, roll, and (to a lesser extent) energy. To obtain more accurate positions for point sources, we use \saotrace\footnote{https://cxc.cfa.harvard.edu/cal/Hrma/SAOTrace.html} to model the X-ray optics and \marx\footnote{https://space.mit.edu/cxc/marx/} to model the detector and simulate a PSF image for each point source for each observation. 

Though we select sources that appear to be point-like, we allow them to be extended by convolving the simulated PSF image with a Gaussian kernel.This practice is also in keeping with the suggested \ciao thread\footnote{\url{https://cxc.cfa.harvard.edu/sherpa/threads/2dpsf/index.html##src}}, which also accounts for aspect uncertainty, and residual systematics in the PSF model as well as in super-resolved binning (unless otherwise stated, we adopt a bin size of $\frac{1}{4}$ sky pixel $\equiv0\farcs123$).

\saotrace can model a point source at the location of the source identified by \texttt{wavdetect} and \marx can apply the detector pixelization and detector response. 
Using these tools, we construct raytraced point sources at the locations of the point-like sources, with $\approx$5000$\times$ the estimated counts in order to minimize simulation fluctuations.  These result in well-sampled images to construct the PSF model for fitting.

In \sherpa, we define a source model using  a 2-D symmetric Gaussian function plus a constant. The source model is convolved with the PSF image to make a ``model image'' for fitting the ``data image''. The ``data image'' is a point source image extracted from the observation using a $31.488\arcsec \times 31.488 \arcsec$ box region.  The data image and model image are both binned to $0.0615\arcsec \times 0.0615\arcsec$ pixels. 
The center, amplitude, and sigma of the Gaussian function, and the constant value are free to change. We use the C-statistic \citep{cash1979} as the fit statistic. The constant value accounts for the detector background level; it is reasonable to assume a constant background within the area of the point source image. The position of the point source is taken to be the fitted center $(x,y)$ of the Gaussian function. The $1 \sigma$ errors of the position, $(\sigma_{x},\sigma_{y})$, are calculated by varying the the $x$ or $y$ center positions along a grid of values while the values of the remaining free parameters are allowed to float to new best-fit values.

\begin{figure}
    \centering
    \includegraphics[width=0.45\textwidth]{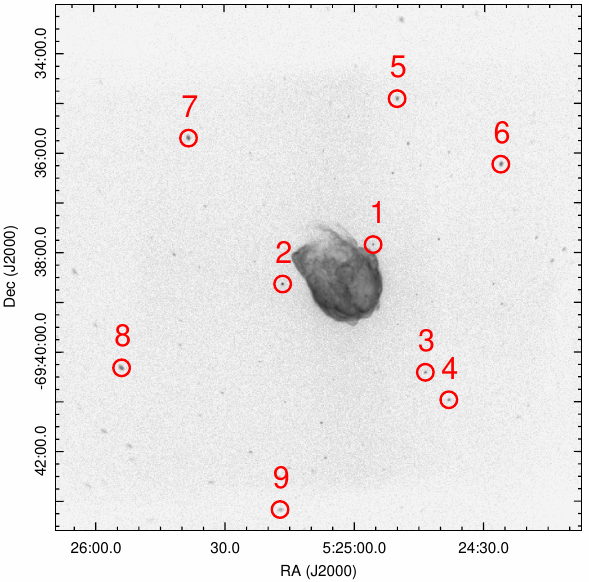}
    \caption{Point sources used in the registration are shown as labelled red circles on a merged \epochb image. The circle labels are the point source IDs (see Table~\ref{tab:source_list}). The image is binned to 2 ACIS sky pixels ($0.964 \arcsec$), and the energy band is 1.2--7.0~keV; logarithmic grey scale.
    }\label{fig:pointsource}
\end{figure}

The accuracy and precision of this method is limited by the point-source counts and the off-axis angle. To characterize this, we simulate point sources with off-axis angles from $1\arcmin$--$6\arcmin$, and point-source counts from 5 to 400 counts. We apply the same PSF fitting method described above to obtain the position and the position error, and exclude sources within each observation which are outside the 5\% and 95\% quantiles of the distribution of distances from the best-fit power-law to the distribution (see Appendix~\ref{sec:app_src_selection} for details).

We also exclude point sources which are located on or near a CCD boundary in an observation; although these point sources have enough counts to yield reasonable fitted positions, the positions may be biased toward the parts of the dithered image fully on the CCD.  Nine point sources common to all of the observations remain after excluding the outliers. These are shown in Figure~\ref{fig:pointsource}, and the coordinates of the point sources are listed in Table~\ref{tab:source_list}.  For each pair of observations, the sources used for registration are listed in Table~\ref{tab:registration}. 

\begin{table}[htb]
\caption{Sources Used in the Registration }
\begin{center}
\label{tab:source_list}
\begin{tabular}{ l c c}
\hline\hline
Source & RA(J2000)  & Dec(J2000) \\  ID& $\alpha$ deg & $\delta$ deg\\
\hline 
1&81.2319050&-69.6308336\\
2&81.3190858&-69.6440358\\
3&81.1814788&-69.6736206\\
4&81.1587763&-69.6828067\\
5&81.2089350&-69.5819136\\
6&81.1091962&-69.6038267\\
7&81.4094288&-69.5950900\\
8&81.4745292&-69.6720078\\
9&81.3218042&-69.7196139\\
\hline
\end{tabular}
\end{center}
\end{table}

\subsubsection{Registration method}\label{sec:regis_method}

We register an observation to a reference observation by shifting the $(x,y)$ positions of point sources in the observation to match the positions of the corresponding point sources in the reference observation.
The shifted positions $(x^\prime, y^\prime)$ for the observation to be matched are
\begin{equation}
 \begin{pmatrix}
 x^{\prime}\\
 y^{\prime}
 \end{pmatrix}
 =
 \begin{pmatrix}
 a_{11} & a_{12}\\
 a_{21} & a_{22}
 \end{pmatrix}
 \begin{pmatrix}
 x\\
 y
 \end{pmatrix}
 +
 \begin{pmatrix}
 t_{1}\\
 t_{2}
 \end{pmatrix}
\end{equation}
where the parameters: $a_{11}$, $a_{12}$, $a_{21}$, $a_{22}$ account for the scale factor and rotation, and $t_{1}$, $t_{2}$ account for the $x$ and $y$ translations.

The position shifts are used to compute a transformation which minimizes the discrepancies.
We use a loss function, $D$, that is robust to outliers; it trends to quadratic for small deviations, but approaches the magnitude of the differences for large deviations\footnote{We use the \texttt{soft\_l1} option of SciPy \citep{virtanen2020} python routine \texttt{scipy.optimize.least\_squares}. This loss function is more robust than least
squares or chi-squared since the weighting of outliers
approaches a linear rather than quadratic penalty.  In the limit $\{x,y\}_\mathit{i}^\prime \rightarrow \{x,y\}_{0,\mathit{i}}$, $D\rightarrow\frac{1}{2}\left[\frac{(x_\mathit{i}^{\prime}
               -x_{0,\mathit{i}})^{2}}
             {\sigma_{x_\mathit{i}}^{2}
             +\sigma_{x_{0,\mathit{i}}}^{2}}
        +\frac{(y_\mathit{i}^{\prime}
             -y_{0,\mathit{i}})^{2}}
             {\sigma_{y_\mathit{i}}^{2}
             +\sigma_{y_{0,\mathit{i}}}^{2}}\right]$
while for large deviations, $D\rightarrow\sqrt{\frac{(x_\mathit{i}^{\prime}
               -x_{0,\mathit{i}})^{2}}
             {\sigma_{x_\mathit{i}}^{2}
             +\sigma_{x_{0,\mathit{i}}}^{2}}
        +\frac{(y_\mathit{i}^{\prime}
             -y_{0,\mathit{i}})^{2}}
             {\sigma_{y_\mathit{i}}^{2}
             +\sigma_{y_{0,\mathit{i}}}^{2}}}$.}
         
\begin{equation}
    D=\sum_{\mathit{i}=1}^\mathit{n} 
      \left(\sqrt{1
        +\frac{(x_\mathit{i}^{\prime}
               -x_{0,\mathit{i}})^{2}}
             {\sigma_{x_\mathit{i}}^{2}
             +\sigma_{x_{0,\mathit{i}}}^{2}}
        +\frac{(y_\mathit{i}^{\prime}
             -y_{0,\mathit{i}})^{2}}
             {\sigma_{y_\mathit{i}}^{2}
             +\sigma_{y_{0,\mathit{i}}}^{2}}}-1\right)
\end{equation}
where $n$ is the number of matched point sources, 
$(x_\mathit{i},y_\mathit{i})$ is the source position of source $i$, and $(x_{0,\mathit{i}},y_{0,\mathit{i}})$ is the position of the corresponding reference source.  The
position error of source $i$ is $(\sigma_{\mathit{x,i}}, \sigma_\mathit{y,i})$  
and $(\sigma_{x_{0,\mathit{i}}}, \sigma_{y_{0,\mathit{i}}})$ is the position error of the corresponding reference source.  
Hereafter, the subscript $i$ will be implicit and not shown unless necessary for disambiguation. 

We used the \ciao tool \texttt{wcs\_update} with the fitted parameters $a_{11}$, $a_{12}$, $a_{21}$, $a_{22}$, $t_{1}$, $t_{2}$ to update the aspect solution file and the WCS (World Coordinate System) of the observation event list. 

The error on the registration is estimated by the registration residual.  The weighted average position residual for the matched point sources of the two observations:

\begin{eqnarray}\label{eq:regis_err}
    r 
    &=& \sqrt{(r_{x})^{2}+(r_{y})^{2}}\\ \nonumber
    &=& \sqrt{\left(\frac{\sum^{n} \frac{(x^{\prime}-x_{0})}{\sigma_{d}^{2}}}{\sum^{n}\frac{1}{\sigma_{d}^{2}}}\right)^{2}+\left(\frac{\sum^{n} \frac{(y^{\prime}-y_{0})}{\sigma_{d}^{2}}}{\sum^{n}\frac{1}{\sigma_{d}^{2}}}\right)^{2}}
\end{eqnarray}

In Eq~\ref{eq:regis_err}, $r_{x}$ and $r_{y}$ are the weighted average position residuals of the matched point sources in the x and y directions. The combined position error for each source is:
\begin{equation}
  \sigma_{d}=\sqrt{\sigma_{x^{\prime}}^{2}+\sigma_{x_{0}}^{2}+\sigma_{y^{\prime}}^{2}+\sigma_{y_{0}}^{2}}
\end{equation}

We registered \epocha observations to \ObsID 5532, and \epochb observations to \ObsID 23317. We separately merged the \epocha observations and the \epochb observations to allow for an expansion measurement by comparing the epochs. The merged \epocha observations are registered to the merged \epochb observations using point sources in common and the registration method described above.  The resulting transformation matrix is applied to each of the \epocha observations, and the \epocha observations are re-merged so that the merged \epocha is registered to the merged \epochb.

After registration the positions and position errors of the point sources were obtained using the method described in Section~\ref{sec:pointsource}.  The registration results are shown in Table~\ref{tab:registration}; the average registration residual of \epocha observations is $1.0$ mas ($\sim2.4\times10^{-4}$~pc at the distance of the LMC), and the average registration residual of later observations is $5.6$ mas. The registration residual of merged \epocha observations and merged \epochb is $1.8$ mas, we quote this value as the systematic error in the expansion measurement in Section~\ref{sec:exp_analysis}. The error of registration residual shown in Table~\ref{tab:registration}, is the standard error of the weighted average position residuals of the point sources used in registration. 

\begin{eqnarray}\label{eq:regis_err_stderr}
    \sigma_{r}       
     &=& \sqrt{(\sigma_{r_{x}})^2+(\sigma_{r_{y}})^{2}} \\
     &=&  \sqrt{
                \left(\frac{\sum^{n}
                      \frac{(x^{\prime}-x_{0}-r_{x})}
                            {\sigma_{d}^2}}
                            {\sqrt{n}\sum^{n}\frac{1}
                            {\sigma_{d}^2}}            
                \right)^2                 
             + \left(\frac{\sum^{n} 
                     \frac{(y^{\prime}-y_{0}-r_{y})}
                          {\sigma_{d}^{2}}}
                          {\sqrt{n}\sum^{n}
                           \frac{1}{\sigma_{d}^{2}}}
        \right)^2 \nonumber
     }
\end{eqnarray}

Here, $\sigma_{\mathit{r}_{\mathit{x}}}$ and $\sigma_{\mathit{r}_{\mathit{y}}}$ are the standard error of the average position residual of matched point sources in x and y directions, separately.

\begin{figure}
    \centering
    \includegraphics[width=0.45\textwidth]{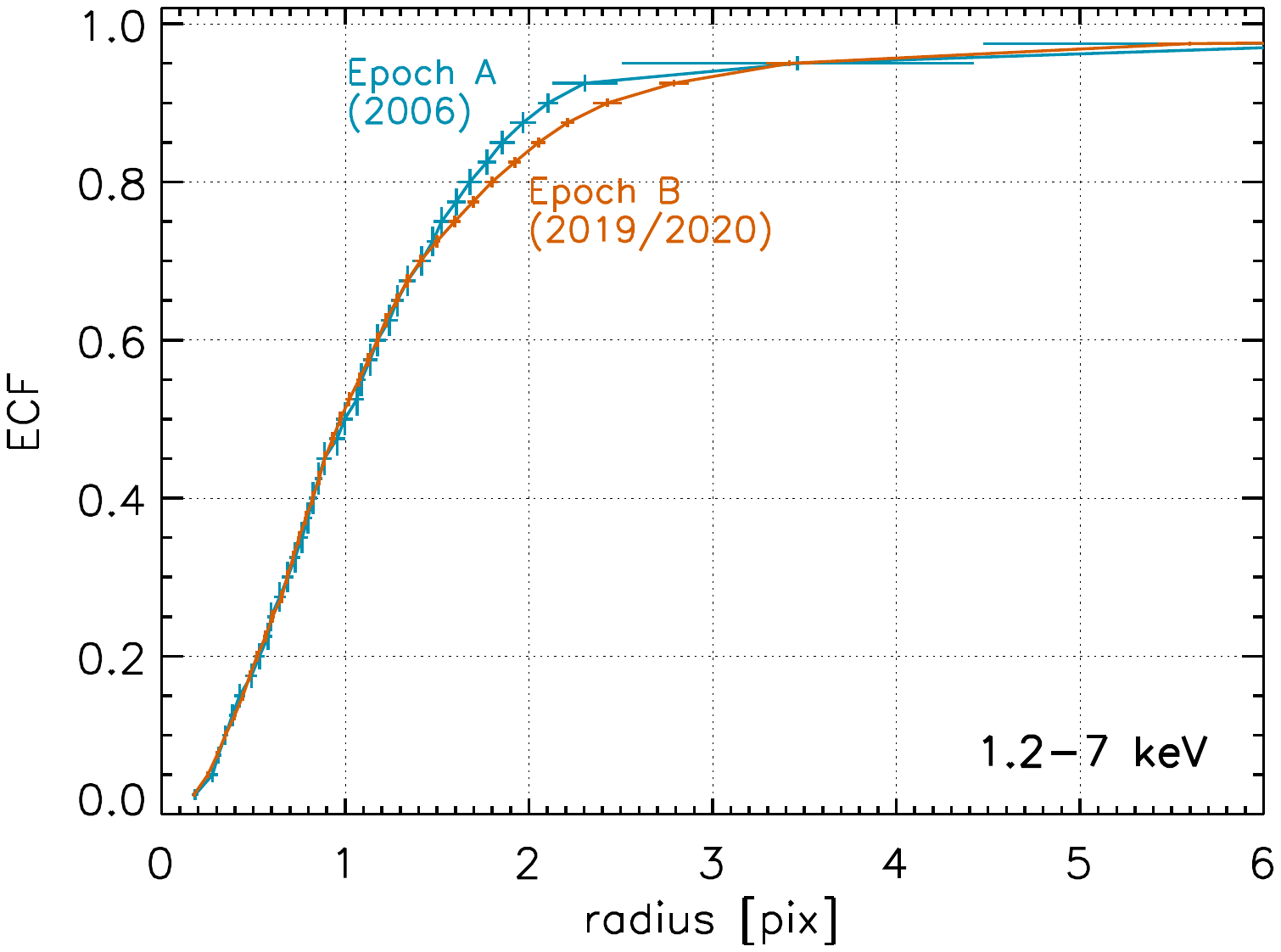}
    \caption{The cumulative radial enclosed counts fraction (ECF) in the two epochs for point source 2 (from Figure~\ref{fig:pointsource}) in energy band 1.2--7.0~keV. The X-axis is the radius in ACIS ``sky'' pixels (``physical coordinates''). The narrower (blue) distribution, showing error bars at ECF intervals of 0.025 are for the merged \epocha observations after registration. The broader (red) line and error bars are for the merged \epochb observations after registration.}\label{fig:pointsourcecdf}
\end{figure}

In Figure~\ref{fig:pointsourcecdf} we show the cumulative counts distribution (1.2--7.0\,keV) for (registered) point source number 2 (see Figure~\ref{fig:pointsource}) for merged \epocha observations (in red) and merged \epochb observations (in blue). The cumulative distributions of the point source from \epocha and \epochb observations are consistent with each other within a radius of 1.5 sky pixels (0.738\arcsec), and this range contains $\sim$ 70\% of the enclosed counts, covering the core of the PSF. This indicates that the observations are well registered. The PSF difference between \epocha and \epochb can result in a possible systematic error in the measurement of expansion of $0.026\arcsec$. We discuss this possible effect in Section~\ref{sec:exp_shock}.

\begin{table*}[htb]
\caption{Registration Results for Each Observation}
\begin{center}
\label{tab:registration}
\begin{tabular}{l c c c l c c c }
\hline
ObsID & $\Delta RA$ & $\Delta Dec$ & Rotation angle & Point Source IDs & \multicolumn{3}{c}{Registration residual [mas]} \\ & [arcsec] & [arcsec] & [degree] & & $r_{x}\pm\sigma_{r_{x}}$ & $r_{y}\pm\sigma_{r_{y}}$ & $r\pm\sigma_{r}$\\
\hline 
\multicolumn{8}{l}{(A) ObsIDs in Epoch \textit{2006} registered to \textit{5532}} \\
\hline
\phantom{0}7259 &-0.0397&0.2776&0.0737&2,3,6,8,9&-0.9 $\pm$ 1.7 & -0.7 $\pm$ 3.4 & 1.2 $\pm$ 3.8\\
\phantom{0}7266 &-0.0621&0.4177&-0.0066&2,5,6,9&0.7 $\pm$ 2.2 & -0.0 $\pm$ 0.1 & 0.7 $\pm$ 2.2 \\
\hline
\multicolumn{8}{l}{(B) ObsIDs in Epoch \textit{2019/2020} registered to \textit{23317}} \\
\hline
21361 &0.7020&-0.0945&0.0366&1,2,3,5,6,7,8&-8.0 $\pm$ 11.6 & \phantom{-0}7.6 $\pm$ 7.4 & 11.0 $\pm$ 13.8\\
21362 &1.3610&0.9308&-0.0060&1,2,3,4,6,7&\phantom{-}9.0 $\pm$ 17.3& \phantom{-0}3.8 $\pm$ 10.5& \phantom{0}9.8 $\pm$ 20.2\\
21363 &0.6201&0.3249&0.0559&1,2,3,4,6,8&\phantom{-}3.7 $\pm$ 6.2& \phantom{-0}0.0 $\pm$ 2.3 & \phantom{0}3.7 $\pm$ 6.6\\
21365 &0.0916&0.2840&0.0031&1,2,3,6,7,8&-0.8 $\pm$ 3.4&\phantom{0}-0.6 $\pm$ 4.1& \phantom{0}1.0 $\pm$ 5.4\\
21882 &0.0214&0.3023&-0.0040&1,2,3,5,6,7,8&-2.7 $\pm$ 7.2& \phantom{-0}1.8 $\pm$ 2.3& \phantom{0}3.3 $\pm$ 7.5\\
21883 &0.0344&0.2665&0.0261&1,2,3,5,6,7,8& -1.0 $\pm$ 3.0 &14.4 $\pm$ 9.2 & 14.4 $\pm$ 9.7\\
21884 &-0.0178&0.2759&0.0108&1,2,4,5,6,7,8&\phantom{-}2.9 $\pm$ 5.0 & \phantom{0}-0.3 $\pm$ 2.3& \phantom{0}3.0 $\pm$ 5.5\\
21886 &-0.0398&0.2609&-0.0088&1,2,3,5,6,7,8&-4.9 $\pm$ 4.9 &\phantom{-0}3.1 $\pm$ 8.1& \phantom{0}5.8 $\pm$ 9.4\\
21887 &0.2173&0.2703&-0.0122&1,2,5,6,7,8&-4.9 $\pm$ 5.9&\phantom{-0}3.7 $\pm$ 5.6& \phantom{0}6.1 $\pm$ 8.2\\
22094 &0.5311&-0.1047&-0.0085&1,2,3,4,6,7&\phantom{-}1.4 $\pm$ 3.6 &\phantom{-0}0.6 $\pm$ 12.0& \phantom{0}1.6 $\pm$ 12.6\\
22687 &0.7102&0.1936&0.0588&1,2,3,4,5,6&\phantom{-}2.7 $\pm$ 5.1 & \phantom{0}-2.4 $\pm$ 4.4 & \phantom{0}3.6 $\pm$ 6.7\\
22841 &0.8491&0.1365&0.0149&1,2,3,4,5,6,7&\phantom{-}6.4 $\pm$ 11.8 & \phantom{0}-2.0 $\pm$ 9.5 & \phantom{0}6.7 $\pm$ 15.1\\
23044 &0.2329&0.3599&0.0148&1,2,3,5,6,7,8& \phantom{-}0.7 $\pm$ 6.1 & \phantom{-0}2.8 $\pm$ 3.8 & \phantom{0}2.9 $\pm$ 7.2\\
\hline
\multicolumn{8}{l}{(C) Combined Epoch \textbf{2006} from A registered to Combined Epoch \textit{2019/2020} from B} \\
\hline
Epoch 2006&0.9039&0.5799&-0.0038&2,3,4,5,6,8,9&-0.9 $\pm$ 2.9 &1.5 $\pm$ 1.3 &1.8 $\pm$ 3.2\\
\hline
\end{tabular}
\end{center}
\end{table*}

\section{Expansion Analysis} \label{sec:exp_analysis}

We extract 1-D radial profiles from the merged \epocha\ and merged \epochb\ event lists for 14 regions of interest around the edge of the remnant (shown in Figure~\ref{fig:n132d}).  The regions are chosen to approximately follow the curvature of the edge of the remnant and have approximately similar angular extents, except for the N4 region.  The southern regions are labeled S1 to S8 and the northern regions are labeled N1 to N6. The dimensions of the regions are determined based on the local counts profile along axes radiating outwards from near the center of the remnant.  First we identify the peak in the profile within $\approx{10}''$ of the edge and place the inner radius where the \epocha\ counts profile falls to 80\% of the peak.  We then place the outer radius of the regions within the range where the counts profile is steepest, excluding any protrusions visible in the broad band image.  The dimensions and locations of the regions are listed in Table~\ref{tab:region_list}.

\begin{table*}[!htb]
\caption{Radial Profile Extraction Regions$^a$ }
\begin{center}
\label{tab:region_list}
\begin{tabular}{l c c c c c c}
\hline\hline
Regions & Center RA(J2000) & Center Dec(J2000) & Start Angle  & Stop Angle & Inner Radius & Outer Radius\\  & $\alpha$ [deg] & $\delta$ [deg] & [deg] & [deg] & [arcsec] & [arcsec] \\
\hline 
S1&81.2588809&-69.6436519&197&220&40\farcs132&42\farcs177\\
S2&81.2559342&-69.6448819&212&237&40\farcs784&42\farcs660\\
S3&81.2566376&-69.6449958&238&266&40\farcs804&42\farcs791\\
S4&81.2567163&-69.6447111&266&290&41\farcs748&43\farcs187\\
S5&81.2540938&-69.6442025&285&310&40\farcs812&42\farcs928\\
S6&81.2586843&-69.6455994&320&345&41\farcs484&43\farcs593 \\
S7&81.2571130&-69.6441644&338&362&41\farcs584&42\farcs871 \\
S8&81.2561512&-69.6440629&\phantom{00}2&\phantom{0}25&40\farcs242&42\farcs103 \\
N1&81.2619398&-69.6390685&\phantom{00}0&\phantom{0}18&43\farcs476&44\farcs672 \\
N2&81.2588729&-69.6436505&\phantom{0}38&\phantom{0}60&48\farcs968&50\farcs337 \\
N3&81.2682702&-69.6499883&115&126&75\farcs976&77\farcs819 \\
N4&81.2552309&-69.6448917&148&152&76\farcs714&78\farcs897 \\
N5&81.2775545&-69.6328541&200&230&40\farcs012&41\farcs400 \\
N6&81.2619398&-69.6390685&192&216&43\farcs984&45\farcs362 \\
\hline
\multicolumn{7}{l}{$a$: A {\sl CIAO} {\tt dmregion} may be constructed for each extraction region of interest as}\\
\multicolumn{7}{c}{{\tt pie}(\textit{RA, Dec, Inner~Radius, Outer~Radius, Start~Angle, Stop~Angle})} \\
\end{tabular}
\end{center}
\end{table*}

Because the quantum efficiency (QE) of ACIS changes with time, we include a correction to the counts profiles so that shifts in the profile due to QE changes are not misinterpreted as expansion shifts.  We do this by assigning each event a weight factor to correct the QE to a reference time, $t_\mathit{ref}$, taken to be the start time of \ObsID\ 23044:
\begin{equation}
t_\mathit{ref} \equiv t[\mathrm{start\ of\ ObsID\ 23044}]
\label{eqn:t_ref}
\end{equation}
The weights are calculated from the QE at the start time of the observation and at $t_\mathit{ref}$.
For the \epocha observations (2006), the weights are 
$w_{t_\mathit{A[i]}}=q(t_\mathit{A[i]})/q(t_\mathit{ref})$, where $i=1,2,3$ are the \epocha Obsid indices:
$t_\mathit{A[1]}$: \ObsID\ 5532, $t_\mathit{A[2]}$: \ObsID\ 7259, $t_\mathit{A[3]}$: \ObsID\ 7266), and $t_\mathit{A[i]}$ is the start time of the $i$th \epocha observation. For the \epochb observations (2019/2020) the weights are $w_\mathit{t_\mathit{B[j]}}=q(t_\mathit{B[j]})/q(t_\mathit{ref})$, with \epochb Obsid indices $j=1,2,3,..,14$ (see Table~\ref{tab:qe_ratio} in Appendix~\ref{sec:app_qe_difference}), and $t_\mathit{B[j]}$ is the start time of the $j$th \epochb observation. Here, $q(t_\mathit{A[i]})$, $q(t_\mathit{B[j]})$ and $q(t_\mathit{ref})$ are the QE values of the events, appropriate for the energy of the event obtained 
using the \ciao tool \texttt{eff2evt}  where the option \texttt{detsubsysmod} is set to the indicated time.  

Because the total exposure times of the \epocha and \epochb observations are different, after the radial profiles are extracted, we scale the radial profiles from \epocha to correct for the exposure time difference by using a scaling factor
$s(t_\mathit{A})=\tau_\mathit{B}/\tau_\mathit{A}$, where $\tau_\mathit{A}$ and $\tau_\mathit{B}$ are the respective total exposure times of the \epocha  and \epochb observations,.

After the QE is corrected for the variation with time and the differences in exposure times are accounted for, 
we find the radial profiles from \epocha and \epochb for a given region can differ by $\sim 3-7\%$ because of residual QE differences between nodes (quadrants of the CCD, see Table~\ref{tab:qe_ratio} in Appendix \ref{sec:app_qe_difference}). 
These small, residual QE differences may be due to errors in the contamination model and/or the charge-transfer inefficiency correction model for ACIS. 
The \epocha observations have almost the same roll angle (330\degree), but the \epochb observations have roll angles ranging from 80\degree to 256\degree. The events for a given region can be on different nodes of the S3 CCD depending on the roll. In our analysis, we have 3 combinations of epoch and node id for the events in a given region: (1) \epochb events on node 1 and \epocha events on node 0, (2) \epochb events on node 0 and \epocha events on node 0, and (3) \epochb events on node 0 and \epocha events on node 1. 

To correct the QE difference to the same level event-by-event, we multiply the weights discussed before with a QE difference normalization according to the node id and time epoch. The QE difference normalizations are: 
\qeratio{B1}{A0} for \epochb events on node 1 and \epocha events on node 0, 
\qeratio{B0}{A0} for \epochb events on node 0 and \epocha events on node 0,
and \qeratio{B0}{A1} for \epochb events on node 0 and \epocha events on node 1.
The QE differences are corrected to the QE on node 0 at \epochb. The QE difference normalizations are determined using the method discussed in Appendix~\ref{sec:app_qe_difference}.

\begin{figure}[!htbp]
    \centering
    \includegraphics[width=0.45\textwidth]{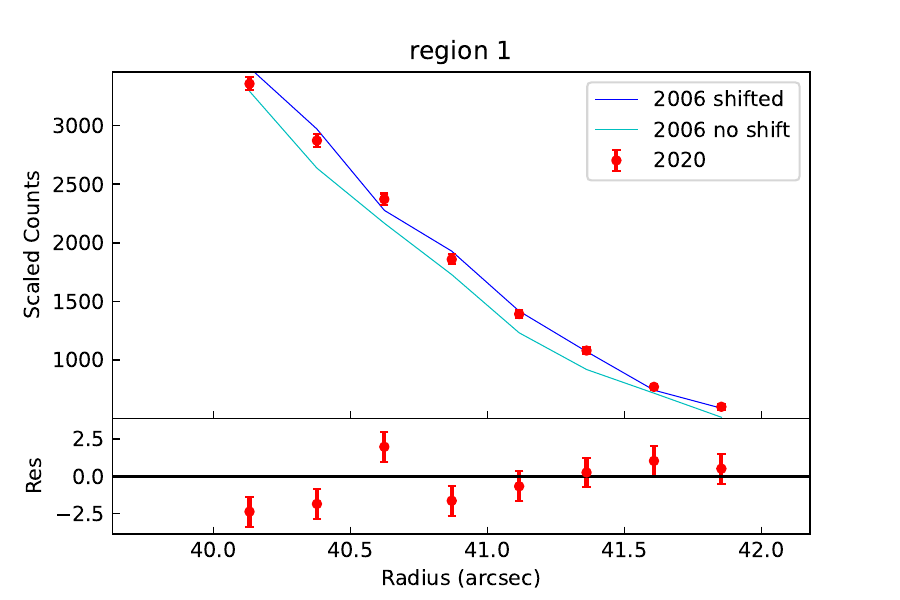}
    \caption{The radial profiles of region S1. The red data points are the profile of the 2019/2020 observations. The cyan line is the profile of the 2006 observations without a shift. The blue line is the 2006 profile shifted by the measured expansion. The lower panel shows the residuals of the 2020 profile and the shifted 2006 profile.\label{fig:profile1}}
\end{figure}
To measure the shift of the radial profile between the two epochs, we shift the events in the merged \epocha dataset in the radial direction, extract a radial profile in a pre-specified radial region, and compare it against the corresponding radial profile obtained for the merged events of \epochb.  As discussed above, the QE change, exposure time differences, and QE difference normalizations are applied while extracting the radial profiles.  We show the radial profiles of region~S1 for illustration in Figure~\ref{fig:profile1}.  All the regions used in this work are shown in Figure~\ref{fig:n132d} and listed in Table~\ref{tab:expansion_regions}. 
\begin{figure}[!htbp]
    \centering
    \includegraphics[width=0.45\textwidth]{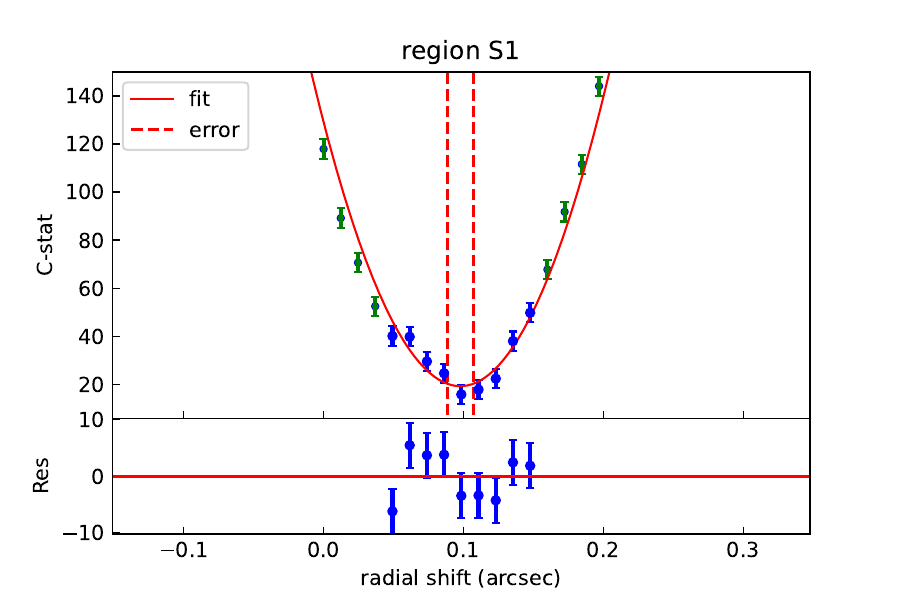}
    \caption{The C-statistic variation 
    with position shift for region S1. The green and blue data points are the C-statistic calculated from the 2020 profile and shifted 2006 profile at the shift positions. The blue data points are included in the quadratic fitting for the shift estimate. The red solid line is the fitted quadratic function. The vertical dashed lines show the $1\sigma$ error range of the estimated shifts.   \label{fig:cstat1}}
\end{figure}
We calculate the C-statistic (hereafter C-stat) between the \epocha radial profile considered as ``model" and the \epochb radial profile considered as ``data".
This results in an array of C-stat values computed as a function of shifts 
of the \epocha radial profile. 
Because this curve is affected by statistical fluctuations, we obtain a smooth representation by fitting it with a quadratic function around the minimum C-stat value. The shift corresponding to the minimum of the quadratic function is our estimate of the shifts between the \epocha and \epochb radial profiles.  We select four data points to the left and right of the  minimum C-stat value to include in the quadratic fitting. Figure~\ref{fig:cstat1} shows an example of such a fit for region~1.
The $1\,\sigma$ error range is determined from the quadratic fit function by finding the range of shifts satisfying  $y-y_{m}\leq 1$, where $y$ is the quadratic function and $y_{m}$ is the minimum value. 

\subsection{Expansion of the Forward Shock} \label{sec:exp_shock}
\begin{figure}[htb]
    \centering
        \includegraphics[width=0.45\textwidth]{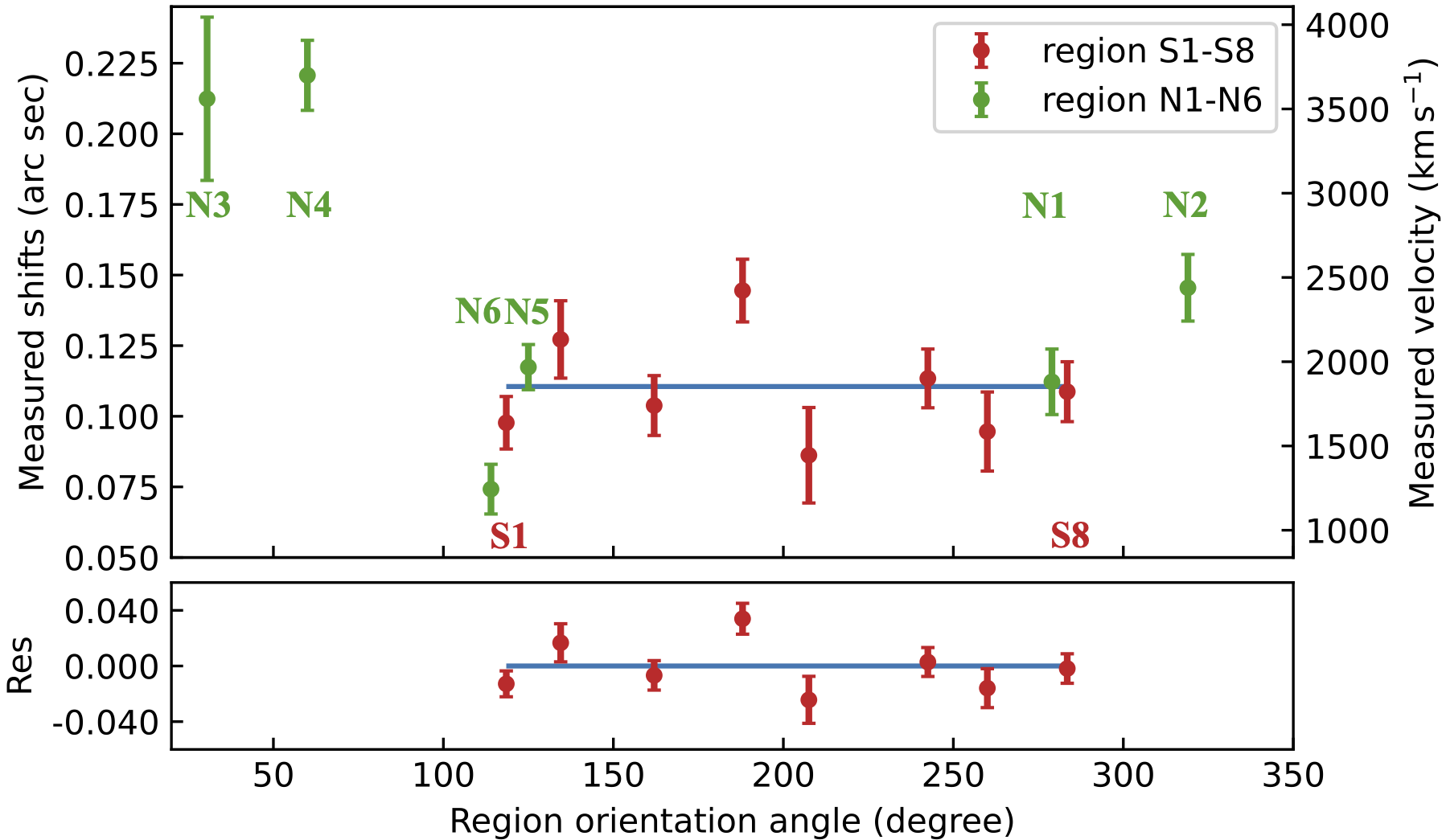}
    \caption{The raw measured expansion of the S1-S8 (labeled sequnetially from left to right) and N1-N6 regions (``raw shifts'' in Table~\ref{tab:expansion_regions}), with the fitted average expansion (blue solid line) from the S1-S8 regions.\label{fig:expansionlinear}}
\end{figure}

We first consider the shifts for each region determined by the method described in the previous section and then consider the systematic errors and possible corrections to those errors.
The expansion results and their uncertainties for the 14 regions indicated in Figure~\ref{fig:n132d} are shown in Table~\ref{tab:expansion_regions} in the column labeled ``raw shifts'' and plotted in Figure~\ref{fig:expansionlinear}.  The expansion results are plotted as a function of the ``region orientation angle'' which we define as the angle between the normal vector to the long dimension of the region and North in Figure~\ref{fig:n132d}.  For most regions (such as the regions S1-S8), this angle is close in value to that of the azimuthal angle with respect to North; but for some regions such as N3 and N5 it can be significantly different. We adopted this representation to more easily distinguish regions for which the normal vector does not point back to the supposed COE.
The eight southern regions (S1-S8) lie on an approximately circular rim and are located at similar radii from the nominal COE \citep{banovetz2023}. It would be reasonable to examine the case in which each of these eight regions had the same or similar expansion and compute an average expansion. Such a case yields an average expansion rate of $0\farcs11\pm0\farcs02$, or equivalently $1850\pm390$~km~s$^{-1}$, for these regions.\footnote{We compute the average expansion for the $M=8$ southern regions and estimate the error on the average by combining the individual statistical errors (the so-called {\sl Within} variance, $W$) and the possible systematic errors based on the observed scatter in the estimates (the {\sl Between} variance, $B$).  We use the method described in \cite{lee2011} (see their equations (5)--(7)) to compute the {\sl Total} variance $$T=W+\frac{M+1}{M}B\,,$$ for a sample size of $M$.  Note that this is a conservative estimate of the error bar on the expansion, and assumes the existence of systematic biases which manifest in the scatter of the expansion estimates for the different regions. \label{foot:MultImp}}
The estimate and associated uncertainties are listed in Table~\ref{tab:expansion_result} and are marked in Figure~\ref{fig:expansionlinear}. A blue horizontal line spans the S1-S8 regions indicating the best fit for the average expansion.  Note that 7 of the 8 regions are consistent with the average at the 1.5$\sigma$ level, with region S4 being the sole outlier at $\approx3\sigma$.  This indicates that the assumed scenario of a common COE yielding the same expansion rate is a reasonable fit to the data for regions S1-S8, and perhaps also for the northern regions N1 and N5.  However, the observed scatter could be an indication of inhomogeneous expansion.  Note that regions N2, N3, and N4 are located at larger distances from the nominal center (see Table~\ref{tab:expansion_regions} for the distances from the centers of the regions to the expansion center) and have significantly larger expansion rates. 
The larger expansion velocities for N2, N3, and N4 are consistent with free (homologous) expansion ($v_\mathit{s} \propto R_\mathit{s}$) for those regions. The spatial velocities for regions N1, N6, and particularly N5 are less certain because their normals deviate significantly from the direction to the C.O.E. Our method measures the velocities perpendicular to the region sectors; the  parallel component is not determined. Thus, the actual velocities for N1, N6, and N5 could be larger than we assess.

We examined several potential systematic errors that could affect our result. One possible source of systematic error is due to the fact that the shape of the \chandra PSF is a function of the off-axis and azimuthal angles. Since our analysis includes data from the same region of the remnant acquired at different off-axis and azimuthal angles, it is possible that the different shape of the PSF could introduce a systematic error if the two data sets to be compared were acquired with different off-axis and azimuthal angles. Based on simulations of the PSF for \epocha and \epochb observations, the differences in the PSF between epochs could mimic a (small) expansion or contraction. To measure this PSF bias effect, we simulated the merged PSF for each region and epoch as described in Appendix~\ref{sec:app_psf_diff}. The differences show an azimuthal variation which we fit with a sinusoid. To correct for this PSF bias, we subtracted the fitted sinusoidal model from the expansion results in Figure~\ref{fig:expansionlinear} to obtain a PSF-bias-corrected expansion as shown in Figure~\ref{fig:expansionpsf} in Appendix~\ref{sec:app_psf_diff}.  The amplitude of this effect can be as large as $\sim40~{\mathrm{mas}}$.

\begin{figure}
    \centering
    \includegraphics[width=0.45\textwidth]{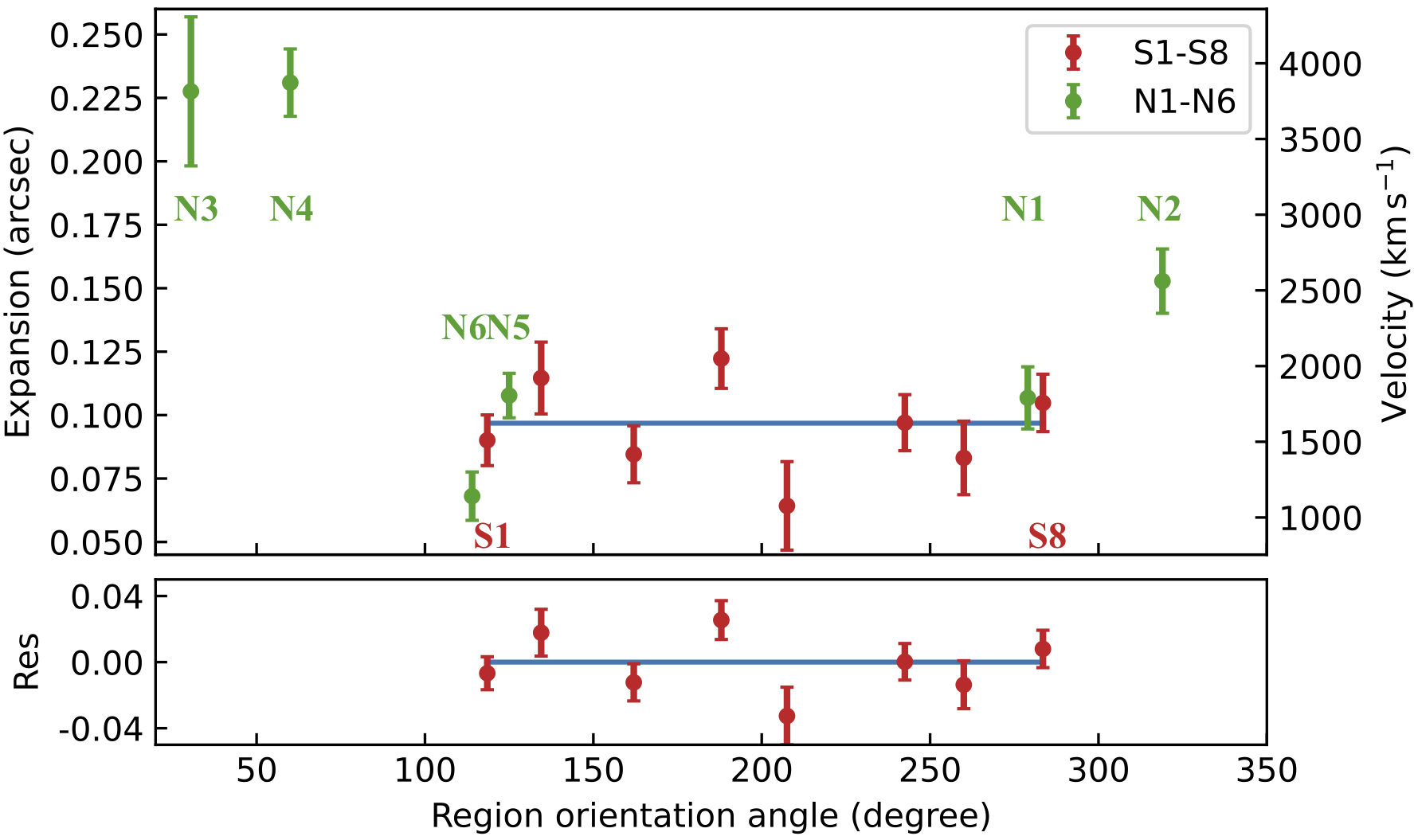}
    \caption{{As in Figure~\ref{fig:expansionlinear}, but for the final expansion estimates.}
    The expansion of each region 
    (``corrected shifts'' in Table~\ref{tab:expansion_regions}) is based on the raw shifts
    shown in Figure~\ref{fig:expansionlinear}, corrected for the PSF bias and COE bias. The blue solid line shows the average expansion of the S1-S8 regions.\label{fig:expansioncorrectregistration}}
\end{figure}

We also consider a systematic error in the effective expansion center we assumed for the southern rim. We initially assume that the southern regions are expanding uniformly.  Any bias in the effective expansion center would tend to increase/decrease the measured expansion in directions 180\degree opposite to each other, resulting in a sinusoidal variation with azimuth. We fit the measured expansion of the southern regions (corrected for PSF bias) with a sinusoidal function as described in Appendix~\ref{sec:app_psf_diff} and shown in the blue curve in Figure~\ref{fig:expansioncorrectbias}. 
This fit indicated that the COE should be shifted by $0\farcs05$ in RA and $0\farcs02$ in Dec if the eight southern regions are assumed to have a uniform expansion.
We subtract this sinusoidal variation 
to obtain the final expansion results listed as the ``corrected shifts'' in Table~\ref{tab:expansion_regions} and shown in
Figure~\ref{fig:expansioncorrectregistration}.

\begin{figure}
    \centering
    \includegraphics[width=0.45\textwidth]{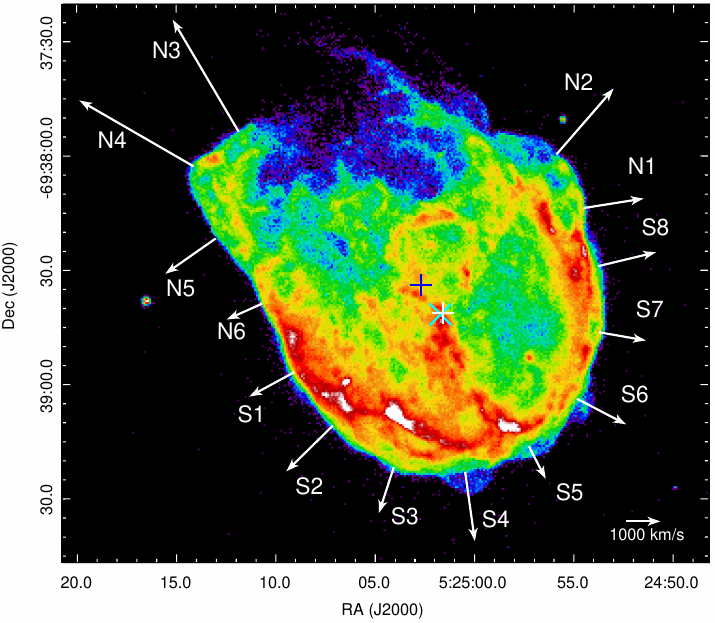}
    \caption{
    The expansion of N132D along selected regions around the rim.  The arrows point in the direction of the orientation of the regions (see Table~\ref{tab:expansion_regions}), and their lengths are proportional to the estimated expansion velocities.  The center of expansion (COE; white cross) is estimated assuming that the regions along the southern rim (S1-S8) lie on a circular rim, and their orientations closely follow the azimuthal locations of the rim (for comparison, the COE of the optical knots are also shown as the cyan cross \citep{banovetz2023} and the blue plus \citep{morse1995}).  The orientations of the regions along the northern rim (N1-N6) are substantially different from their azimuthal locations relative to the COE.
    \label{fig:expansion.vector}}
\end{figure}

The comparison between Figures~\ref{fig:expansionlinear} and~\ref{fig:expansioncorrectregistration} shows that the combined effect of these two corrections is small and the average expansion for the eight southern regions only changes by  $\sim13\%$ after the corrections are applied.
The average expansion of the 8 southern regions is $0\farcs10 \pm 0\farcs02$ after applying both corrections (see ``corrected'' shift in Table~\ref{tab:expansion_result}), giving an average expansion velocity of $1620 \pm 400\, \mathrm{km\,s^{-1}}$.  
The relation of the expansion of the southern regions to the average  expansion is similar to the ``raw shifts" results in that seven of the eight regions are consistent with a constant expansion rate at the $2.0\sigma$ level, with region S4 being the sole outlier at $2.5\sigma$.  Regions N1 and N5 are also consistent with the average expansion rate while region N2 is more than $3\sigma$ higher and N6 is $3\sigma$ lower.  The lower expansion for the N6 region might be due to the fact that our method is only sensitive to the velocity component that is in the same direction as the normal to the long axis of the region or a possible impedance by the molecular material seen in the ${\mathrm {^{12}CO}}$ maps of \cite{sano2017}.
The north eastern regions N3 and N4 have similar expansion values and given their proximity, we combine them to get an average expansion for the northeast rim.  The average raw expansion for regions N3 and N4 is  $0\farcs22\pm0\farcs02$
corresponding to a velocity of $3630\pm260\,\mathrm{km\,s^{-1}}$ (see Table ~\ref{tab:expansion_result}) and the corrected average expansion is $0\farcs23\pm0\farcs02$, with an average expansion velocity  of $3840\pm260\,\mathrm{km\,s^{-1}}$, clearly much larger than the average expansion rate of the eight southern regions. Regions N3 and N4 also have the largest radii with respect to the COE, consistent with their larger expansion values.  It is worth noting that our estimate of the COE shown in Figure~\ref{fig:expansion.vector} and listed in Table~\ref{tab:expcenter} assumed a symmetric explosion for the S1-S8 regions, while the COE analysis of \cite{banovetz2023} did not. Our COE estimate differs by $1\farcs5$ or 0.4~pc from the \cite{banovetz2023} COE; both of which difer from the \cite{morse1995} estimate by  $9\farcs2$ or 2.2~pc. The picture that emerges from this expansion analysis is that regions S1-S8 have a similar expansion history and similar current expansion rate while some of the regions N1-N6 have significantly different expansions histories and/or current expansion rates. 

\begin{table*}
\caption{Expansion Measurements of the Regions S1-S8 and N1-N6 } 
\begin{center}
\label{tab:expansion_regions}
\begin{tabular}{ l c c c c c c }
\hline\hline
 Region & Orientation$^a$ & Distance$^b$ & \multicolumn{2}{c}{Raw$^c$} & \multicolumn{2}{c}{Corrected$^d$} \\
 \cmidrule(lr){4-5} \cmidrule(lr){6-7}
        &                 &     &            shift & velocity & shift & velocity \\
 & [degree] & [arcsec] & [arcsec] & [$\mathrm{km\,s^{-1}}$] & [arcsec] & [$\mathrm{km\,s^{-1}}$] \\
\hline 
S1& 118.5 & 41.99 &$0\farcs098 \pm 0\farcs009$ & $1640 \pm 160$ & $0\farcs090 \pm 0\farcs009$ & $1510 \pm 160$\\
S2& 134.5 & 41.36 &$0\farcs127 \pm 0\farcs014$ & $2130 \pm 230$ & $0\farcs115 \pm 0\farcs014$ & $1920 \pm 230$\\
S3& 162.0 & 42.53 &$0\farcs104 \pm 0\farcs011$ & $1740 \pm 180$ & $0\farcs085 \pm 0\farcs011$ & $1420 \pm 180$\\
S4& 188.0 & 42.19 &$0\farcs145 \pm 0\farcs011$ & $2420 \pm 180$ & $0\farcs122 \pm 0\farcs011$ & $2050 \pm 180$\\
S5& 207.5 & 41.64 &$0\farcs086 \pm 0\farcs017$ & $1450 \pm 290$ & $0\farcs064 \pm 0\farcs017$ & $1080 \pm 290$\\
S6& 242.5 & 41.78 &$0\farcs113 \pm 0\farcs010$ & $1900 \pm 170$ & $0\farcs097 \pm 0\farcs010$ & $1630 \pm 170$\\
S7& 260.0 & 41.29 &$0\farcs095 \pm 0\farcs014$ & $1590 \pm 240$ & $0\farcs083 \pm 0\farcs014$ & $1390 \pm 240$\\
S8& 283.5 & 42.42 &$0\farcs109 \pm 0\farcs011$ & $1820 \pm 180$ & $0\farcs105 \pm 0\farcs011$ & $1760 \pm 180$\\
N1& 279.0 & 45.98 &$0\farcs112 \pm 0\farcs012$ & $1880 \pm 190$ & $0\farcs107 \pm 0\farcs012$ & $1790 \pm 190$\\
N2& 319.0 & 51.11 &$0\farcs146 \pm 0\farcs012$ & $2440 \pm 200$ & $0\farcs153 \pm 0\farcs012$ & $2560 \pm 200$\\
N3& \phantom{0}30.5 & 71.64 &$0\farcs212 \pm 0\farcs029$ & $3560 \pm 480$ & $0\farcs228 \pm 0\farcs029$ & $3810 \pm 480$\\
N4& \phantom{0}60.0 & 76.11 &$0\farcs221 \pm 0\farcs012$ & $3700 \pm 210$ & $0\farcs231 \pm 0\farcs012$ & $3870 \pm 210$\\
N5& 125.0 & 62.71 &$0\farcs117 \pm 0\farcs008$ & $1970 \pm 130$ & $0\farcs108 \pm 0\farcs008$ & $1810 \pm 130$\\
N6& 114.0 & 47.53 &$0\farcs074 \pm 0\farcs009$ & $1240 \pm 150$ & $0\farcs068 \pm 0\farcs009$ & $1140 \pm 150$\\
\hline
\multicolumn{7}{l}{
  \parbox[t]{0.7\textwidth}{
    $a:$ The counter-clockwise angle of the direction of the normal vector of the regions from North, as in Figure~\ref{fig:expansion.vector}.
  }
} \\
\multicolumn{7}{l}{$b:$ The distance of region centers to the center of expansion in Table~\ref{tab:expcenter}} \\
\multicolumn{7}{l}{$c:$ Raw estimates of expansion, prior to correction for PSF bias and expansion center shift} \\
\multicolumn{7}{l}{$d:$ PSF bias and effective expansion center bias corrected result} \\
\end{tabular}
\end{center}
\end{table*}

\begin{table*}[htb]
\caption{Average Expansion for the Southern Rim and NE Rim: Raw and Corrected Estimates}
\begin{center}
\label{tab:expansion_result}
\begin{tabular}{l c c c c c }
\hline\hline
 Measurement & Average expansion (velocity) & {\sl Within} error$^a$ & {\sl Between} error$^a$ & {\sl Total} error & $\chi^{2}$ (DoF)\\
\hline 
Southern(raw) & $0\farcs111$ (1850 $\mathrm{km\,s^{-1}}$)&$0\farcs012$ &$0\farcs019$&$0\farcs023$& 16.60(7)\\
Southern(corrected) & $0\farcs097$ (1620 $\mathrm{km\,s^{-1}}$)&$0\farcs013$ &$0\farcs019$&$0\farcs024$& 12.80(5)\\
\hline
N3+N4(raw) & $0\farcs217$ (3630 $\mathrm{km\,s^{-1}}$)       & -  & - & $0\farcs016$ & -\\
N3+N4(corrected) & $0\farcs230$ (3840 $\mathrm{km\,s^{-1}}$) & -  & - & $0\farcs016$ & - \\
\hline
\multicolumn{6}{l}{$a$: The scatter of measurements within and the mean estimated uncertainties, see Footnote~\ref{foot:MultImp}.}
\end{tabular}
\end{center}
\end{table*}

\begin{table}[htbp]
\caption{Center of Expansion Estimates}
\begin{center}
\label{tab:expcenter}
\begin{tabular}{l l l}
\hline\hline
Expansion centers & RA(J2000) & DEC(J2000) \\
\hline
\citet{morse1995}           & 05h\,25m\,02.70s & $-69^{\circ}\, 38\arcmin\, 34\farcs00$\\
\citet{banovetz2023}           & 05h\,25m\,01.71s & $-69^{\circ}\, 38\arcmin\, 41\farcs64$\\
This work$^a$ & 05h\,25m\,01.59s & $-69^{\circ}\, 38\arcmin\, 41\farcs20$\\ 
\hline
\end{tabular}
\parbox{0.9\linewidth}{
\footnotesize $a:$ Estimated expansion center based on regions S1-S8}
\end{center}
\end{table}

\section{Spectral Analysis} \label{sec:spec_analysis}

We extracted spectra from the regions used for the expansion analysis as indicated in Figure~\ref{fig:n132d}. 
These narrow regions near the outer extent of the remnant should be dominated by emission from the ISM swept up by the forward shock and should have little or no contribution from ejecta emission. 
We used the set of 14 observations (summed exposure time of 574.8~ks) from \epochb that were registered to each other for the expansion analysis to ensure that the spectra were extracted from as close to the same region as possible.
Off-remnant background 
spectra were extracted for each observation individually given the different roll angles of the observations and point sources were excluded in these background regions.
Spectral files and the necessary response files for spectral fitting were generated with the \ciao tool \texttt{specextract}.

We fit the source spectrum in \xspec~version 12.11.0k \citep{arnaud1996} with a \texttt{vpshock} model \citep{borkowski2001} appropriate for a plane-parallel shock with a constant temperature. The \texttt{wilms}  \citep{wilms2000} abundances and the \texttt{vern} \citep{verner1995} photoelectric cross sections are used. We used a two-component absorption model, with one component (\texttt{tbabs}) for the Galactic line-of-sight absorption, fixed at $N_{\mathrm{H,Galactic}}= 0.047\times 10^{22}\,\mathrm{cm^{-2}}$ \citep{2016A&A...594A.116H}, and another (\texttt{tbvarabs}) for the LMC absorption, $N_{\mathrm{H,LMC}}$,  with the elemental abundances set to 0.5 $\times$ solar abundances.  
A background model was constructed consisting of detector and sky background components and fit to the background spectra simultaneously with the source spectra.  The detector background model was based on the 
ACIS instrumental background 
described in \cite{suzuki2021}.  The sky background model includes two thermal components  
for the foreground emission from the Galaxy and the LMC
and also a component resulting from N132D produced as the CCD is being read out
(the so-called ``transfer streak'').
We also included 
a non-thermal component for the extragalactic background.
The data were unbinned for the fitting process and have only been 
binned for display purposes. The C-stat was used as the fit statistic to determine the best fitted values of the parameters. 

\begin{figure}[htb]
    \centering
    \includegraphics[width=0.45\textwidth]{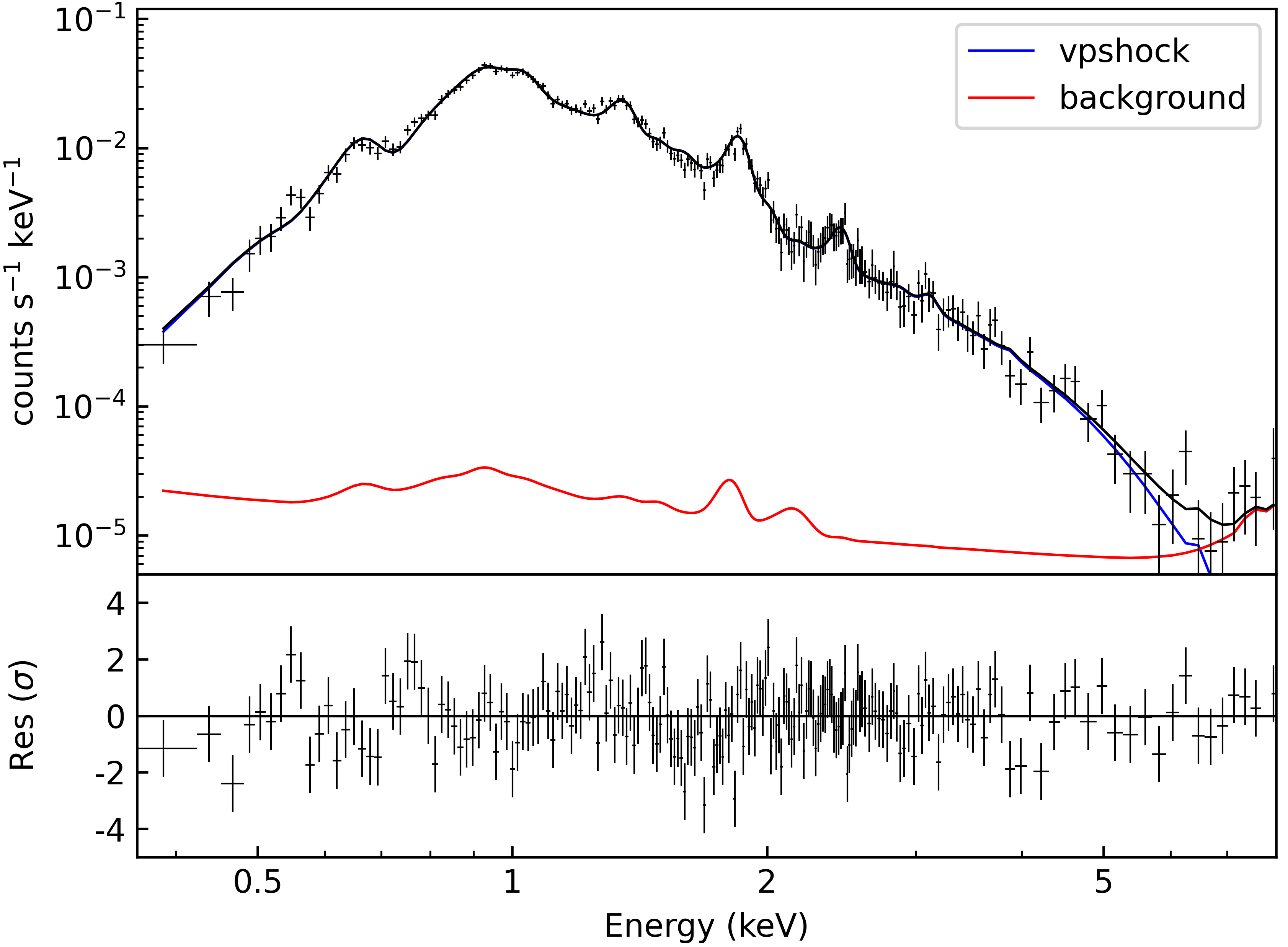}
    \caption{The spectrum of region S2 from the \epochb data fit with a vpshock model and the background model. The black points and line are the spectral data and model. The red line is the background model. The vpshock model parameters are listed in Table~\ref{tab:spec}.\label{fig:spec.s2}}
\end{figure}

A representative spectrum from region S2 from the \epochb data is displayed in Figure~\ref{fig:spec.s2} with the source and summed background models overplotted.
This figure shows that the source flux dominates the background in the 0.5--5.0~keV band given the brightness of N132D. Therefore most of the parameters of the source model are relatively insensitive to the background model.
The ACIS effective area has changed significantly between 2006 and 2019/2020 owing to the buildup of a contamination layer on the optical blocking filter (OBF) in front of the CCDs \citep{paul2016}. At the energy of the bright \ion{O}{8} Ly$\alpha$ line at 0.654\,keV, the effective area decreased by about a factor of four between the two epochs.  We therefore decided to fix the O abundance and $N_\mathrm{{H,LMC}}$ for \epochb as follows: we fit the \epocha spectrum for each region with a \texttt{vpshock} model allowing  the $N_\mathrm{{H,LMC}}$,
the O, Ne, Mg, Si, S, and Fe abundances, 
the ionization time scale, the temperature, and the normalization to vary in order to obtain the $N_{\mathrm{H,LMC}}$ and O abundance for each region. 
We then fit the \epochb spectrum for each region using a \texttt{vpshock} model with the O abundance and the $N_{\mathrm{H,LMC}}$ fixed at the \epocha values.  The \texttt{vpshock} abundances for Ne, Mg, Si, S, and Fe, the ionization timescale, the temperature, and the normalization were free to vary for the \epochb spectral fits. As seen in Figure~\ref{fig:spec.s2}, the \epochb spectra peak around 1.0~keV close to the energy of the Ne lines.  Therefore, useful constraints on the Ne, Mg, Si, S, and Fe abundances may be obtained from the \epochb spectra given the ACIS effective area at the time of those observations and the relatively large summed exposure of the \epochb observations.

\begin{table*}[t]
\begin{flushleft}
\rotatebox{90}{
\begin{minipage}{\textheight}
\begin{longtable}{ccccccccccccc}
\caption{Spectral Fit Parameters for the \epochb\ Spectra with the $1 \sigma$ Uncertainties Assuming a {\tt vpshock} Model.} \label{tab:spec} \\
\toprule
Region & $N_{H_{LMC}}$$^{\mathrm{a}}$ & $kT_{e}$ & $\tau$ & Norm & O$^{\mathrm{a}}$ & Ne & Mg & Si & S & Fe & C-stat & dof \\
 & ($10^{22} \mathrm{cm}^{-2}$) & (keV) & ($10^{11} \mathrm{cm}^{-3}\ \mathrm{s}$) & ($10^{-4}$) & & & & & & & & \\
\midrule
\endfirsthead
\toprule
Region & $N_{\mathrm{H,LMC}}$$^{\mathrm{a}}$ & $kT_{e}$ & $\tau$ & Norm & O$^{\mathrm{a}}$ & Ne & Mg & Si & S & Fe & C-stat & dof \\
 & ($10^{22} \mathrm{cm}^{-2}$) & (keV) & ($10^{11} \mathrm{cm}^{-3}\ \mathrm{s}$) & ($10^{-4}$) & & & & & & & & \\
\midrule
\endhead
\bottomrule
\multicolumn{13}{l}{\footnotesize$^{\mathrm{a}}$ Parameter value fixed to the value determined from the spectral fit to the \epocha\ spectra.} \\
\endfoot

S1  & 0.12 & $0.98_{-0.02}^{+0.02}$ & $1.48_{-0.10}^{+0.10}$ & $6.16_{-0.24}^{+0.24}$ & $0.45$ & $0.52_{-0.03}^{+0.03}$ & $0.36_{-0.02}^{+0.02}$ & $0.49_{-0.03}^{+0.03}$ & $0.52_{-0.05}^{+0.06}$ & $0.37_{-0.02}^{+0.02}$ & 12824 & 14634 \\
S2  & 0.07 & $1.20_{-0.04}^{+0.04}$ & $1.39_{-0.12}^{+0.17}$ & $1.97_{-0.11}^{+0.15}$ & $0.66$ & $0.51_{-0.06}^{+0.06}$ & $0.47_{-0.05}^{+0.05}$ & $0.61_{-0.06}^{+0.05}$ & $0.40_{-0.07}^{+0.08}$ & $0.51_{-0.04}^{+0.04}$ & 12174 & 14634 \\
S3  & 0.15 & $1.14_{-0.04}^{+0.05}$ & $0.88_{-0.09}^{+0.09}$ & $1.44_{-0.10}^{+0.09}$ & $0.35$ & $0.40_{-0.05}^{+0.06}$ & $0.45_{-0.04}^{+0.05}$ & $0.55_{-0.06}^{+0.06}$ & $0.59_{-0.11}^{+0.12}$ & $0.51_{-0.05}^{+0.06}$ & 11787 & 14634  \\
S4  & 0.17 & $1.01_{-0.07}^{+0.10}$ & $1.54_{-0.33}^{+0.39}$ & $0.80_{-0.11}^{+0.11}$ & $0.60$ & $0.39_{-0.08}^{+0.09}$ & $0.51_{-0.07}^{+0.08}$ & $0.49_{-0.08}^{+0.08}$ & $0.66_{-0.15}^{+0.17}$ & $0.53_{-0.08}^{+0.09}$ & 11397 & 14634 \\
S5  & 0.11 & $0.94_{-0.03}^{+0.04}$ & $1.36_{-0.17}^{+0.11}$ & $2.12_{-0.15}^{+0.11}$ & $0.68$ & $0.45_{-0.05}^{+0.05}$ & $0.52_{-0.05}^{+0.05}$ & $0.61_{-0.05}^{+0.06}$ & $0.72_{-0.10}^{+0.12}$ & $0.62_{-0.05}^{+0.06}$ & 11945 & 14634 \\
S6  & 0.17 & $0.76_{-0.01}^{+0.02}$ & $1.09_{-0.09}^{+0.12}$ & $3.43_{-0.20}^{+0.19}$ & $0.38$ & $0.54_{-0.02}^{+0.03}$ & $0.45_{-0.03}^{+0.03}$ & $0.50_{-0.05}^{+0.04}$ & $0.65_{-0.10}^{+0.13}$ & $0.38_{-0.02}^{+0.03}$ & 11777 & 14634  \\
S7  & 0.05 & $0.79_{-0.02}^{+0.02}$ & $2.23_{-0.20}^{+0.20}$ & $2.04_{-0.12}^{+0.19}$ & $0.54$ & $0.64_{-0.05}^{+0.05}$ & $0.63_{-0.05}^{+0.04}$ & $0.62_{-0.06}^{+0.06}$ & $0.47_{-0.10}^{+0.10}$ & $0.49_{-0.05}^{+0.02}$ & 11276 & 14634  \\
S8  & 0.07 & $0.91_{-0.03}^{+0.04}$ & $0.98_{-0.12}^{+0.10}$ & $1.18_{-0.09}^{+0.07}$ & $0.48$ & $0.60_{-0.06}^{+0.06}$ & $0.46_{-0.05}^{+0.06}$ & $0.58_{-0.07}^{+0.08}$ & $0.46_{-0.12}^{+0.16}$ & $0.49_{-0.02}^{+0.06}$ & 11221 & 14634  \\
N1  & 0.09 & $0.81_{-0.04}^{+0.05}$ & $1.14_{-0.21}^{+0.19}$ & $1.20_{-0.13}^{+0.11}$ & $0.54$ & $0.64_{-0.05}^{+0.06}$ & $0.49_{-0.05}^{+0.06}$ & $0.57_{-0.07}^{+0.08}$ & $0.64_{-0.17}^{+0.18}$ & $0.40_{-0.05}^{+0.05}$ & 11078 & 14634  \\
N2  & 0.02 & $0.94_{-0.05}^{+0.06}$ & $0.61_{-0.09}^{+0.10}$ & $0.60_{-0.05}^{+0.05}$ & $0.43$ & $0.78_{-0.06}^{+0.07}$ & $0.50_{-0.06}^{+0.07}$ & $0.65_{-0.10}^{+0.11}$ & $0.84_{-0.25}^{+0.29}$ & $0.33_{-0.05}^{+0.05}$ & 10853 & 14634  \\
N3  & 0.05 & $0.77_{-0.03}^{+0.05}$ & $1.37_{-0.25}^{+0.24}$ & $1.25_{-0.13}^{+0.11}$ & $0.28$ & $0.40_{-0.05}^{+0.04}$ & $0.28_{-0.04}^{+0.05}$ & $0.44_{-0.07}^{+0.08}$ & $0.50_{-0.16}^{+0.18}$ & $0.35_{-0.04}^{+0.05}$ & 11220 & 14634 \\
N4  & 0.04 & $0.77_{-0.04}^{+0.06}$ & $1.15_{-0.21}^{+0.23}$ & $0.83_{-0.10}^{+0.09}$ & $0.32$ & $0.52_{-0.05}^{+0.06}$ & $0.26_{-0.05}^{+0.06}$ & $0.31_{-0.07}^{+0.08}$ & $0.52_{-0.20}^{+0.22}$ & $0.28_{-0.04}^{+0.05}$ & 10661 & 14634  \\
N5  & 0.01 & $0.72_{-0.03}^{+0.03}$ & $1.03_{-0.15}^{+0.12}$ & $1.14_{-0.09}^{+0.09}$ & $0.45$ & $0.67_{-0.05}^{+0.05}$ & $0.46_{-0.05}^{+0.06}$ & $0.53_{-0.08}^{+0.09}$ & $1.31_{-0.27}^{+0.30}$ & $0.49_{-0.05}^{+0.05}$ & 11045 & 14634  \\
N6  & 0.04 & $0.74_{-0.02}^{+0.03}$ & $1.65_{-0.18}^{+0.18}$ & $3.28_{-0.21}^{+0.20}$ & $0.40$ & $0.55_{-0.04}^{+0.03}$ & $0.51_{-0.04}^{+0.05}$ & $0.53_{-0.04}^{+0.04}$ & $0.66_{-0.12}^{+0.14}$ & $0.49_{-0.02}^{+0.05}$ & 11609 & 14634  \\
\end{longtable}
\end{minipage}
}
\end{flushleft}
\end{table*}

The fit results for the regions are listed in Table~\ref{tab:spec}. All of the fits are acceptable as determined by the value of the C-statistic and the degrees of freedom. There is no reason to explore a more complicated spectral model as a single \texttt{vpshock} component is sufficient to represent these moderate-resolution 
(${\mathrm{E/\Delta E\sim15}}$)
spectra to the statistical precision afforded by the number of counts in the spectra. The fitted values of the $N_\mathrm{{H,LMC}}$ indicate that the absorbing column is higher in the southern regions than the northern regions, a pattern which was described in \cite{sharda2020}. 
The fitted values of the electron temperature 
range from 0.76 to 1.20 $\mathrm{keV}$ in the southern regions and from 0.72 to 0.94 $\mathrm{keV}$ in the northern regions.  The highest temperatures are observed for regions S1-S4, while the lowest
temperatures are observed for regions S6 \& S7 and N3-N5.
The fitted values of the ionization timescale range from a minimum value of $0.61 \times 10^{11}\mathrm{cm^{-3}\,s}$ for the N2 region to a maximum value of $2.23 \times 10^{11}\mathrm{cm^{-3}\,s}$ for the S7 region. Nevertheless, ten of the fourteen fits are clustered around an ionization timescale of $1.0\pm0.5 \times 10^{11}\mathrm{cm^{-3}\,s}$ suggesting a similarity in the underlying spectra.
The element abundances are consistent with the LMC ISM abundances ($\approx0.5\times$solar).   This implies that these regions are free of a significant contribution from ejecta emission and are representative of the forward shock conditions. In general, these spectral fit results are in good agreement with those in  \cite{sharda2020} which were based on the \epocha data alone. 
Given that the southern regions S1-S8 have a similar expansion with respect to each other and the regions N3 and N4 have a similar expansion with respect to each other, it is instructive to explore the average properties of these regions.
The southern regions have an emission-measure-weighted average temperature and ionization timescale of $0.95\pm 0.17\,\mathrm{keV}$ 
and $1.39\pm 0.49\times 10^{11}\, \mathrm{cm}^{-3}\,\mathrm{s}$ respectively, where the uncertainties are the multiple-imputation total variances 
as described earlier. The regions N3 and N4 have an 
emission-measure-weighted average temperature and  ionization time scale of $0.77\pm0.04\,\mathrm{keV}$ and $1.28\pm 0.17\,\times 10^{11}\, \mathrm{cm}^{-3}\,\mathrm{s}$ respectively. 
One would naively expect the temperatures of the northern regions to be higher given that our measured shock velocities are higher in the north.  In \S{\ref{sec:temperature}} we discuss the relation of the temperature estimates based on fits to the X-ray spectra to the temperatures implied by the shock expansion estimates, and consider possible explanations for discrepancies.

\section{Temperature Comparison from Expansion versus Spectrum Model}
\label{sec:temperature}

In Section \ref{sec:exp_shock} we evaluated the velocities of the outer shocks
based on proper motion measurements for various regions on the rim: S1-S8 along the
roughly circular rim in the south, and N1-N6 at various positions in the
northeast and northwest.
In Section \ref{sec:spec_analysis}, we extracted spectra for those same regions
and evaluated the electron temperature based on the thermal plasma X-ray emission.

Because energy equilibration and ionization processes take time, the X-ray spectra
give the electron temperature for conditions far from complete equilibration.
In this section we discuss the degree to which the electron thermal pool has
equilibrated, and whether the X-ray temperatures are consistent with the 
ion temperatures estimated from the proper motion velocity.

For a steady, planar shock with velocity $v_\mathit{s,n}$ perpendicular to the shock interface, in an ideal
gas with $\gamma=5/3$ and particles of mass $m$,
the Rankine-Hugoniot conditions imply a temperature
\begin{equation}
kT = \frac{3}{16} m v_\mathit{s,n}^2 .
\end{equation}
In an electron-ion plasma, the mass-proportional electron temperature $kT_\mathit{e,mp}$, proton temperature $kT_\mathit{p,mp}$, and average temperature $kT_\mathit{av}$ are obtain by substituting $m$ with $m_\mathit{e}$, $m_\mathit{p}$, and $\mu m_\mathit{p}$, respectively. Here, $\mu$ is the mean mass per free particle in units of the proton mass (1 amu). For a fully-ionized plasma with ``cosmic'' abundances \citep{wilms2000}, $\mu \approx 0.61$.
The mass-proportional electron temperature is $T_\mathit{e,mp} = (m_\mathit{e}/m_\mathit{p}) T_\mathit{p,mp} \lll T_\mathit{e,mp}$, that is, the electrons start out cold (a few eV for typical SNR shock speeds). Some form of electron-ion equilibration is needed for there to be electron thermal temperatures of $\sim 1\,\mathrm{keV}$.
In a collisionless plasma, the equilibration between electron and ion temperatures proceeds via Coulomb interactions at least.
This process is slow  and the effective collisional mean-free-path can exceed the dimensions of a
SNR. In such cases,  collisionless equilibration in the shock
may be needed to explain the excess of electron temperature over the Coulomb-equilibrated value.

In Section \ref{sec:exp_shock} we found the average proper-motion expansion
velocity for the eight southern regions to be
$v_\mathit{s,n} = 1620\pm 400\,\mathrm{km}\,\mathrm{s}^{-1}$, while for N3 and N4 in the northeast, the average expansion velocity is higher at $v_\mathit{s,n} = 3840\pm 260 \, \mathrm{km}\,\mathrm{s}^{-1}$.
In Table~\ref{tab:proper.motion.ktav.ktp.kte} we show the resulting mean temperature based on proper motion, $kT_\mathit{pm,av}$, assuming a fully-ionized plasma with $\mu=0.61$, and the corresponding mass-proportional proton and electron temperatures, $kT_\mathit{p,mp}$ and $kT_\mathit{e,mp}$, respectively.
The measured X-ray-based temperatures in Section~\ref{sec:spec_analysis} (see Table~\ref{tab:spec})
are clearly far from full equilibration, but also elevated above the mass-proportional electron temperatures $kT_\mathit{e,mp}$.  Additional collisionless heating is not indicated since it would make the electron temperature discrepancy worse.

\begin{table*}
\caption{Temperatures derived from proper motion velocity (see text).}
\begin{center} 
\label{tab:proper.motion.ktav.ktp.kte}
\begin{tabular}{ l c c c c }
\hline
Regions & $v_\mathit{s,n}$ 
        & $kT_\mathit{pm,av}$ 
        & $kT_\mathit{p,mp}$ 
        & $kT_\mathit{e,mp}$
        \\
        & [$\kms$]
        & [$\kev$]
        & [$\kev$]
        & [$\kev$]
        \\
        \hline
S1-S8   & $1620\pm 400$
        & \phantom{0}$3.14^{+1.75}_{-1.36}$
        & \phantom{0}$5.15^{+2.86}_{-2.23}$
        & $2.80^{+0.16}_{-0.12}\times 10^{-3}$
        \\
N3-N4   & $3840\pm 260$
        & $17.6^{+2.5}_{-2.31}$
        & $28.0^{+4.1}_{-3.8}$
        & $1.57^{+0.22}_{-0.21}\times 10^{-3}$
        \\
\hline
\end{tabular}
\end{center}
\end{table*}

\begin{table*}[!htbp]
\caption{Temperatures based on proper motion \textit{vs} corrected temperatures}
\begin{center} 
\label{tab:kt.pmcorr.vs.kt.xr}

\begin{tabular}{ l c c c c c c c }
\hline
Regions & Velocity
        & \multicolumn{3}{c}{Derived from expansion velocity$^a$}
        & \multicolumn{2}{c}{Derived from X-ray fits}
        & Ratio
        \\
        \cmidrule(lr){2-2}
        \cmidrule(lr){3-5} \cmidrule(lr){6-7} \cmidrule(lr){8-8}
        & $v_\mathit{pm}$ 
        & $kT_\mathit{e,ce+ad}$
        & $kT_\mathit{p,ce+ad}$
        & $kT_\mathit{av,ce+ad}$
        & $kT_\mathit{e,x}$ 
        & $n_\mathit{e}\,t$
        & $kT_\mathit{e,x}/(kT_\mathit{e,ce+ad})$
        \\
        & [$\kms$]
        & [$\kev$]
        & [$\kev$]
        & [$\kev$]
        & [$\kev$]
        & [$\mathrm{cm}^{-3}\,\mathrm{s}$]
        &
        \\
        \hline
S1-S8   & $1620\pm 400$
        & $0.80\pm 0.39$
        & $\phantom{0}3.90\pm 1.90$
        & $\phantom{0}2.87\pm 0.62$
        & $0.95\pm 0.17$
        & $(1.39\pm 0.49)\times 10^{11}$
        & $1.19\pm 0.62$
        \\
N3-N4   & $3840\pm 260$
        & $1.74\pm 0.22$
        & $25.84\pm 3.27$
        & $16.82\pm 2.13$
        & $0.77\pm 0.04$
        & $(1.28\pm 0.17)\times 10^{11}$
        & $0.42\pm 0.06$
        \\
\hline\multicolumn{5}{l}{$^a$ `ce+ad': after Coulomb equilibration and  adiabatic expansion}
\end{tabular}
\end{center}
\end{table*}



There must be at least Coulomb equilibration in the plasma,
which increases $T_\mathit{e}$ from the mass-proportional value
$T_\mathit{e0} = (m_\mathit{e}/m_\mathit{p})\,T_\mathit{p0}$ and correspondingly reduces the proton temperatures $T_\mathit{p}$.
Because the flows are expanding, adiabatic expansion will also slightly reduce
$T_\mathit{e}$.  In the following, we examine these effects.

The temperatures of the electrons and protons will equilibrate via Coulomb collisions. To estimate this equilibration, we assume a {fully ionized H plasma, $n_e = n_p$.} A pair of coupled differential equations 
determines the time evolution of the proton and electron temperatures:
\begin{equation}
\frac{d T_\mathit{e}}{d (n_\mathit{e}t)}
  = -\frac{d T_\mathit{p}}{d (n_\mathit{e} t)} = 0.13\left(\frac{T_\mathit{p}-T_\mathit{e}}{2T_{e}^{3/2}}\right)
 \label{eqn:te_tp}
\end{equation}
\citep[see their Eqs.\ (B7), (B8) for $Z=1$]{laming2003}, where densities are in units of $\mathrm{cm}^{-3}$, $t$ is in s and $T_{p}$ and $T_{e}$ are in units of K.
We compute the plasma temperature in thin shells behind the shock and integrate Equation~\ref{eqn:te_tp} from $n_\mathit{e}\,t = 0$ at the  shock to $1.39\times10^{11}\,\mathrm{cm}^{-3}\,\mathrm{s}$ at the innermost shell, with $\Delta (n_\mathit{e} \,t)$ constant for each shell. We solve the equations using Runge-Kutta integration \texttt{scipy.integrate.solve\_ivp}, and compute the emission-measure-weighted temperatures across the width of the shock.  Combined with the estimated shock velocity of $v_\mathit{pm} = 1620 \pm 400$~km~s$^{-1}$ for the southern rim, this yields an emission-measure-weighted electron temperature $kT_\mathit{e} = 0.89 \pm 0.43$~keV.  The uncertainty is estimated by propagating the fractional uncertainty due to the uncertainty on the velocity estimate.  A similar calculation for the northeastern regions, adopting a forward shock velocity of $v_\mathit{pm} = 3840 \pm 260$~km~s$^{-1}$, and carrying out the integration over a range  $n_\mathit{e}t=0$ to $1.28\times10^{11}$~cm$^{-3}$~s$^{-1}$, yields an emission-measure weighted electron temperature $kT_\mathit{e}=1.84 \pm 0.23$\,keV.

For adiabatic cooling, the integration limits for each shell are set by assuming homologous expansion from a common center (see Table~\ref{tab:expcenter}), and that the radial distance determines the time scale for the cooling.
Including the effects of adiabatic cooling reduces the temperature to $kT_\mathit{e}=0.80 \pm 0.39\,\mathrm{keV}$ for the southern rim and $kT_\mathit{e}=1.74 \pm 0.22\,\mathrm{keV}$ for the northeastern regions.

In Table~\ref{tab:kt.pmcorr.vs.kt.xr}, we compare the electron temperature $kT_{e,ce+ad}$, based on the proper motion velocity, $v_\mathit{pm}$, and corrected for Coulomb equilibration and adiabatic expansion, to the electron temperature, $kT_\mathit{e,x}$, based on the X-ray spectral fits. The $kT_\mathit{e,x}$ are emission-weighted values for regions S1 to S8 and for N3+N4. The temperature $kT_{e,ce+ad}$ for the southern rim is consistent with the temperature estimated from spectral fitting, but for the northeastern regions, $kT_{e,ce+ad}$ is significantly higher than $kT_\mathit{e,x}$.  

We discuss some possible origins of this discrepancy in the next sections.
In Section~\ref{subsubsec:shock.obliquity} we examine shock obliquity effects (see \citet{Shimoda2015}).  In Section~\ref{subsubsec:okada-iontenp}, we examine the temperatures and shock velocities obtained by \citet{okada25} for a set of N132D regions using the \texttt{IONTENP} \texttt{Xspec} table model developed by \citet{ohshiro24} which provides a self-consistent 1-D planar shock model. Finally, in Section~\ref{subsubsec:cr-acceleration}, we consider possible energy losses due to cosmic-ray acceleration.

\subsection{Rippled-Shock Obliquities}
  \label{subsubsec:shock.obliquity}
  
As discussed in \citet{Shimoda2015}, as a SNR blast wave propagates into a clumpy ISM the shock slows in the denser portions and moves faster in the lower density portions between so that the shock refracts around the denser clumps. This results in a shock interface which is rippled, with locally oblique shocks. For an oblique shock, the velocity component parallel to the shock interface is conserved, so that the temperature jump depends only on the velocity component normal to the shock.  If the angle between the flow velocity and the local shock normal is $\theta$, the component normal to the shock is $v_\textit{s,n} = v_\mathit{s} \cos\theta$. The downstream temperature is proportional to $v_\mathit{s,n}^2$, so the immediate downstream temperature is reduced by a factor of $\cos^2\theta$ :
\begin{equation}
\label{eqn:temperature.oblique.shock}
    kT_\mathit{ds} = {kT_\mathit{pm}} \cos^2\theta
                 \le kT_\mathit{pm}
\end{equation}
where $kT_\mathit{pm}$ is a temperature obtained from the proper-motion velocity, and $kT_\mathit{ds}$ is the temperature immediately downstream of a shock with angle $\theta$. For a shock propagating through a clumpy medium, the postshock temperature is reduced by a factor of $\langle\cos^2 \theta\rangle$ which is an average of the deflections within the region under consideration.

\citet{Shimoda2015} assume mean interstellar gas density $\langle\rho\rangle_0$ with dispersion $\Delta \rho = (\langle\rho\rangle^2 -\langle\rho\rangle^2_0)^{1/2}$. To estimate the scales, they note that the turbulence in the interstellar medium is likely driven by SNRs with an injection length scale of $L_\mathit{inj} \sim 100\,\mathrm{pc}$ with density fluctuations at the injection scale of $\Delta \rho_{L_\mathit{inj}} \sim 1$.
In the turbulent medium these fluctuations cascade to smaller scales.
They argue that the typical dispersion at scale $\lambda$ is $\Delta\rho_\lambda/\langle\rho\rangle \simeq (\lambda/L_\mathit{inj})^{1/3}$ and suggest that for $\lambda \sim 2\,\mathrm{pc}$, $\Delta\rho/\langle\rho\rangle_0 \sim 0.3$, a typical ISM value.  To simplify the notation, we define $w \equiv \Delta\rho / \langle\rho\rangle_0$.

\begin{figure*}
    \centering
    \includegraphics[width=1.0\textwidth]{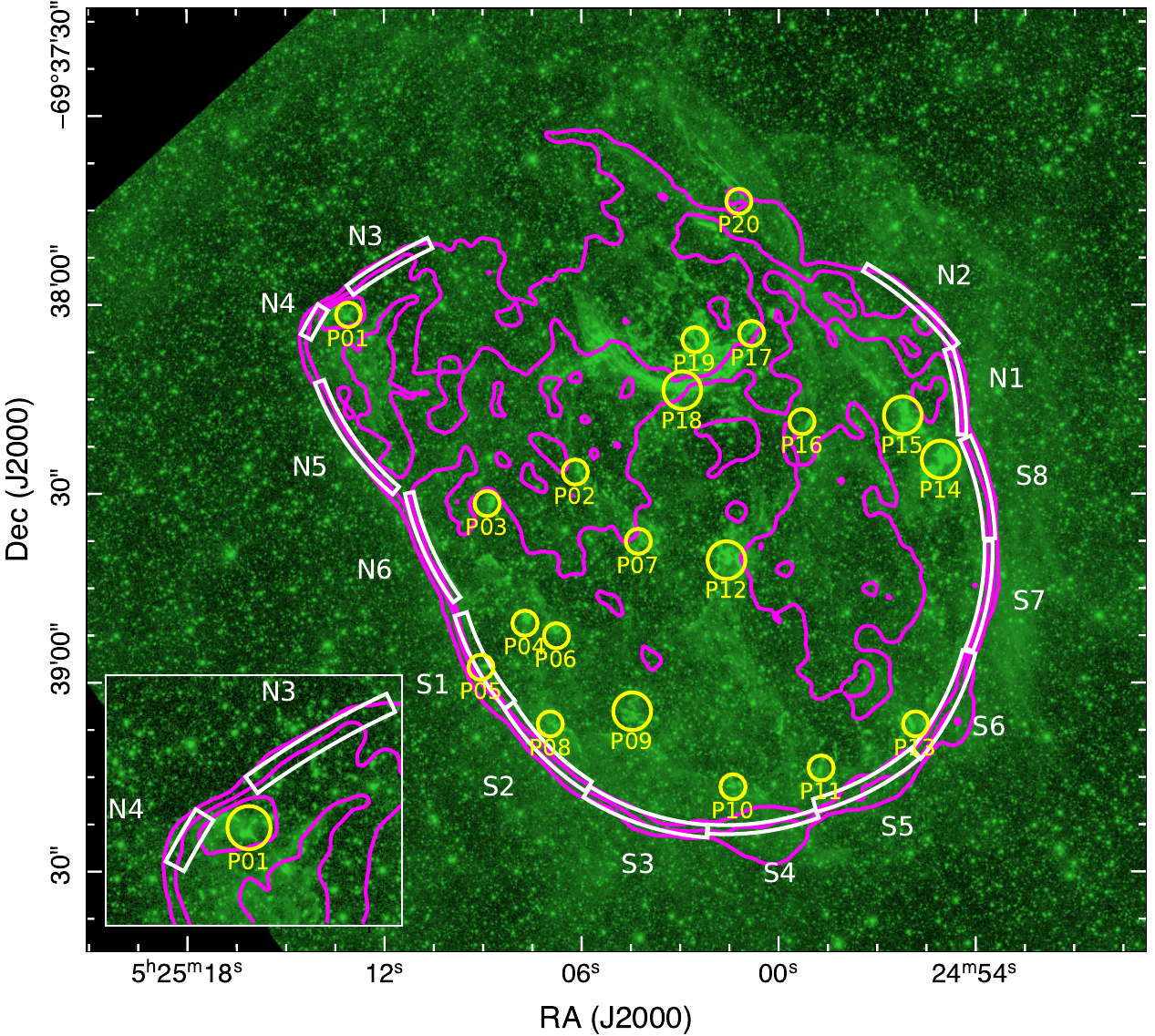}
    \caption{\citet{dopita2018} shocked H$\alpha$-cloud regions displayed on the \hst H$\alpha$ (F658N) image with \chandra 2019/2020  (1.2--7.0 keV) contours in purple. Our expansion and spectral analysis regions 
    are plotted in white. The inset shows an expanded view of the northeast.  Cloud P01 shows indications of having interacted with the blastwave in the northeast -- see the indentation in the X-ray contours in the inset and the X-ray enhancement in Figure~\ref{fig:n132d}. The small cloud P05 shows no indication of significant interaction in region S1. \label{fig:hi.clouds}}
\end{figure*}

Direct measurements of the fluctuation scales in the neighborhood of N132D are not available.
For the southern parts of N132D the tangential length scales are $\sim 3\,\mathrm{pc}$ for the faint protrusions beyond regions S4, S5, and S6 (see Figure~\ref{fig:n132d}), suggesting fluctuation scales of that length in the ambient medium. The scales for regions N3 and N4 are 3.5\,pc and 1.2\,pc, respectively.
Using \hst imaging followed up with Wide Field Spectrograph (WiFeS) spectroscopy, \citet{dopita2018} have identified a number of \ion{H}{1} clouds of about 1\,pc in size interacting with N132D (see their Figures 1 and 2, and our Figure~\ref{fig:hi.clouds} in which the clouds and our extraction regions are plotted on the HST H$\alpha$ (F658N) image). Based on the above scales one might expect $w \sim 0.2$ or 0.3.

\citet{Shimoda2015} obtain an analytical estimate
to constrain
$\eta \equiv (T_\mathit{pm} - T_\mathit{ds}) / T_\mathit{pm}=1 - T_\mathit{ds}/T_\mathit{pm}$
based on $w$ 
where $T_\mathit{pm}$  and $T_\mathit{ds}$ are the average temperatures of all particles:
\begin{equation}
  w^2 \lesssim \eta \lesssim 2 w
\end{equation}
We thus estimate

\parbox{7cm}{
\begin{eqnarray*}
\label{eqn:eta.ranges}
w &=& 0.2: \qquad 0.04 \lesssim \eta \lesssim 0.4 \\
w  &=& 0.3: \qquad 0.09 \lesssim \eta \lesssim 0.6
\end{eqnarray*}
} \hfill
\parbox{1cm}{\begin{eqnarray}\end{eqnarray}}

For the N3+N4 regions in the northeast, the mean temperature based on the proper motion velocity is $kT_\mathit{pm,av}=17.6^{+2.5}_{-2.3}\,\mathrm{keV}$ 
(see Table~\ref{tab:proper.motion.ktav.ktp.kte}). If we take into account Coulomb equilibration and adiabatic expansion, the mean temperature based on the proper motion velocity is reduced to $16.8\pm2.1\,\mathrm{keV}$, with electron temperature becoming $1.74\pm0.22\,\kev$ (see Table~\ref{tab:kt.pmcorr.vs.kt.xr}).  This electron temperature still  greatly exceeds the $0.77\pm0.04\,\mathrm{keV}$ measured from the X-ray spectra.

To assess the degree to which shock obliquities could account for the difference, we estimate the initial mean temperature, $kT_\mathit{av,0} = 2.85\,\mathit{keV}$, such that the mass-proportional electron temperature, $kT_\mathit{e0} = 2.5\times 10^{-3}\,\mathrm{keV}$ reaches the X-ray based electron temperature after Coulomb equilibration and adiabatic expansion: $kT_\mathit{e0,ce+ad} \equiv kT_\mathit{e,x} = 0.77\,keV$.
Taking $kT_\mathit{av,0} = 2.85\,\mathrm{keV}$ as $kT_\mathit{ds}$ and comparing to  the temperature based on proper motion velocity, $kT_\mathit{pm,av} = 17.6\,\mathrm{keV}$ implies $\eta = 0.84$, 
which is well beyond the ranges for shock obliquity effects given in Equation~\ref{eqn:eta.ranges} for $w=0.2$ or $w=0.3$. Propagation into a clumpy medium and the resulting shock obliquities may explain part, but not all, of the discrepancies seen in the northeast.

Besides the likely shock obliquities due to propagation into a clumpy medium, N132D also shows larger-scale oblique shocks such as regions N6 and N5 in the northeast and N1 in the northwest, whose shock normals are at an angle to the direction to the expansion center. Note that our approach to measuring expansion necessarily provides the normal shock velocity $v_\mathit{s,n}$ -- it is insensitive to  velocity components tangent to the shock surface. Thus, for these shocks, additional corrections for shock obliquity are not needed.  An exception is the N4 region, which does not fully follow the shock surface.  The edge of the remnant within the N4 region has obliquities on the sides of 20\degree to 30\degree, resulting in a temperature reduction of 12\% to 25\% based on Equation~\ref{eqn:temperature.oblique.shock}. Assuming that the region approximately divides into three azimuthal pieces with obliquities of 20\degree, 0\degree, 30\degree, respectively, 
the net temperature reduction would be $\sim$12\%, not enough to explain the temperature discrepancy.

Finally, the northeast is the site of an \ion{H}{1} cloud/shock interaction \citep[cloud P01, see their Figs. 2 and 3]{dopita2018}. In Figure~\ref{fig:hi.clouds}, we plot the HST H$\alpha$ image (F658N) with the Dopita et al. cloud positions plotted as yellow circles. We also plot Chandra X-ray contours (1.2-7.0 keV) and our extraction regions S1..S8 and N1..N6. The inset figure shows a blowup of the N3 and N4 region. The P01 cloud coincides with an X-ray enhancement, likely produced by the increased pressure in the reflected shock region. The overpressure drives a slower shock into the cloud, producing the optical emission (H$\alpha$ emission, and [\ion{O}{3}] emission as the cloud shock becomes radiative). The X-ray rim to the north also shows a slight concavity between regions N3 and the eastern part of N4, indicative of oblique shocks generated as the shock refracts around the \ion{H}{1} cloud.
In selecting the N4 and N3 extraction regions, we attempted to avoid 
the cloud interaction, but it is possible that they still include some 
of the oblique shocks in the cloud interaction. If the cloud had a diffuse
\ion{H}{1} envelope, this would also decelerate the shock and might
explain part of the lower temperatures obtained in the X-ray spectral fitting.


\subsection{Shock Model Fits}
  \label{subsubsec:okada-iontenp}
  
\citet{ohshiro24} have implemented an \texttt{Xspec} table model, \texttt{IONTENP}, for a self-consistent one-dimensional shock model in which the changing electron density and temperature states are followed downstream of the shock, including the effects of Coulomb equilibration between ions and electrons. The shock velocity is a parameter in the fitting.  This is in contrast to the \texttt{Xspec} \texttt{NEI} class of models (e.g., \texttt{VNEI} or \texttt{PSHOCK})) where the postshock electron temperature and the density are held constant.  \citet{ohshiro24} compared synthesized X-ray spectra (0.5-2\,keV) for an \texttt{IONTENP} model with an \texttt{NEI} model with the same ionization timescale ($10^{11}\,\mathrm{cm}^{-3}\,\mathrm{s}$). The \texttt{IONTENP} model assumed a velocity of $1000 \,\mathrm{km}\,\mathrm{s}^{-1}$ and $kT_\mathit{e}/kT_\mathit{ion} = 0.01$, where $kT_\mathit{ion}$ is the mean ion temperature immediately behind the shock. The \texttt{NEI} model was used to synthesize a corresponding spectrum for $kT_\mathit{e} = 0.91\,\mathrm{keV}$, which is the temperature at $10^{11}\,\mathrm{cm}^{-3}\,\mathrm{s}$ in the \texttt{IONTENP} model.
They find that $n_\mathit{e}\,t$ is about 30\% smaller for \texttt{IONTENP} than for \texttt{NEI}, and the emission differs at lower energies since the plasma is cooler as energy is taken by the ionization processes.

\citet{ohshiro24} examined a few regions of N132D and \citet{okada25} conducted a more extensive analysis of the rim of N132D using merged \chandra LP data totaling 838\,ks.  Compared to the current work, the
data were not spatially co-aligned before merging, and the selection of regions differ. In the present work, the regions are spatially narrower ($1\farcs 3 - 2\farcs 1$) extending past the shock surface in order to capture the edge of the shock structure, avoiding the brighter material interior to the shock.   \citet{okada25} use regions which are thicker ($3\arcsec$) and the regions appear to be somewhat inside the edge (see their Figure 1). In addition, there are partial azimuthal overlaps with our regions, and some of their regions (r16, r17) do not overlap our regions at all. Our N4 region has no significant overlap with their region r1. Their regions r5-r10 approximately coincide with our S1-S5, and r11-r13 approximately coincide with S6-S7.  

For r5-r10 \textit{vs.} S1-S5, the emission-weighted temperatures are consistent: $kT_\mathit{e} = 1.11\pm 0.10\,\mathrm{keV}$ and $1.03\pm 0.03\,\mathrm{keV}$, respectively. For S6-S7 compared to r11-r13, our temperature is about 14\% lower. For r2+r3 \textit{vs.} S5, we are about 7\% higher, but within the uncertainties. 

For r1 (approximately overlapping our N4 region and extending into the gap between N4 and N3), the temperature agrees well with our N4 and N3 regions. They also compare $v_\mathit{th,xray}$ with $v_\mathit{pm}$ where the $v_\mathit{th,xray}$ is the shock velocity estimated from spectral fitting, and $v_\mathit{pm}$ is the proper motion velocity. They find that region r1 is the most discrepant. However, this region has significant overlap with the \citet{dopita2018} P01 \ion{H}{1} shock/cloud interaction (see section \ref{subsubsec:shock.obliquity}). 
The region includes multicomponent plasmas including the reflected shock from the cloud interaction and the oblique shocks as the blast wave refracts around the cloud. As noted in section \ref{subsubsec:shock.obliquity}, a possible cloud envelope and the oblique refracted shocks may also result in cooler X-ray temperatures   which may be responsible for some of the r1 discrepancy.
The reflected shock region (see Figure~\ref{fig:hi.clouds} inset) includes double-shocked material (blast wave, then reflected shock) which further compresses and heats the plasma. This double-shocked material and oblique shocks complicate the interpretation of the $v_\mathit{th,xray}$ parameter of the \texttt{IONTENP} model; the reflected shock results in multiple temperatures from the two shocks (blast wave and shock reflection).

\subsection{Cosmic Ray Acceleration}
  \label{subsubsec:cr-acceleration}

\begin{figure}[htb]
    \centering
    \includegraphics[width=0.45\textwidth]{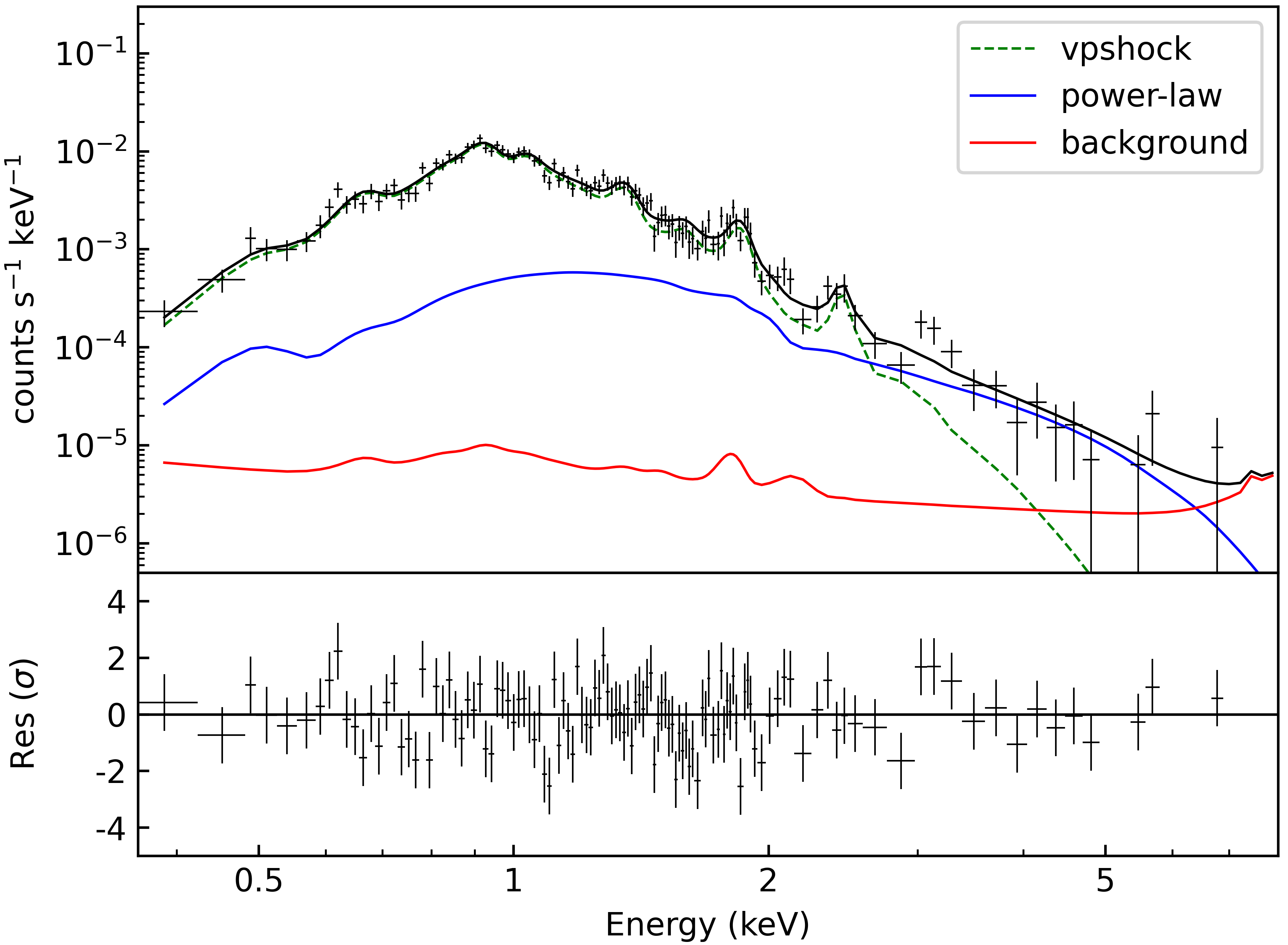}
    \caption{The spectrum of shock region N4 fit with a vpshock + power-law model and the background model. The black points and line are the spectral data and model. The green dashed line is the vpshock component. The blue line is the power-law component with an index of 3. The red line is the background model. \label{fig:spec.n4} }
\end{figure}

SNRs in the Galaxy younger than 2,000~yr with shock speeds larger than $2,500~\kms$ have been observed to produce detectable X-ray synchrotron emission \citep{helder2012}.  If this synchrotron emission is produced by efficient cosmic ray acceleration by the shock, it would remove energy from the shock and reduce the shock temperature \citep{decourchelle2000,ellison2007,patnaude2009}.
We consider the possibility of electron synchrotron emission by adding a power-law component to the spectral model for region N4.  
We assume a power-law index of 3.0, and the resulting fit is shown in Figure~\ref{fig:spec.n4}. The c-stat(dof) for the original model is 10661(14634), and when the power-law component is added, we get c-stat(dof) 10655(14633). The power-law component does not significantly improve the fit and is not needed to explain the data. 
Given the power-law normalization, we can integrate the model flux over the energy band 0.35--8.0\,keV, to obtain
an upper limit on the power-law component flux of $1.68^{+0.21}_{-0.67}\times 10^{-14}\,\mathrm{erg}\,\mathrm{s}^{-1}\,\mathrm{cm}^{-2}$.  This compares to a flux in the thermal component of 
$1.76^{+0.08}_{-0.08}\times 10^{-13}\,\mathrm{erg}\,\mathrm{s}^{-1}\,\mathrm{cm}^{-2}$ in the  0.35--8.0\,keV band.  Therefore, the upper limit of any nonthermal contribution is less than $10\%$ of the total flux.
The intensity of the X-ray synchrotron emission depends on the density of electrons responsible for the emission and the strength and structure of the magnetic field \citep{houck2006,allen2008} in the emitting region.  The density is in general low in the northeast part of N132D  ($\sim0.01~\mathrm{amu\,cm^{-3}}$, see Section~\ref{subsec:remnant.evolution}) 
and our measured shock speed of $3620~\kms$ for the northeast should result in a cutoff frequency for the synchrotron emission around 0.5~keV \citep{helder2012}. Therefore, it seems plausible that any X-ray synchrotron emission is too faint to be detected in this region.
It is unlikely that cosmic-ray acceleration can explain the discrepancy in region N4 between $kT_\mathit{pm,ce+ad}$ and $T_\mathit{e,x}$.

\section{Supernova Remnant Models} 
\label{sec:discussion}

\subsection{One Dimensional Shock Models}
\label{sec:one.dimensional.shock.models}
 
We generate analytic solutions following the work of \citet{truelove99,truelove99erratum} (hereafter TM99), as extended by \citet{laming2003}, \citet{hwang2012} and \citet{micel2016}.
\citet{sedov}
noted that the Euler equations do not contain any dimensioned constants, and that dimensional aspects are introduced through initial and boundary conditions. 
Here, the initial conditions introduce three dimensioned parameters: 
$E$ (explosion energy in erg),
$M_\mathit{ej}$ (mass of the ejecta in solar masses), and $\rho_0$ (mass density of the preshock medium in units of $\mathrm{g}/\mathrm{cm}^3$).
For the ejecta distribution, the core is taken to have constant density for simplicity, surrounded by an envelope described by a power-law distribution with index $n$.
TM99 note that the distribution of mass and energy depends on $n$; for
$n < 3$, mass and energy are concentrated in the outer (high speed) ejecta,
while for $n > 5$, mass and energy are concentrated in the inner (low speed)
ejecta.  \cite{chevalier1994} suggest indices $n \ge 7$ for CCSNe and lower indices for Type~Ia SNe.
Our explorations showed that the results were relatively insensitive to $n$ for $n > 7$; in our treatment below we set $n = 9$ to be in the middle of the range of the values suggested by \cite{chevalier1994}.

The preshock ambient medium is also assumed to follow a power-law distribution
\begin{equation}
\rho \propto \rho_0(r) / r^\mathit{s}
\label{eqn:ambient.medium.profile}
\end{equation}
where $s$ is 0 or 2, and $\rho_0(r)$ is the ambient medium mass density just ahead of the shock. For the $s=0$ case (constant density ambient medium) the value of $\rho_0(r)$ is constant. For the $s=2$ case (ambient density falls as $1/r^2$, appropriate for a constant stellar wind), the value of $\rho_0(r)$ pertains to a particular choice for the blast wave radius, $R_\mathrm{b,0}$, so that $\rho(r) = \rho_0/(r/R_\mathrm{b,0})^2$.
The introduction of a power-law ambient
medium density distribution does not introduce additional dimensioned constants
and the ejecta structure function is a dimensionless function,
so asymptotic similarity solutions can still be constructed. Note that the $r^{-2}$ ambient medium is assumed to be stationary ($v_\mathit{wind} \equiv 0$). This implies that the velocity of the constant wind should be much smaller than the forward shock velocity. Incorporating a significant wind velocity would introduce an additional dimensioned parameter, significantly complicating the solution.  These two assumed profiles for the ambient medium are simplifications of the true 3-D structure which is undoubtedly more complex due to variations in the stellar wind properties over the life of the star and multidimensional effects that can produce hydrodynamical instabilities \citep{dwarkadas2005,dwarkadas2007}. More sophisticated models use a lognormal \citep{padoan2011ApJ} or lognormal and power-law \citep{burkhart2018} probability distribution function for the density of the medium.
Nevertheless, the global properties of the bubble as predicted by 1-D simulations should be similar to those from multidimensional simulations \citep{dwarkadas2007} as the differences will manifest themselves on smaller spatial scales.

TM99 produced a detailed treatment of the $s=0$ case, but only a limited treatment of the $s=2$ case. \cite{hwang2003}, \cite{laming2003}, and \cite{micel2016}
extend the treatment to a more detailed consideration of the $s=2$ case. We refer the reader to these papers for the details. Here, we touch mainly on those aspects relevant to our analysis.

\subsection{Remnant Evolution}
\label{subsec:remnant.evolution}

We define the blast-wave radius as $R_\mathit{b,0}$ and the preshock density as $\rho(r)$ for the following discussion. For $s=0$, $\rho(r) = \rho_0$ is constant, while for $s=2$, $\rho(r) = 
\rho_0/(r / R_\mathit{b,0})^\mathit{s}$.
For the $s=2$ case we follow \cite{micel2016},
while for the $s=0$ case, we follow TM99. 
The $s=2$ (constant wind) case is an oversimplification of the evolution of the star before the explosion; however, it is useful to contrast this with the uniform medium case (itself a simplification since the ambient medium is likely to be inhomogeneous).

Our models  depend on three dimensioned parameters, 
$E$, $M_\mathit{ej}$, and $\rho_0$, plus two dimensionless structure parameters, $n$ and $s$, for a total of five adjustable parameters.
The ambient medium mass density is
related to the hydrogen number density, $n_0$, by $n_0 = (\rho_0/\mu_\mathrm{H})/(1\,\mathrm{cm}^{-3})$ where $\mu_\mathrm{H}$ is the mean mass per hydrogen nucleus, assuming cosmic abundances.
As noted in \S\ref{sec:one.dimensional.shock.models}, the solutions are relatively insensitive to $n$ for $n\ge 7$, and we adopt $n=9$. We consider the $s=0$ and $s=2$ cases separately. For each $s$ value, we 
consider ejecta masses of 2, 4, 6, and 8\,\msol. 
We chose $E=1.5\times 10^{51}\,\mathrm{erg}$ as the explosion energy for these initial calculations but explore different values later in this section. 
Our observational constraints are the measured blast wave velocity, $v_\mathit{b}$,
and the observed blast wave radius, $R_\mathit{b}$.
The ejecta mass, $M_\mathit{ej}$, explosion energy, 
$E$, and structure parameters $n$ and $s$ are fixed, leaving a single free parameter, $\rho_0$, to be varied.  We vary $\rho_0$ until the blast-wave radius and velocity, $R_\mathit{b}$ and $v_\mathit{b}$, are matched.

Once the match is found, the remnant age can be evaluated based on the parameters of the model.
For the southern rim and the $s=0$ case, a circumstellar density of $\rho_0$ of $0.33^{+0.25}_{-0.12}~\mathrm{amu\,cm^{-3}}$ 
matches the observed $v_\mathit{b}$ and $R_\mathit{b}$.
The remnant ages of the $s=0$ case are consistent with the age estimates from the optical observations\citep{law2020,banovetz2023} as shown in Table~\ref{tab:model.results}. 
The calculated reverse shock radius ranges from $5.60^{+0.95}_{-2.33}\,\mathrm{pc}$ to $7.27^{+0.07}_{-0.28}\,\mathrm{pc}$, for ejecta mass, $M_\mathit{ej}$,
from 2 to 8~\msol, respectively. From the X-ray morphology it appears that the reverse shock is relatively close in radius to the forward shock in the southern rim of the remnant.
For the southern rim and the $s=2$ case, a circumstellar density of $\rho_0$ of $0.22^{+0.17}_{-0.08}~\mathrm{amu\,cm^{-3}}$ matches the observed $v_\mathit{b}$ and $R_\mathit{b}$.  However, the estimated ages are $\sim4200$~yr which are discrepant with the estimates from the optical at the $1.5\sigma$ level. In addition, the predicted location of the reverse shock is farther from the forward shock and closer to the center of the remnant with values ranging from $2.30^{+1.97}_{-2.30}\,\mathrm{pc}$ to $5.98^{+0.16}_{-0.71}\,\mathrm{pc}$.  The uniform ambient medium profile is more consistent with the estimated age from the optical and the apparent position of the reverse shock assuming this  1-D shock evolution model and our measured values of $v_\mathit{b}$ and $R_\mathit{b}$.
The relatively low value of the ambient medium density of $\rho_0$ of $0.33^{+0.25}_{-0.12}~\mathrm{amu\,cm^{-3}}$ is consistent with a lower than average density of the ISM which could have been produced by the stellar winds of the progenitor sweeping out a cavity in the molecular cloud complex and is much lower than the density of molecular H in the clouds surrounding N132D which range from $10^2-10^3~\mathrm{cm^{-3}}$ \citep{sano2020}.  For the northeast rim, a density of $0.010^{+0.002}_{-0.001}~\mathrm{amu\,cm^{-3}}$ for the s=0 case and $0.007^{+0.001}_{-0.001}~\mathrm{amu\,cm^{-3}}$ for the s=2 case matches the observed $v_\mathit{b}$ and $R_\mathit{b}$. These densities are significantly lower than what we derived for the southeastern rim and are consistent with the shock propagating into a low density medium in the northeast.

\begin{table*}[!htbp]
\caption{Shock Models for the Southern Rim Region}
\begin{center} 
\label{tab:model.results}
\begin{tabular}{ l l c c c c c c c c}
\hline\hline
Parameters & Symbol(units) & \multicolumn{4}{c}{$s=0$} & \multicolumn{4}{c}{$s=2$}\\
\hline
Ejecta mass & $M_\mathit{ej}(\msol)$& 2 & 4 & 6 & 8 & 2 & 4 & 6 & 8\\
Reverse shock radius &$\mathrm{R_{r}}$(pc)& $5.60^{+0.95}_{-2.33}$ & $6.88^{+0.31}_{-0.95}$ & $7.17^{+0.13}_{-0.49}$ & $7.27^{+0.07}_{-0.28}$ & $2.30^{+1.97}_{-2.30}$ & $4.99^{+0.76}_{-2.03}$ & $5.71^{+0.35}_{-1.16}$ & $5.98^{+0.16}_{-0.71}$ \\
Age & $\mathrm{yr}$ & $2700^{+790}_{-470}$ & $2860^{+760}_{-440}$ & $3010^{+740}_{-420}$ & $3140^{+710}_{-400}$ & $4190^{+1370}_{-820}$ & $4210^{+1360}_{-810}$ & $4250^{+1340}_{-790}$ & $4290^{+1320}_{-770}$ \\
\hline
\end{tabular}
\parbox{0.9\linewidth}{
\footnotesize $NOTE:$ Models for ejecta profile $n=9$, $R_{b}=10.4$~pc, and $v_{b}= 1620\pm400\,\mathrm{km~s}^{-1}$ assuming an explosion energy of $E=1.5 \times 10^{51}\,erg$ which corresponds to a circumstellar density ($\rho_{0}$) of $0.33^{+0.25}_{-0.12}$ $(\mathrm{amu~cm}^{-3})$ and $0.22^{+0.17}_{-0.08}$ $(\mathrm{amu~cm}^{-3})$ for the southern rim region for the $s=0$ and $s=2$ cases respectively.}
\end{center}
\end{table*}

\begin{figure*}[ht!]
\gridline{
  \fig{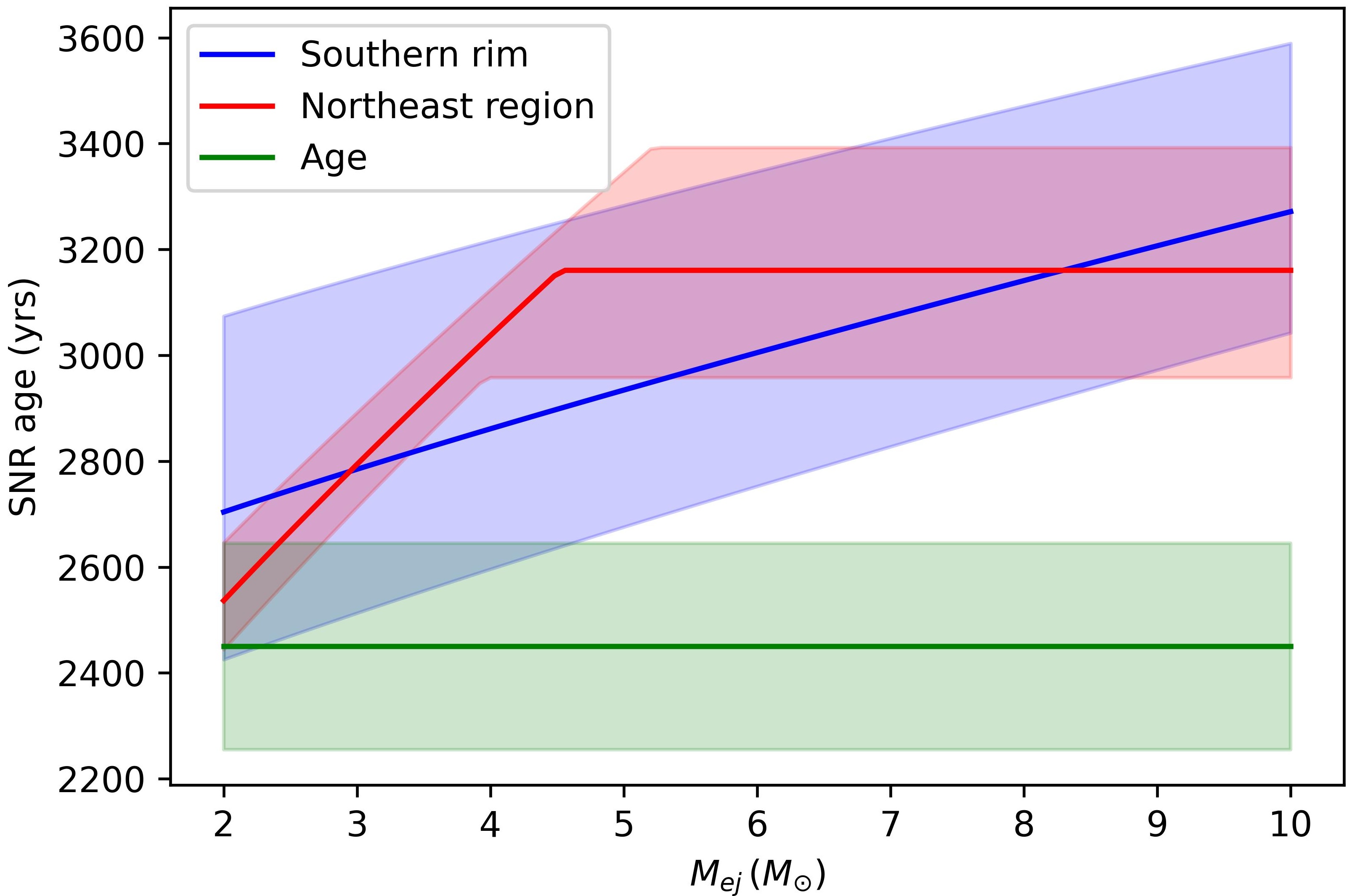}{0.48\textwidth}{(a) $E = 1.5 \times 10^{51}$ erg}
  \fig{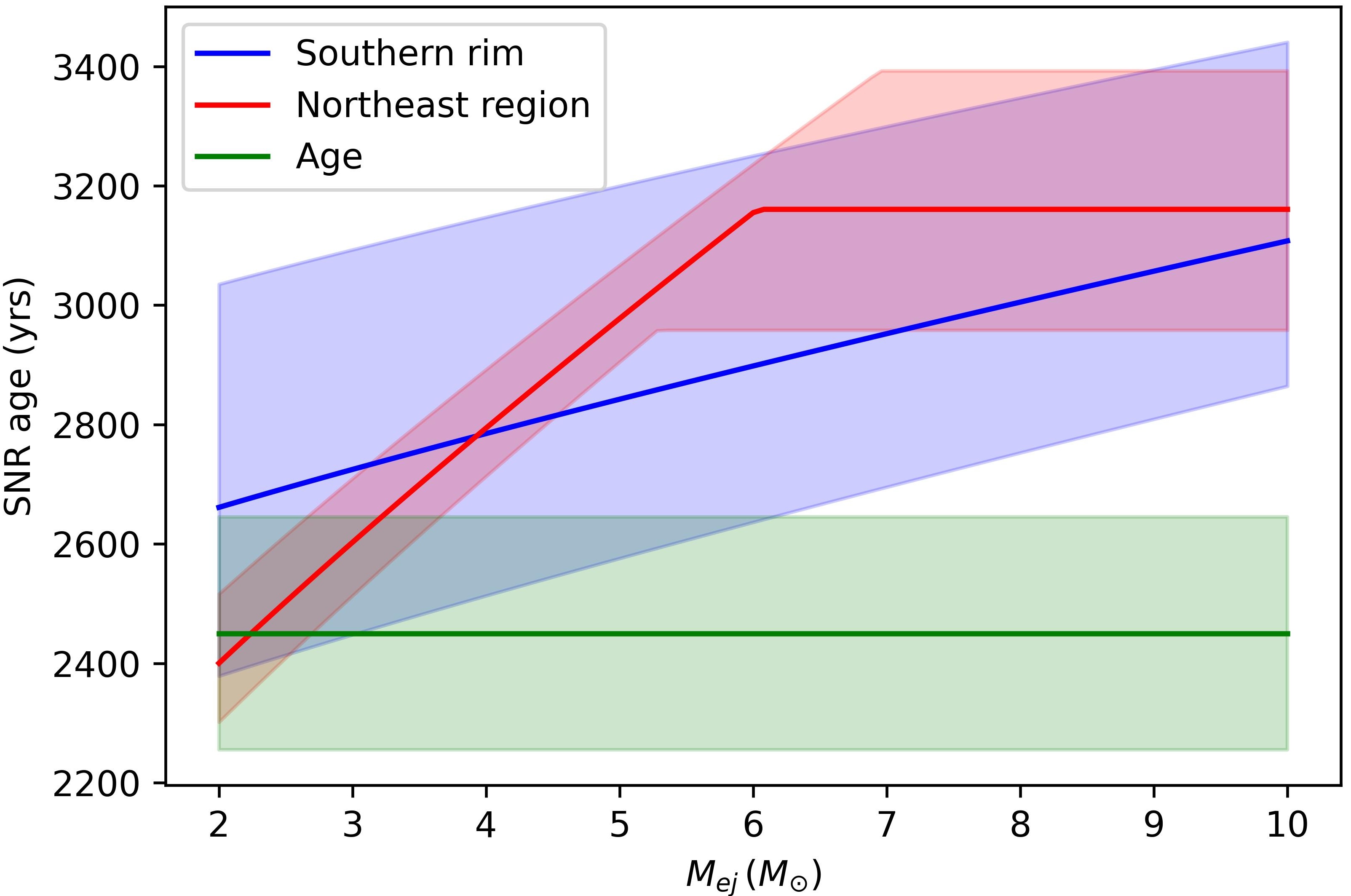}{0.48\textwidth}{(b) $E = 2.0 \times 10^{51}$ erg}
}
\vspace{-0.3cm}
\gridline{
  \fig{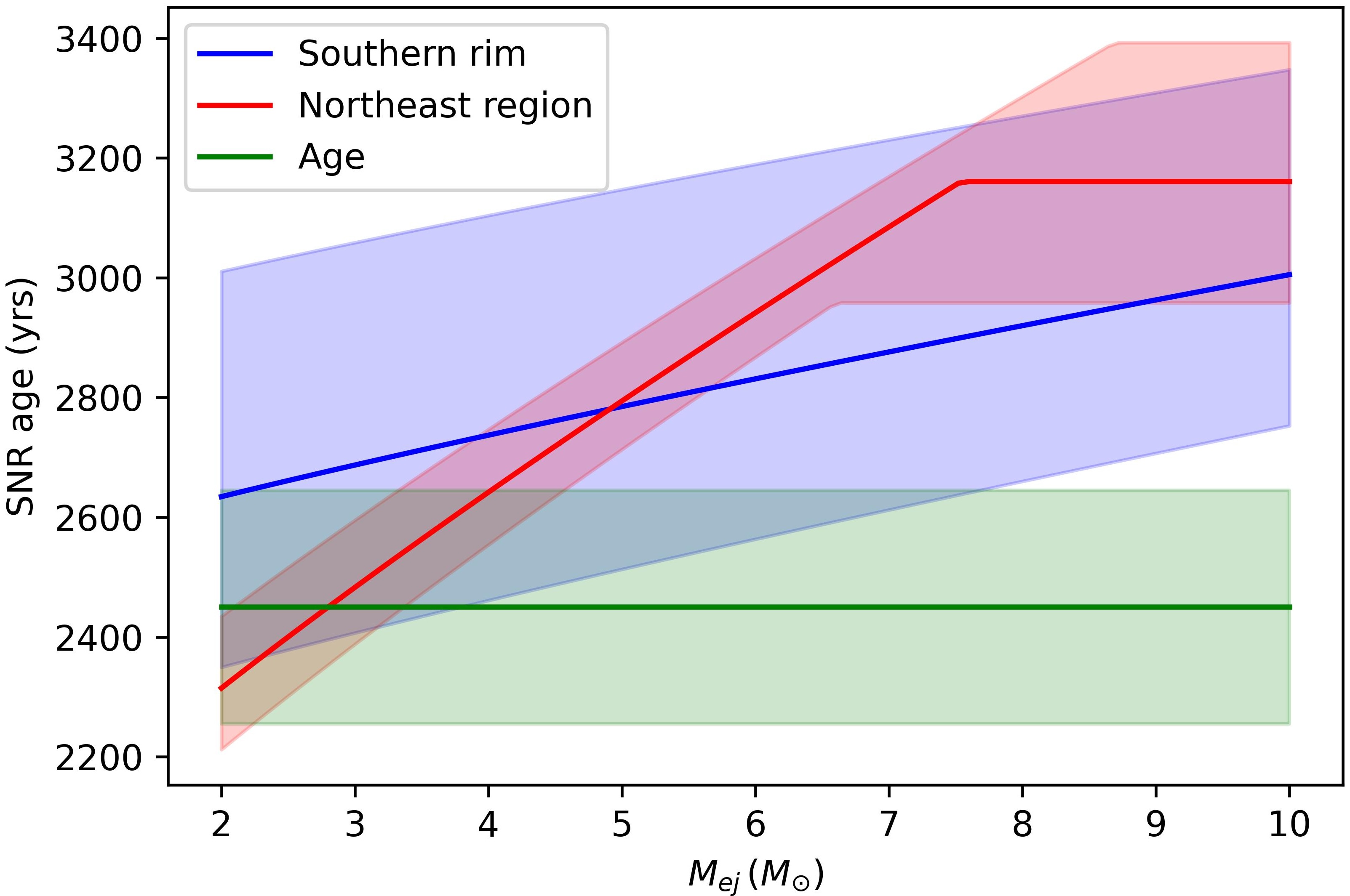}{0.48\textwidth}{(c) $E = 2.5 \times 10^{51}$ erg}
  \fig{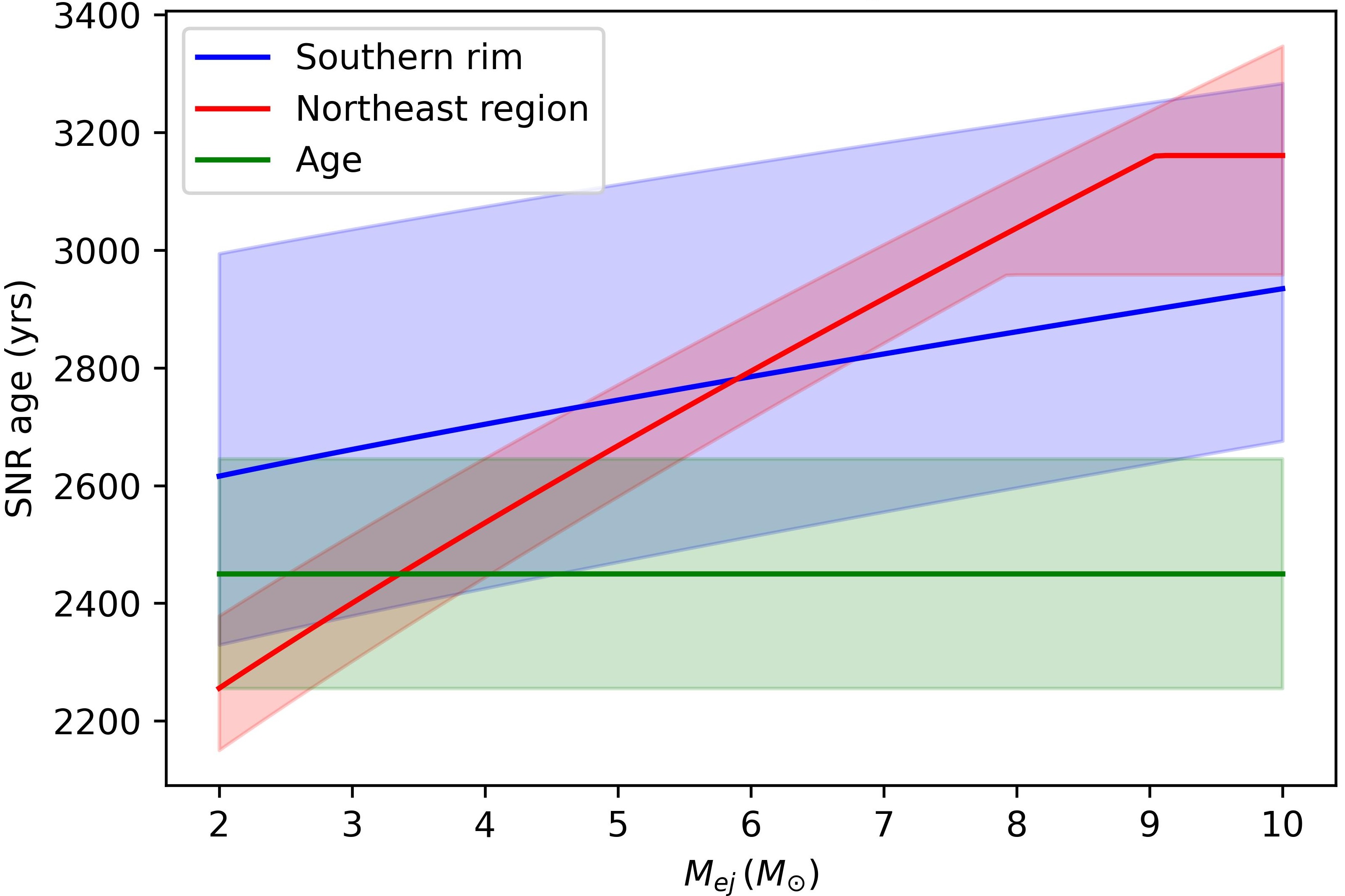}{0.48\textwidth}{(d) $E = 3.0 \times 10^{51}$ erg}
}
\caption{The 1-D model results of the age versus ejecta mass assuming explosion energies of 1.5 -- 3 $\times 10^{51} \mathrm{erg}$ (panels a-d) and an $s=0$ ambient medium profile
for the Southern rim and Northeastern regions. The green line and shaded area is the age estimated by \cite{law2020} from optical data. The blue/red line and shaded area are the results of the 1-D model for the Southern rim and for the Northeastern region respectively.
}\label{fig:1d.model}
\end{figure*}

An additional constraint is that the northeast region and the southern rim should have the same age since they originated from the same explosion.  
We calculated model results for the southern rim and northeastern regions for a grid of explosion energies, with energies from $E = 0.5\times 10^{51}\,\mathrm{erg}$ to $3.0\times 10^{51} \,\mathrm{erg}$ in steps of $0.5\times 10^{51}\, \mathrm{erg}$ and for ejecta masses ranging from $\mathrm {2-10~\msol}$ assuming an $s=0$ ambient medium profile.   
We found that explosion energies of $1.5-3.0\times 10^{51}\,\mathrm{erg}$ result in ages of the southern rim and northeastern regions that are consistent with each other and the age from the optical result \citep{law2020}, as shown in Figure~\ref{fig:1d.model}. Ejecta masses of $\mathrm {2-6~\msol}$ provide consistent solutions for both regions.
Models with explosion energies less than $1.5\times 10^{51}\, \mathrm{erg}$ do not provide an age that is consistent with the optical result. 
The ambient medium densities range from  $\sim0.33-0.66$ $\mathrm{amu~cm}^{-3}$  in the south and $\sim0.01-0.02$ $\mathrm{amu~cm}^{-3}$ in the northeast for these solutions.  These correspond to estimates of the swept-up mass of $\sim22.0 - 42.5\,M_{\odot}$.

Explosion energies larger than $3.0\times 10^{51}\ \mathrm{erg}$ and ejecta masses up to $\mathrm {\sim9~\msol}$  also provide consistent solutions given the relatively large uncertainty on the model ages. Our models can not rule out such a ``hypernova'' explanation for N132D but it would be difficult to explain the radius of  $R_{b}=10.4$~pc which is in good agreement with the expected size of a stellar-wind bubble created by a $15\pm5\,M_{\odot}$ star. 
 Furthermore, explosion energies larger than $3.0\times 10^{51} \,\mathrm{erg}$ are unlikely for normal CCSNe \citep{sukhbold2016,burrows2024} and ejecta masses larger than $\mathrm {10~\msol}$ are unlikely given that \cite{sharda2020} estimate a progenitor mass of $15\pm5\,M_{\odot}$.  
 We consider the normal CCSNe explanation more likely for N132D. 

We also calculated model results for the $s=2$ profile for explosion energies from $E = 0.5\times 10^{51}\,\mathrm{erg}$ to $3.0\times 10^{51} \,\mathrm{erg}$ in steps of $0.5\times 10^{51}\, \mathrm{erg}$ and for ejecta masses ranging from $\mathrm {2-10~\msol}$ and found that all of these solutions resulted in an age that is discrepant with the optical result.
Therefore, a consistent solution for the remnant age can be found for the southern rim and the northeast regions for explosion energies of $1.5-3.0\times 10^{51}\,\mathrm{erg}$ and ejecta masses of $\mathrm {2-6~\msol}$ for different, but constant density media in the south and the northeast.

\cite{2003ApJ...595..227C} applied their semianalytic model assuming the thin shell approximation for a shock crossing a density jump to N132D.  They adjusted the free parameters in their model (the density contrast at the jump and the evolutionary state of the shock relative to a Sedov solution) to match an age of 3,150~yr, an X-ray temperature of $T_x=0.8~\mathrm{keV}$, and an ambient density of $n_o=3.0~\mathrm{cm^{-3}}$ to derive an explosion energy of  $3.0\times 10^{51}~\mathrm{erg}$.  They suggest that the shock was propagating at $1,900~\kms$ before encountering the density jump and has decelerated to $800~\kms$ in the last 700~yr.  Our results agree broadly with their results in that an explosion energy larger than $1.0\times 10^{51} \mathrm{erg}$ is required to match the data but our measured shock velocity is significantly larger than theirs. We note that they did not consider Coulomb equilibration and adiabatic expansion in their comparison to the X-ray temperature after assuming a shock velocity and they assumed a density jump described by a square wave which is clearly a simplification.

\section{Summary} \label{sec:conclusions}

We have analyzed 878~ks of \chandra\ data of the LMC~SNR~N132D from two epochs separated by $\approx$14.5~years to measure the expansion of the forward shock. This measurement in X-rays can only be done with the high angular resolution and sensitivity of \chandra.
We carry out a comprehensive registration of the pointings using several point-like sources serendipitously detected around the remnant resulting in a relative astrometric precision of $1.8{\pm}3.2$~mas between the epochs (see Table~\ref{eq:regis_err}). We extract radial profiles and spectra from fourteen narrow regions ($\sim2.0\arcsec$) at the rim of the remnant that match the curvature of the shock front.  We measure the expansion of the forward shock by comparing intensity profiles between the two epochs (see Section~\ref{sec:exp_analysis}) at these fourteen locations.  
We measure an average proper motion of $0\farcs11\pm0\farcs02$ for eight regions from the bright southern rim and $0\farcs22\pm0\farcs02$ for two regions in the northeast rim, corresponding to velocities of $1850{\pm}390$~km~s$^{-1}$ and $3630{\pm}260$~km~s$^{-1}$ respectively.  We correct for a bias introduced by the fact that the shape of the \chandra\, PSF varies with azimuthal angle which affects comparisons of observations at different azimuthal angles. 
We also correct for an offset in the adopted COE assuming that the eight southern rim regions have a uniform expansion.  After making these corrections, 
 we estimate the shock velocity along the southern rim to be $1620\pm 400$~km~s$^{-1}$, and that of the northeast rim to be $3840{\pm}260$~km~s$^{-1}$ (see Figure~\ref{fig:expansioncorrectregistration} and Table~\ref{tab:proper.motion.ktav.ktp.kte}).

 We extract and fit spectra from the fourteen regions (see Section~\ref{sec:spec_analysis}) with a {\texttt{vpshock}} model that provided adequate fits with typical LMC abundances. The fitted electron temperature is $kT_{e,x}=0.95\pm0.17~\mathrm{keV}$ and $kT_{e,x}=0.77\pm0.04~\mathrm{keV}$ for the southern rim and northeast rim respectively.  After accounting for Coulomb equilibration and adiabatic expansion the electron temperature for the southern rim is consistent with the temperature one would infer from the measured shock velocity but the electron temperature for the northeast rim is significantly lower than what one would infer from the measured shock velocity.  We explore possible explanations for this discrepancy in Section~\ref{sec:temperature}, but none of these by themselves can explain the difference.  We suggest that this discrepancy warrants further investigation to determine if a combination of effects or effects which we did not consider could explain the discrepancy. We also note that we are limited by the quality of the data and future high angular resolution and high spectral resolution X-ray observations have the potential to resolve the discrepancy. Future missions such as the recently proposed {\em {Line Emission Mapper}} (LEM) \citep{kraft2024,orlando2024} would provide more than an order of magnitude improvement in spectral resolution albeit with worse angular resolution than \chandra, the {\em{New Athena}} mission \citep{cruise2025} would provide an order of magnitude improvement in effective area and spectral resolution but with worse angular resolution than \chandra, and the {\em{Lynx}} flagship X-ray observatory concept \citep{gaskin2019} would provide more than an order of magnitude improvement in effective area and spectral resolution with comparable angular resolution to \chandra. 

 We investigate the evolutionary state of the forward shock by comparing our measured shock radii and shock velocities to the 1-D SNR models of \cite{truelove99} assuming an age determined by \cite{law2020}. We further require that there be a consistent solution for the southern shock and northeast shock which have shock velocities and positions that differ by more than a factor of two. We find consistent solutions for explosion energies of $1.5-3.0\times 10^{51}\,\mathrm{erg}$ and ejecta masses of $\mathrm {2-6~\msol}$ for a constant ambient medium density of $0.33-0.66~\mathrm{amu\,cm^{-3}}$ for the southern rim and $0.01-0.02~\mathrm{amu\,cm^{-3}}$ for the northeast region.  Consistent solutions exist for larger values of the explosion energy and ejecta mass, however, we consider these solutions less likely given that the progenitor mass was most likely $15\pm5\,M_{\odot}$ as estimated by \cite{sharda2020} and \cite{foster2025}.  The relatively low and constant value of the ambient density is consistent with the cavity explanation in which the progenitor created a low density environment in or around the molecular cloud complex to the south. The elongated X-ray morphology to the north and the higher shock velocities in the north are consistent with shock propagation into a low density medium or a blowout opposite to the molecular cloud complex.  In summary, the \chandra\, expansion results are consistent with an energetic SN in a pre-existing cavity created by the progenitor.

\begin{acknowledgments}

Based on observations with the NASA Chandra X-ray Observatory operated by the Smithsonian Astrophysical Observatory.
This work was supported by NASA Chandra X-ray Center grants GO9-20068 and GO2-23045X.  P.P.P., T.J.G., V.L.K., and D.J.P\ acknowledge support from the Smithsonian Institution and the Chandra X-ray Center through NASA contract NAS8-03060. X.L.\ is supported by a GRF grant of the Hong Kong Government under HKU 17304524.
Support for C.J.L.\ was provided by NASA through the NASA Hubble Fellowship grant No. HST-HF2- 51535.001-A awarded by the Space Telescope Science Institute, which is operated by the Association of Universities for Research in Astronomy, Inc., for NASA, under contract NAS5-26555. D.M. acknowledges support from the National Science Foundation (NSF) through grants PHY-2209451 and AST-2206532.

\end{acknowledgments}

%

\vspace{15mm}
\facilities{NASA Chandra X-ray Observatory (CXO)}

\newpage 





\appendix
\section{Selecting Point Sources for Registration}\label{sec:app_src_selection}

{We carry out spatial registrations of the various observations to each other using point-like sources found serendipitously around the remnant (see Table~\ref{tab:source_list}).  These sources are primarily selected as having reliable position determinations and for also being present in specific \ObsID\ pairs (see the Point Source IDs column of Table~\ref{tab:registration}).}

\begin{figure}
    \centering
    \includegraphics[width=0.45\textwidth]{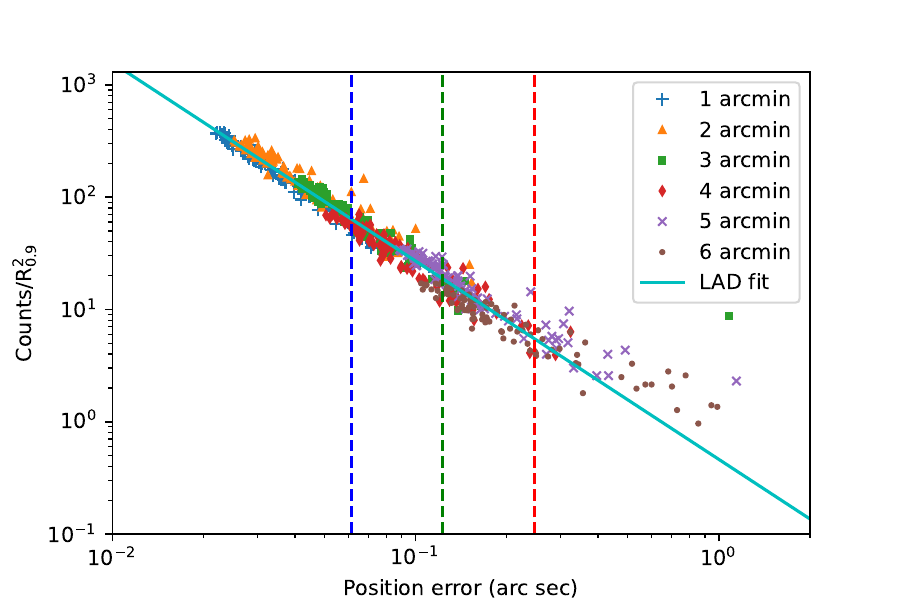}
    \caption{
    {Trend of position errors of spatial fits vs.\ brightness of simulated point sources.  Several point sources are placed at randomly selected azimuthal positions at different off-axis locations (see inset legend for point types corresponding to specific off-axis angles), with counts drawn from a uniform distribution ranging from 5 to 400, and are raytraced, and fitted.  The narrow trend is well-described by a power-law, which we fit using a least absolute deviation (LAD) fit, which is robust to outliers.  Vertical dashed lines mark $\frac{1}{8}$, $\frac{1}{4}$, and $\frac{1}{2}$ of the ACIS pixel size to illustrate scale.} 
    \label{fig:positionerror}}
\end{figure}

{We estimate the reliability of the position determination by carrying out spatial fits to the data using a raytraced PSF generated separately at each location, and using the position uncertainties as a determining factor for selection for registration.
We calibrate the quality of the spatial fitting by carrying out simulations of 100 
point sources each at several off-axis angles ranging from 1 to 9~arcmin, which covers the range of all detectable point sources, 
and fitting their locations for counts ranging from 5 to 400 in each.  We add a nominal background to the simulation that corresponds to that seen in \ObsID~5532.  The results of the fits for the simulated sources are shown in Figure~\ref{fig:positionerror}, which plots the effective brightness in each simulated source (obtained as the counts from a circular region centered on the best-fit location and within the 90\% enclosed counts fraction radius) and its corresponding position error, computed as the square-root of the sum of the squared errors along each axis. 
Note that as is expected, weaker sources have larger position errors.  Notice also that the values fall on a narrow locus, which is well-fit by a power-law function.  The deviations of each point from this fitted power-law is a useful diagnostic of the quality of the fit; we consider any source that falls within the 5\%-95\% range of the distribution of these deviations to be an adequate spatial position fit.}

\begin{figure}
    \centering
    \includegraphics[width=0.45\textwidth]{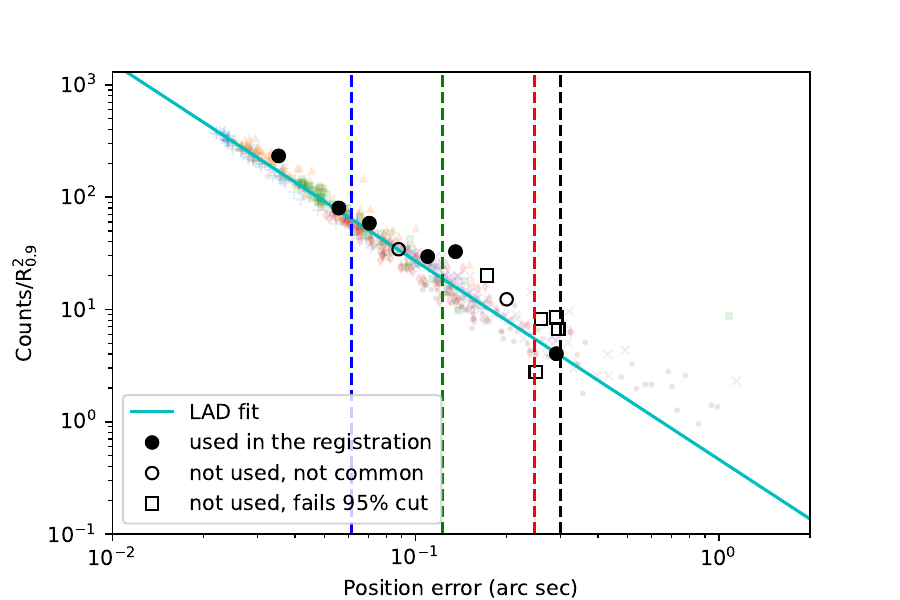}
    \caption{
    {Similar to Figure~\ref{fig:positionerror}, but an illustration of how sources are selected for registration, using the example of \ObsID~21365.  Only the detected sources with fitted position errors $<0.3$\arcsec\ (rightmost dashed vertical line) are shown.  Neither the sources that fall outside the 5\%-95\% range of the distribution of deviations from the power-law LAD fit (empty squares), nor sources that are not present in other observations (empty circles), are used in the registration.  The remaining sources are used for registration (filled circles; see Table~\ref{tab:source_list}).  All the simulation points from Figure~\ref{fig:positionerror} are also shown as grey shaded points.  A similar analysis is done for all \ObsID's.} }
    \label{fig:positionerrorexample}
\end{figure}

{We next identify sources in each observation using \texttt{wavdetect} \citep{2002ApJS..138..185F} and carry out independent spatial fitting, generating a separate raytraced PSF for each source location.  We exclude any source with position error $\ge$0.3\arcsec\ from further consideration, since deviations larger than that are unlikely according to aspect astrometric error.  This limit is shown for the case of \ObsID~21365 in Figure~\ref{fig:positionerrorexample} as the rightmost dashed vertical line.  From the sample of sources thus selected, we further exclude those which fall outside the 5\%-95\% bounds of the distribution of deviations (denoted as empty squares in the plot) obtained via simulations.  Among the remaining sources (denoted by circular symbols), only those that are also present in other {\ObsID}s (filled circles) are kept and those that are not (empty circles) are discarded.  This method is used for all {\ObsID}s, and the full list of sources used for registration are listed in Table~\ref{tab:source_list}, and the results of the registration based on these sources are in Table~\ref{tab:registration}.}
  
\section{QE Difference Normalization}
\label{sec:app_qe_difference}
\begin{figure}
    \centering
    \includegraphics[width=0.45\textwidth]{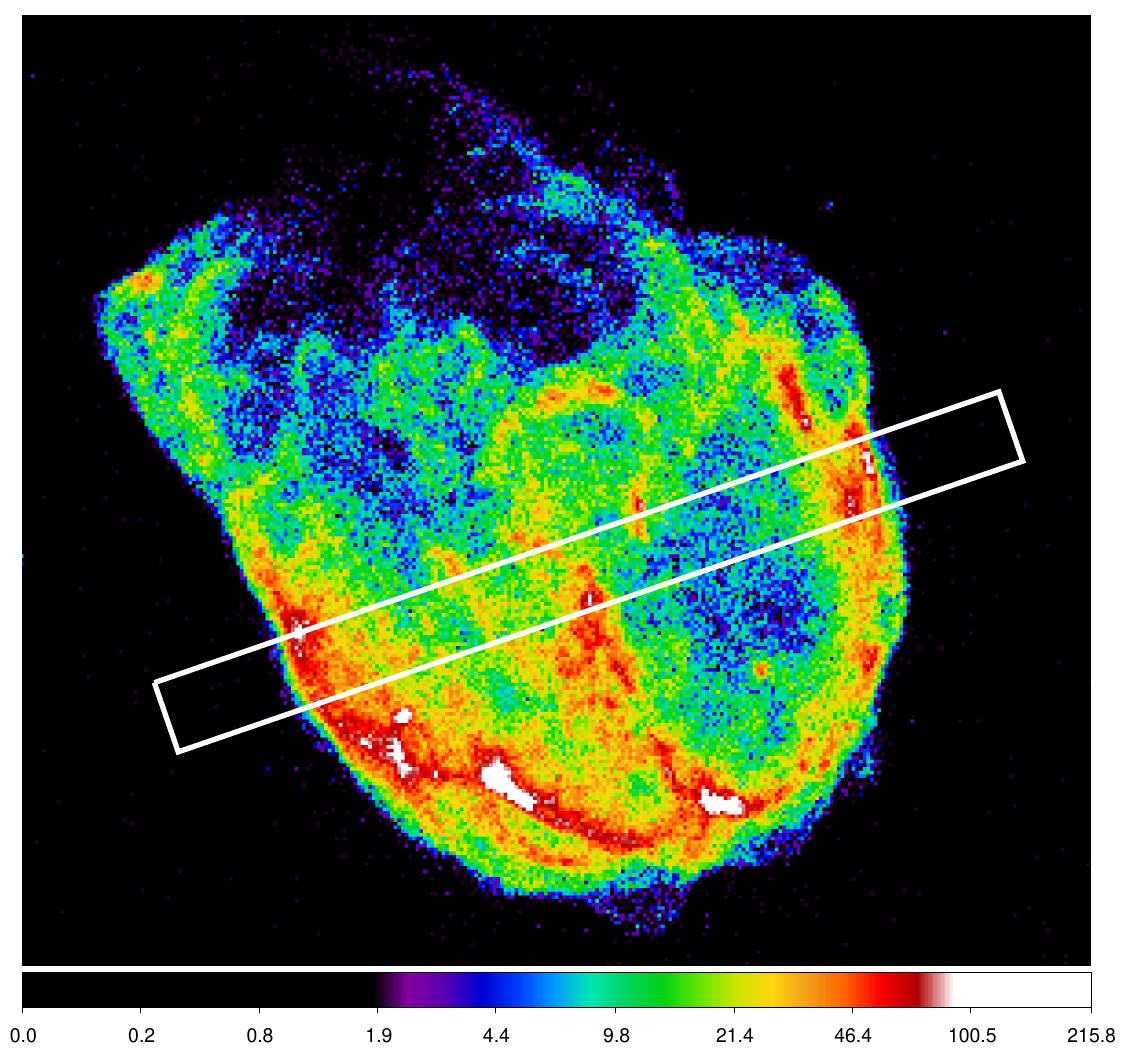}
    \caption{The white rectangle shows the profile extraction region for calculating the initial values of QE difference normalization, plotted on the merged \epocha observations (1.2-7\,keV) in ACIS ``sky'' coordinates.}
    \label{fig:rectangle}
\end{figure}

\begin{figure}
    \centering
    \includegraphics[width=0.45\textwidth]{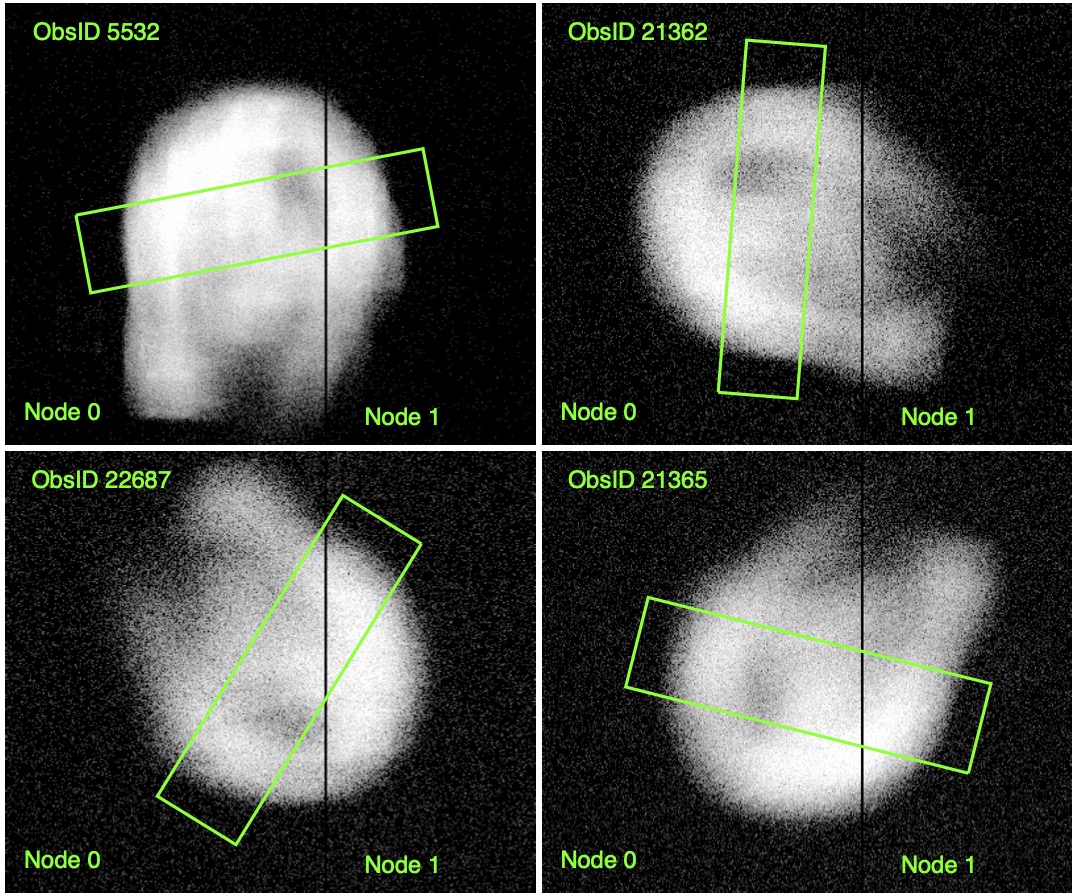}
    \caption{The rectangular profile extraction regions in detector coordinates (\texttt{CHIPX}, \texttt{CHIPY}) for representative ObsIDs from each of the four roll groups.  The upper left panel corresponds to \epocha. The remaining panels show the \epochb groups.
}
\label{fig:rollangle}
\end{figure}

We extracted profiles from a $120\arcsec \times 9.84 \arcsec$ east/west rectangular region  in the 1.2-7.0~keV band across the remnant (see Figure~\ref{fig:rectangle}). The rectangle was split into 732 thin $0.164\arcsec \times 9.84\arcsec$ strips. We extracted profiles for the merged \epocha observations; the three \epocha observations were at nearly the same roll. We extracted profiles for the individual \epochb observations. Because observations were at different rolls, the rectangular strip in physical ``sky'' coordinates covered different regions on the detector depending on the roll. The observations fell into four roll-angle groups (see Table~\ref{tab:obslist} for the roll angles and Figure~\ref{fig:rollangle} which shows a representative ObsID from each group). This resulted in three combinations of epoch and node id: \epocha node 0 and \epochb node 1 ($A0B1$), \epocha  node 0 and \epochb node 0 ($A0B0$), and \epocha node 1 with \epochb node 0 ($A1B0$).

\begin{figure}
    \centering
    \includegraphics[width=0.45\textwidth]{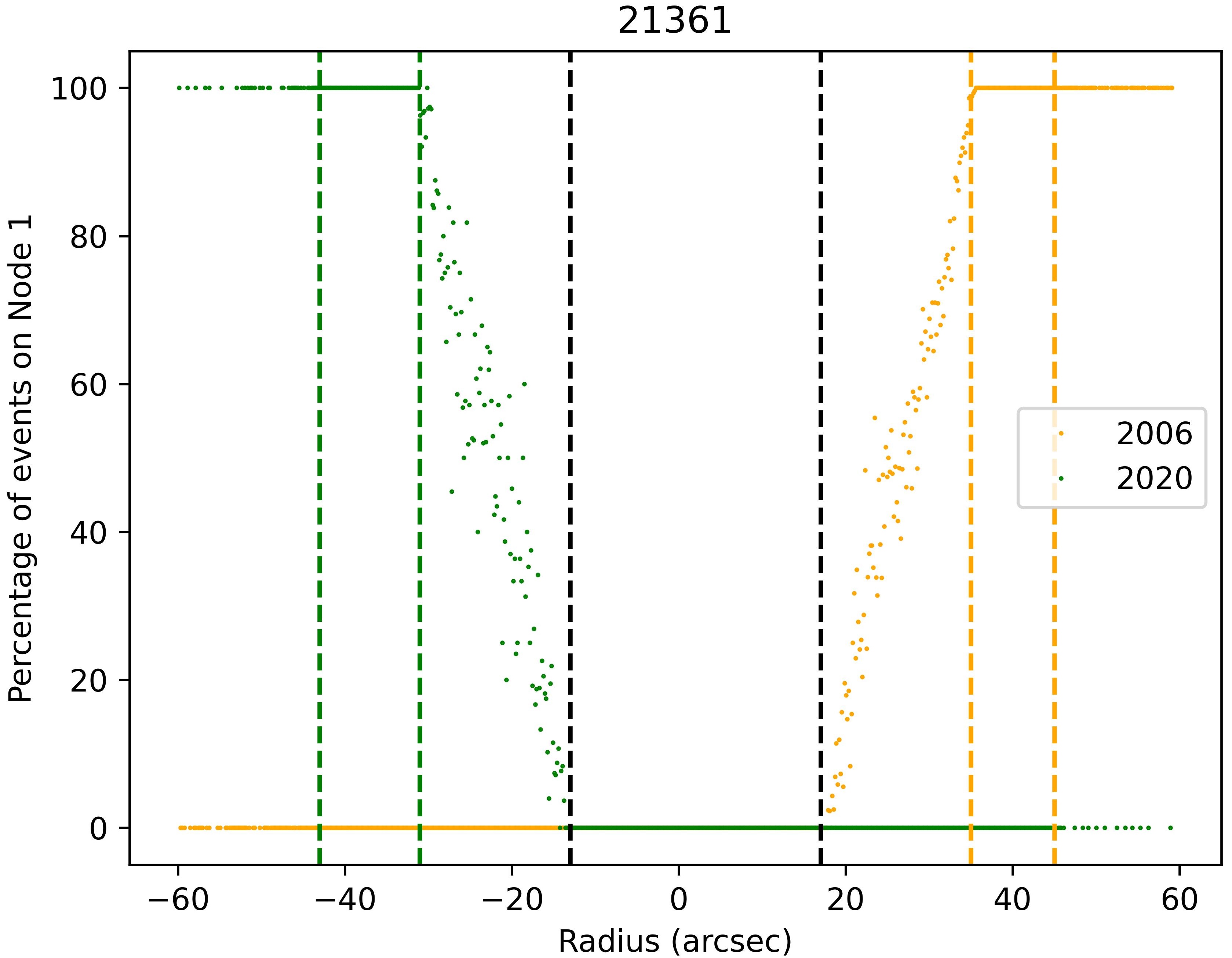}
    \caption{The percentage of events on node 1 for \epocha observations (2006) and one of the \epochb observations (ObsID 21361, 2020). The range between green dashed lines has events combination $A0B1$, the range between black dashed lines has events combination $A0B0$, and the range between orange dashed lines has events combination $A1B0$. \label{fig:nodeid}}
\end{figure}

\begin{figure}
    \centering
    \includegraphics[width=0.45\textwidth]{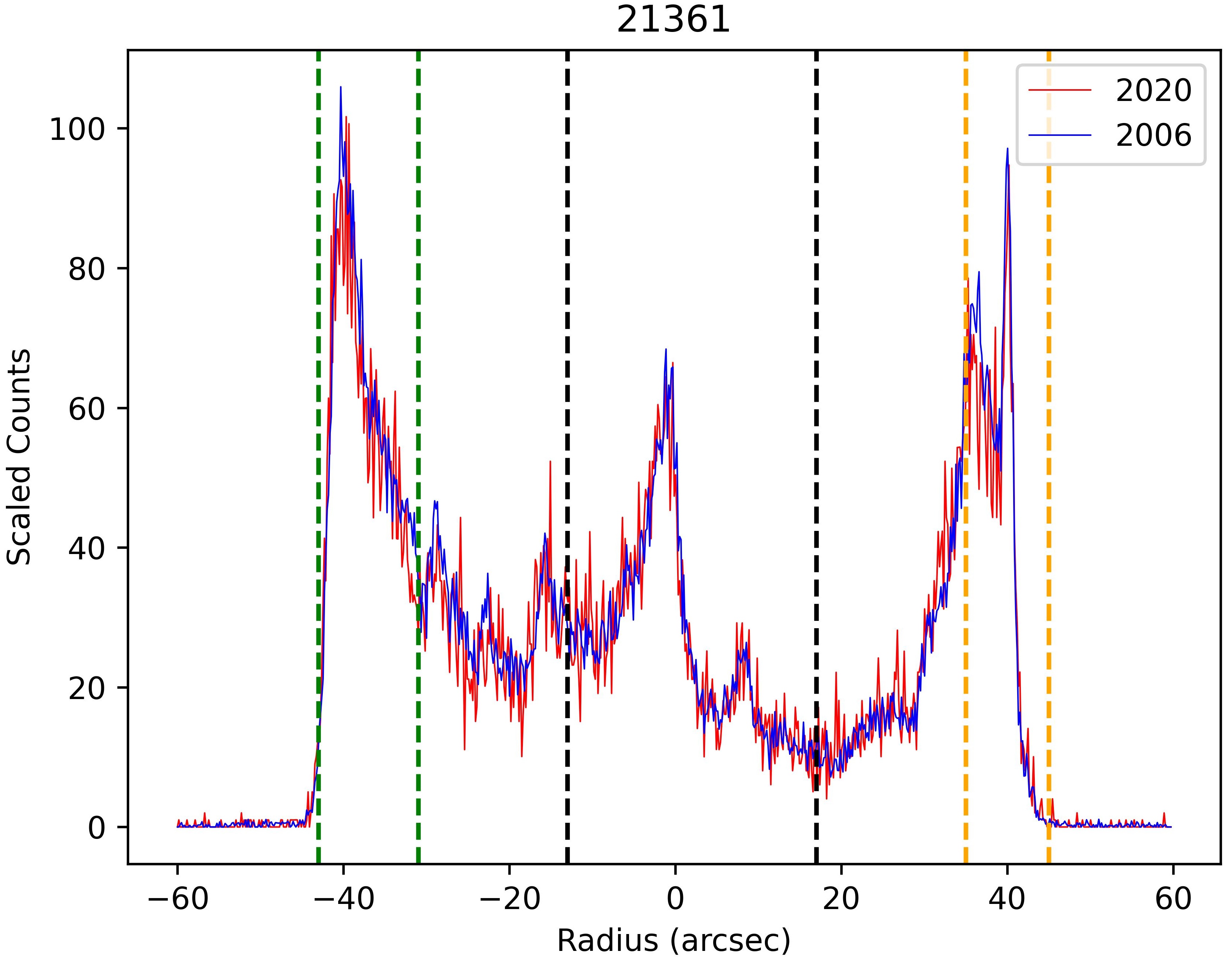}
    \caption{The QE and exposure time corrected profiles from \epocha observations (2006) and one of the \epochb observations (ObsID 21361, 2020). The range between green dashed lines has the events combination $A0B1$, the range between black dashed lines has the events combination $A0B0$, and the range between orange dashed lines has the events combination $A1B0$. \label{fig:profiles}}
\end{figure}

To calibrate the QE differences between different nodes and time epochs, we find regions within the rectangle containing only node 0 events or only node 1 events so that the QE for the $\langle epoch\rangle\langle node\rangle$ combinations can be examined (a given region of the remnant that is close to the node boundary can be observed on node 0 and node 1 in an observation because \chandra dithers during observations). Figure~\ref{fig:nodeid} displays an example of the distribution of counts between the nodes for the merged \epocha data and one ObsID (21361) from the \epochb data.
We accumulated the total scaled counts under the profile for the merged \epocha observations and for each of the \epochb observations. The profiles were corrected for the QE change relative to the reference time as discussed in section~\ref{sec:exp_analysis}. To correct for the exposure time differences, the profile of the \epocha observations were scaled by a factor $s(t_\mathit{A})=\tau_\mathit{B[j]}/{\tau_\mathit{A}}$. Here $\tau_\mathit{A}$ is the total exposure time for \epocha and $\tau_\mathit{B[j]}$ is the exposure for the $j$th observation of \epochb. The ObsIDs for \epochb (and the associated $j$ indices) are provided in Table~\ref{tab:qe_ratio}. 
Figure~\ref{fig:profiles} shows the profiles of \epocha observations and one of the \epochb observations (ObsID 21361).
The ratio of the accumulated scaled counts between the green dashed vertical lines and between the orange vertical lines gives the QE difference between the node and epoch combinations.   We make the assumption that the intrinsic flux from these regions of the remnant has changed by a negligible amount over the time interval considered.

For example, in Figures~\ref{fig:nodeid} and \ref{fig:profiles} , 
the region between the green vertical dashed lines has scaled counts for
\epochb, node 1 (observation $j$) and (merged) \epocha, node 0;
the total of the scaled counts in this region are $C_\textit{B1[j]}$ and
$C_\textit{A0}$.
The region between the orange vertical dashed lines has scaled counts for
\epochb, node 0 (observation $j$) and \epocha, node 1;
the total of the scaled counts in this region are 
$C_\textit{B0[j]}$ and $C_\textit{A1}$.
Finally, the region between the black vertical dashed lines has scaled 
counts for 
\epochb, node 0 (observation $j$)
and (merged) \epocha, node 0;
the total of the scaled counts in this region is $C_\textit{B0[j]}$ and
$C_\textit{A0}$.
From these total counts, we can evaluate the relative QEs as
$\qeratio{B1[j]}{A0} = C_\mathit{B1[j]} / C_\mathit{A0]}$,
$\qeratio{B0[j]}{A1} = C_\mathit{B0[j]} / C_\mathit{A0]}$, and
$\qeratio{B0[j]}{A0} = C_\mathit{B0[j]} / C_\mathit{A0]}$.
This procedure is carried out for each of the \epochb observations.
We list the QE ratios for each  \epochb observation in Table~\ref{tab:qe_ratio}.

For some \epochb observations, the profile did not have any range with events only from node 1 (observation indices $j=1$ to 5; see Table~\ref{tab:qe_ratio}).  We therefore use QE ratios from observations which have data for all three combinations: $B1A0$, $B0A0$, and $B0A1$. The average QE ratios are calculated as averages over the nine observation indices $j=6$ to 14: $\qeratio{x}{y} = (1/9) \sum_{j=6}^{14} \qeratio{x[j]}{y}$ where $x$, $y$ are $B0$, $B1$, or $A0$. The resulting average QE ratios are given in the second part of Table~\ref{tab:qe_ratio}.

To correct the QE difference to the same level, we chose the averaged node 0, \epochb as the reference. 
The QE difference normalizations from the rectangle region were used to scale the weight for each events according to the node ID and epoch.
The correction factors range from $\sim3\%$ to $\sim7\%$ indicating that the QE is lower in \epochb than the current calibration files estimate.

\begin{table}[htb]
\caption{QE ratio between node 0 and node 1 at different epochs, and QE ratios averaged over $j=6$ to 14.}
\begin{center}
\label{tab:qe_ratio}
\resizebox{\columnwidth}{!}{%
\begin{tabular}{l c | c c c }
\hline\hline
ObsID(\phantom{0}j) & Roll
    & $\qeratio{B1[j]}{A0}$ 
    & $\qeratio{B0[j]}{A0}$ 
    & $\qeratio{B0[j]}{A1}$ \\
\hline 
21362(\phantom{0}1) & 255.62 &  & $0.956\pm 0.017$ & $0.892\pm 0.024$\\
22094(\phantom{0}2) & \phantom{0}93.15 &  & $0.900\pm 0.040$ & $0.897\pm 0.023$\\
22841(\phantom{0}3) & \phantom{0}93.15 &  & $0.927\pm 0.041$ & $0.908\pm 0.023$\\
22687(\phantom{0}4) & 102.66 &  & $0.923\pm 0.042$ & $0.887\pm 0.023$\\
21363(\phantom{0}5) & 105.15 &  & $0.956\pm 0.021$ & $0.904\pm 0.021$\\
21361(\phantom{0}6) & 160.14 & $0.956\pm 0.021$ & $1.000\pm 0.020$ & $0.930\pm 0.026$\\
23317(\phantom{0}7) & 160.14 & $0.952\pm 0.021$ & $0.967\pm 0.019$ & $0.920\pm 0.022$\\
21365(\phantom{0}8) & 175.14 & $0.966\pm 0.018$ & $0.974\pm 0.016$ & $0.960\pm 0.019$\\
21884(\phantom{0}9) & 183.14 & $0.971\pm 0.026$ & $0.949\pm 0.019$ & $0.931\pm 0.022$\\
23044(10) & 175.15 & $0.946\pm 0.019$ & $0.965\pm 0.016$ & $0.950\pm 0.020$\\
21886(11) & 175.14 & $0.983\pm 0.021$ & $0.963\pm 0.018$ & $0.910\pm 0.021$\\
21883(12) & 178.14 & $0.947\pm 0.019$ & $0.952\pm 0.021$ & $0.931\pm 0.025$\\
21882(13) & 178.14 & $0.999\pm 0.020$ & $0.988\pm 0.020$ & $0.919\pm 0.024$\\
21887(14) & 183.14 & $0.962\pm 0.019$ & $0.950\pm 0.016$ & $0.907\pm 0.020$\\
\hline \\
\hline\hline
\multicolumn{2}{l|}{}
& \qeratio{B1}{A0} 
& \qeratio{B0}{A0} 
& \qeratio{B0}{A1} \\ \cline{3-5}
\multicolumn{2}{l|}{\raisebox{1.5ex}[0pt]{Averages ($j=6..14)$}}
& $0.964\pm 0.007$ 
& $0.967\pm 0.006$ 
& $0.929\pm 0.007$ \\ \hline
\end{tabular}
}
\end{center}
\end{table}

\section{Accounting for PSF and COE Bias}\label{sec:app_psf_diff} 
\begin{figure}
    \centering
    \includegraphics[width=0.45\textwidth]{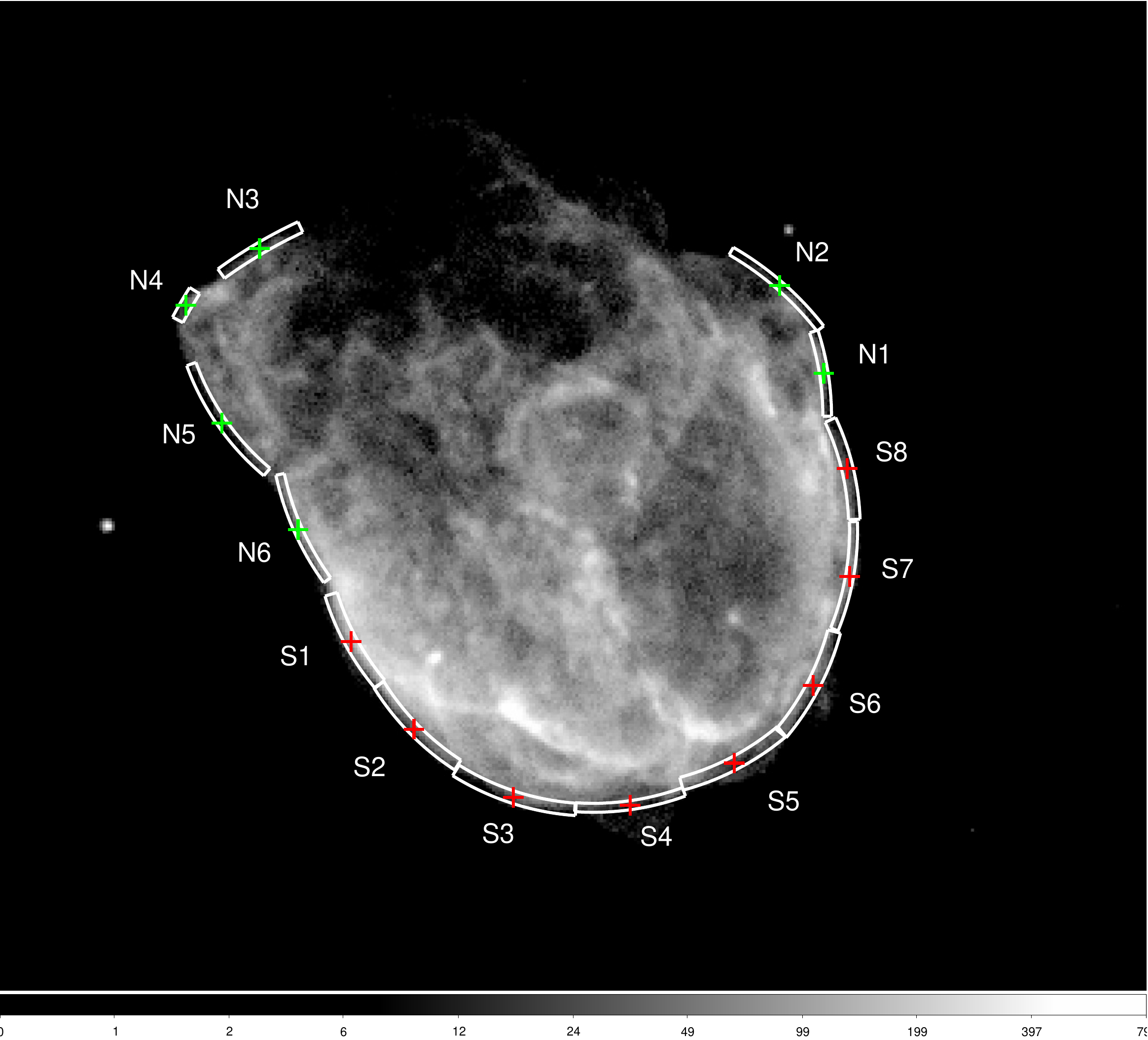}
    \caption{The positions for the PSF simulations for the S1-S8 and N1-N6 regions. The red and green crosses are the PSF simulation centers and match the region color coding in Figure~\ref{fig:expansionpsf}. The white sectors are the shock regions where we measure the expansion (see Table~\ref{tab:expansion_regions}). }
    \label{fig:shockpsf} 
\end{figure}

\begin{figure}
    \centering
    \includegraphics[width=0.45\textwidth]{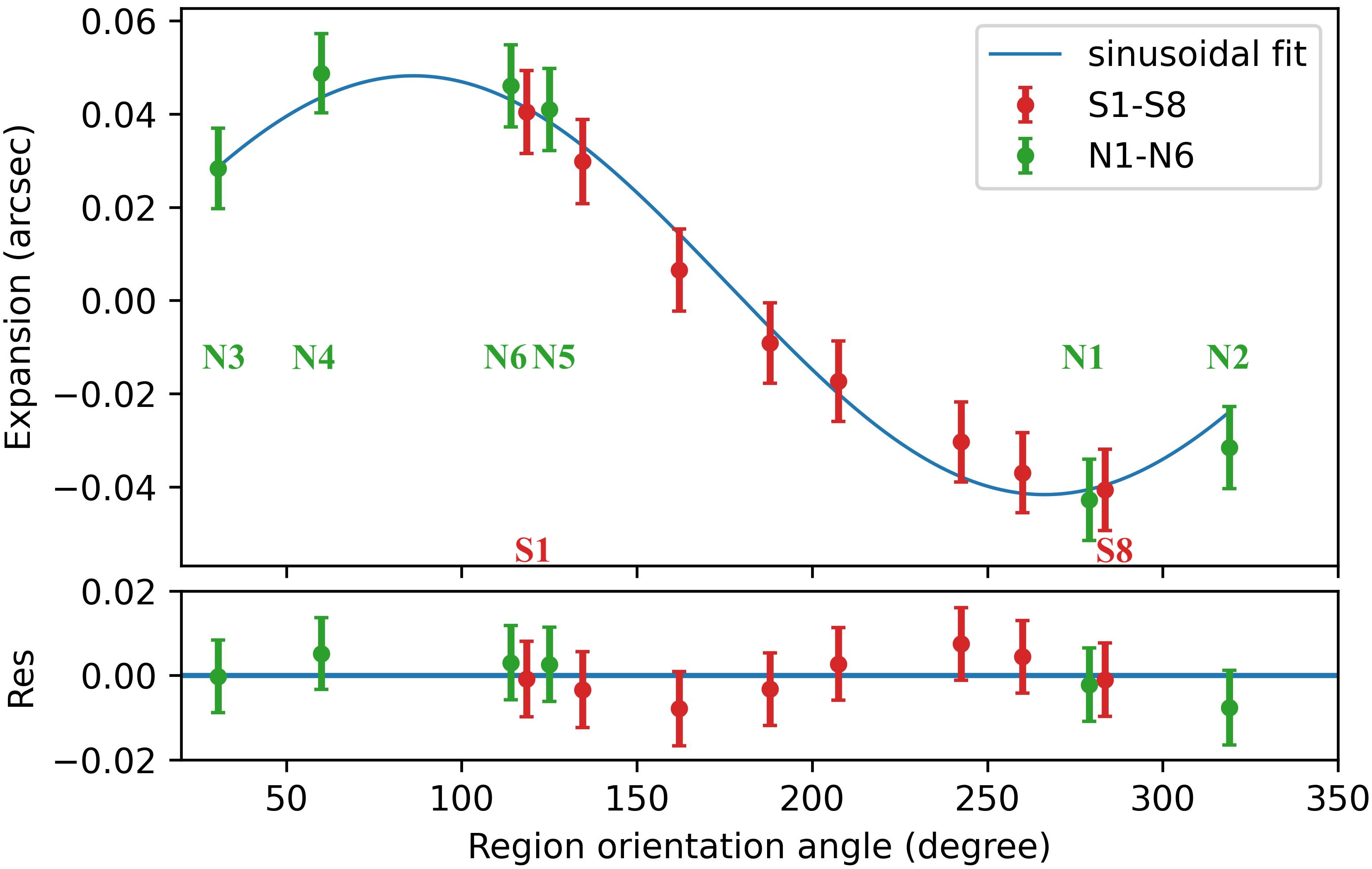}
    \caption{Estimate of the systematic bias in the expansion measurement due to PSF differences.  The data points are arranged similarly to Figure~\ref{fig:expansionlinear} with regions S1-S8 labeled sequentially from left to right. The apparent expansion derived from PSF simulations with the appropriate off-axis and azimuthal angles of the \epocha\ and \epochb\ observations of a non-moving shock front at the locations marked in Figure~\ref{fig:shockpsf} are shown, along with the sinusoidal fit (solid blue line) to all the points.  This sinusoidal variation is attributable to the ``PSF-bias''.}
    \label{fig:expansionpsf}
\end{figure}

\begin{figure}
    \centering
    \includegraphics[width=0.45\textwidth]{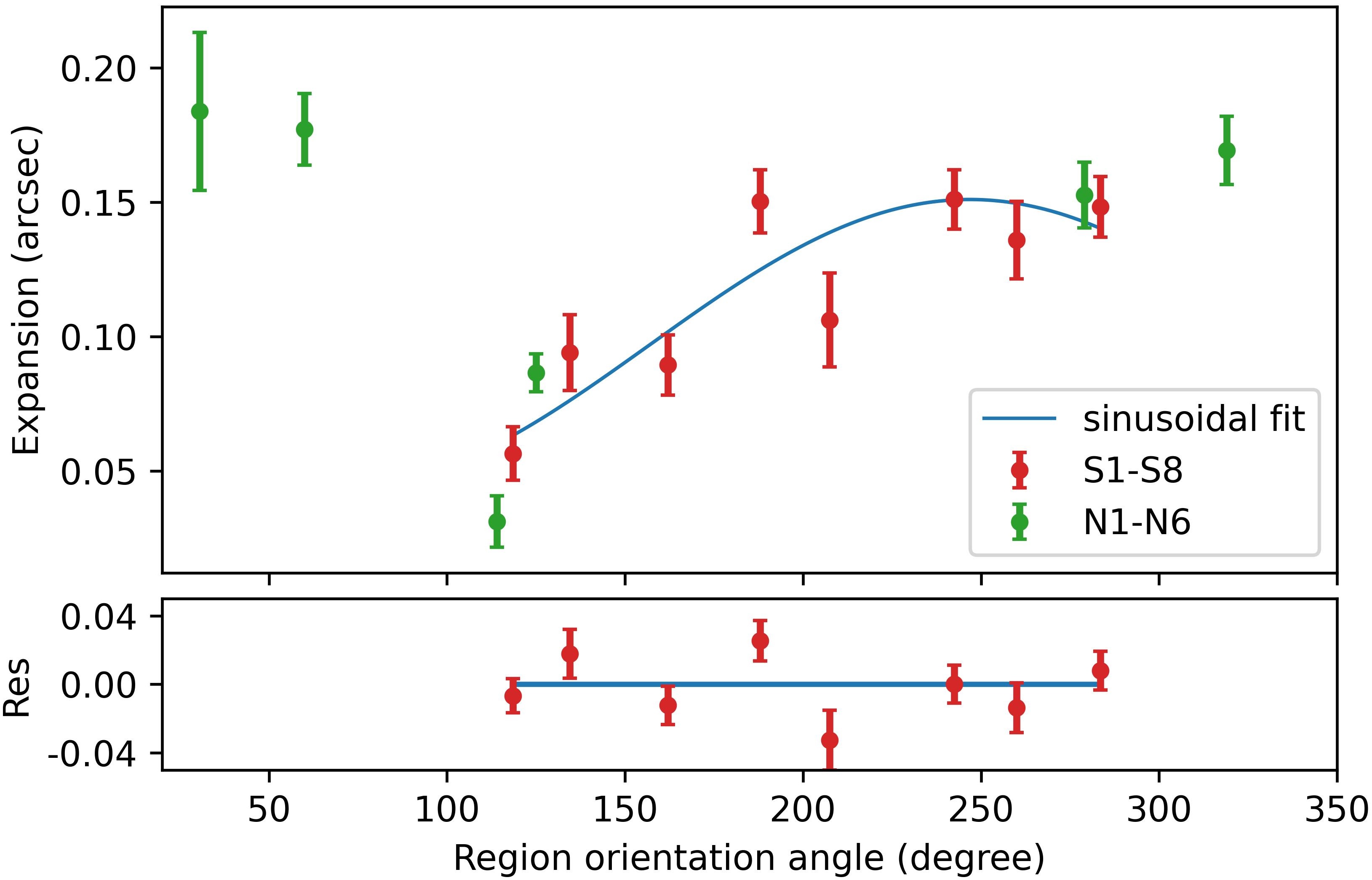}
    \caption{
    Estimate of the systematic bias in the expansion measurement due to an error in the assumed COE. 
    The apparent expansion from Figure~\ref{fig:expansionpsf} is subtracted from the measured raw expansion shown in Figure~\ref{fig:expansionlinear}, and a sinusoid is fit to account for an offset in the assumed expansion center (see Equation~\ref{eqn:sinusoid}) (solid blue line).  The final estimate of the expansion in Figure~\ref{fig:expansioncorrectregistration} accounts for this offset.
    \label{fig:expansioncorrectbias}}
\end{figure}

Based on PSF simulations, we find that differences in the PSF as a function of off-axis and azimuthal angle between the \epocha and \epochb observations could result in an apparent ``shift'' of the forward shock.  
This is the result of the same region of the remnant being observed at different azimuthal angles in \epocha and \epochb.
Given the small range in off-axis angles (less than $1.0\arcmin$) of the regions considered in this analysis, the variation in azimuthal angle is the dominant effect.
We investigate these systematics 
by using \saotrace and \marx to simulate PSFs at the centers of the shock regions as shown in Figure~\ref{fig:shockpsf}. The simulated PSFs were convolved  with a 2-D theoretical shock model, based on a line-of-sight 
projection through a spherical rim ``cap'',
to simulate the observed shock front. We extracted a radial profile in the 1.2--7~keV passband of simulated non-moving shock fronts using the merged \epocha and merged \epochb events lists based on the PSF simulations separately,
then measured the apparent displacements in the shock profiles using the same method described in Section~\ref{sec:exp_analysis}. The ``expansions'' resulting from the PSF bias are shown in Figure~\ref{fig:expansionpsf}, displaying a sinusoidal-type behavior with region orientation angle. Such a sinusoidal behavior with azimuthal angle is expected given the PSF aberration of the \chandra mirrors, and we therefore fit these data with a sinusoid to determine the bias as a function of angle.
To correct the expansion for PSF bias, we subtract the fitted model values shown in 
Figure~\ref{fig:expansionpsf} from the raw shifts shown in Figure~\ref{fig:expansionlinear} to obtain the PSF-bias-corrected expansion in Figure~\ref{fig:expansioncorrectbias}.

We also explore the possibility that our assumed COE introduces a bias in our expansion measurement. 
Under the assumption that the southern rim is expanding uniformly, any error in the assumed COE would introduce residuals with a sinusoidal pattern. If that is the case, a correction to the expansion center can be obtained by fitting the eight southern regions with a sinusoid (equivalent to an offset eccentric circle):
\begin{equation}
\delta R_\mathit{n} = \Delta X \sin\theta_\mathit{n}
         - \Delta Y\cos\theta_\mathit{n}
         + \Delta R_\mathit{S}
 \label{eqn:sinusoid}                     
\end{equation}
where $\delta R_\mathit{n}$ are the 
expansions of each of the southern regions (corrected for the PSF bias effect; see Figure~\ref{fig:expansioncorrectbias}), and $\theta_\mathit{n}$ are the region orientation angles (see Table~\ref{tab:expansion_regions}).  The parameters $\Delta X$ and $\Delta Y$ define the shift of the COE in the RA and Dec directions respectively, and
$\Delta R_\mathit{S}$ defines the average expansion of the southern regions.
Fitting such a model to the data in Figure~\ref{fig:expansioncorrectbias} results in a shift in RA of $\Delta X = -0\farcs05{\pm}0\farcs007$ and Dec of $\Delta Y = 0\farcs02{\pm}0\farcs012$, with $\Delta R_\mathit{S} = 0\farcs097{\pm}0\farcs008$.  The COE given in Table~\ref{tab:expcenter} includes the corrections $(\Delta X, \Delta Y)$.  We provide a conservative assessment of the errors on $\Delta R_\mathit{S}$ in Table~\ref{tab:expansion_result} accounting for the uncertainty in estimates of each $\delta R_\mathit{n}$ as well as the scatter therein.  
The COE correction is applied to each of the regions 
in Figure~\ref{fig:expansioncorrectbias} 
as $\Delta X \sin\theta_\mathit{n} - \Delta Y\cos\theta_\mathit{n}$, and the result is shown in
Figure~\ref{fig:expansioncorrectregistration}.

\newpage
\bibliography{n132d_exp}{}
\bibliographystyle{aasjournal}



\end{document}